\documentclass[10pt]{article}
\usepackage{amsmath,amsthm,latexsym,amssymb,amsfonts,epsfig,psfrag}
\usepackage{subfigure}
\usepackage{tikz}
\usepackage{graphicx,overpic}

\def\be{\begin{equation}}
\def\ee{\end{equation}}
\def\ba{\begin{eqnarray}}
\def\ea{\end{eqnarray}}

\newcommand\nn{\nonumber}
\newcommand\q{\quad}

\newcommand{\cc}{\mathcal C}

\newcommand{\sig}{\sigma}

 \newcommand{\Fj}[6]{  \begin{bmatrix}
  #1 & #2 & #3 \\
  #4 & #5 & #6 
 \end{bmatrix} }

  \newcommand{\Tj}[6]{ \begin{Bmatrix}
  #1 & #2 & #3 \\
  #4 & #5 & #6 
 \end{Bmatrix} }

\title{Topological lattice field theories from intertwiner dynamics}
\author{Bianca Dittrich$^1$,  Wojciech Kami{\'n}ski$^{1,2}$\\
\small  $^1$ Perimeter Institute for Theoretical Physics,\\
 \small 31 Caroline Street North, Waterloo, Ontario, Canada N2L 2Y5,\\
 \small $^2$ Instytut Fizyki Teoretycznej, Uniwersytet Warszawski,
  ul. Ho\.{z}a 69, 00-681 Warszawa, Poland
}

\date{}

\begin{document}

\maketitle

\begin{abstract}

We introduce a class of 2D lattice models that describe the dynamics of intertwiners, or, in a condensed matter interpretation, the fusion and splitting of anyons. We identify different families and instances of triangulation invariant, that is, topological, models inside this class. These models give examples for symmetry protected topologically ordered 1D quantum phases with quantum group symmetries. Furthermore the models provide realizations for  anyon condensation into a new effective vacuum. We explain the relevance of our findings for the problem of identifying the continuum limit of spin foam and spin net models.

\end{abstract}



\section{Introduction}

We introduce a class of 2D lattice models, describing the dynamics of (quantum) group intertwiners, and present families of  triangulation invariant, that is topological,  models inside this class. The models, and the investigations of their fixed point structure under coarse graining, are motivated by a research program to understand the phase diagram and continuum limit of spin foam models \cite{bdetal13}, which are candidates for a quantum theory of gravity \cite{alexreview,cbook}. There are however several additional tantalizing connections to quantum gravity as well as other areas of physics. One is the theory of anyon condensation \cite{bais1,bais2}, for which we provide Hamiltonians. The condensate states appear as ground states of these Hamiltonians and are given by the topological models. Finding all possible topological models leads to a classification problem of module categories over the category of representations of a quantum group \cite{kirillov,kong, bw}. This connects to the task of classifying phases for condensed matter \cite{wenclass,schuch}, for which we here provide a broad class of examples based on quantum group symmetries.

Besides these topics there are other reasons why these models and its fixed points are of interest for quantum gravity:

\begin{itemize}\parskip -1mm
\item For the construction of spin foam models itself, in particular the intertwiners defining these models. 
The fixed points of our models define naturally intertwiners for spin foam vertices of arbitrary valency. Such a connection between fixed points 
and a consistent construction of spin foam vertices for arbitrary valence has been first pointed out by Reisenberger in \cite{reisenberger} in connection with the Barrett--Crane model \cite{bc}
and will be explored in detail in the forth coming work \cite{toappearBMS}. 
\item The fixed points for the intertwiner models will also determine fixed points for spin net models, introduced and investigated in \cite{finite,eckert,bahretal12,bdetal13}. The motivation for these models is the construction of analogue models for spin foams. The hope is that statistical properties of 2D spin nets and 4D spin foams are similar. This is based on a similar property for 4D lattice gauge models and related  edge (Ising like) 2D models, and the fact that spin foams can be seen as generalized lattice gauge models, \cite{bahretal12} and references therein.

\item  
The 2D triangulation invariant models constructed in this work also allow for a geometrical interpretation of the underlying variables. Similar to the 3D quantum gravity models these models can be used to study and illustrate conceptual questions such as  a notion of diffeomorphism invariance in the discrete \cite{bd08,bd12a}, uniqueness of such diffeomorphism invariant models, the relation between covariant and canonical formalism and a derivation of Hamiltonian constraints, a dynamical notion of cylindrical consistency \cite{bd12b} and the expansions of theories around different vacua (here fixed points). More precise relations of the mentioned topics to the models we introduce here will emerge in the course of this work.

\item We introduce techniques to construct triangulation invariant models via recursion relations. In 2D these recursion relations are derived from the 2--2 Pachner move invariance or crossing symmetry. Furthermore we will argue that the 2--2 Pachner move invariance leads to Hamiltonian and Diffeomorphism constraints in these 2D models.  
We believe that these techniques can be also applied to higher dimensional models. The technique illustrates nicely how larger building blocks are constructed from some basic (smallest) building block using the principle of triangulation invariance. In this sense the microscopic theory determines macroscopic physics even for single building blocks.   
Furthermore such recursion relations lead to the Hamiltonian constraints \cite{cranebarrett,bonzomfreidel}. The methods here are restricted to triangulation invariant models with local couplings only and therefore  topological field theories. We however believe that the recursion relations, representing (the Ward identities of) diffeomorphism symmetry, admit a generalization to theories with propagating degrees of freedom.
\end{itemize}

In the following section we will elaborate on the relation to spin net models and comment more in sections \ref{interpretation},\ref{condense} and the discussion section \ref{discuss} on the relation to anyon condensation, the classification of phases and the role of diffeomorphism symmetry. The reader not interested in spin net models can directly go to the next section \ref{models}, where the models will be introduced.

This section will also specify the fixed point conditions -- which  leads here to the  requirement of  triangulation invariance of the model. We will express this requirement as conditions on the amplitudes or weights of the model.  From these conditions we  derive in section \ref{recursion} recursion relations for the amplitudes, for which we find a family of solutions in section \ref{solving}. In section \ref{examples} we apply an alternative method to find fixed point amplitudes and give additional examples. 

We then explore the physical interpretation of these fixed point models, first by defining and computing the partition function of the torus in section \ref{torus}. This will give the number of ground states of the Hamiltonians associated to a given fixed point model.
In section \ref{cdl} we in particular reconsider the fixed point models obtained by solving the recursion relations and show that these models originate from a so called  Corner Double Line  structure or valence bond construction. We comment on the general structure of the fixed point amplitudes, which in particular depends on the ground state degeneracy,  and relate to the classification of phases for 1D  quantum systems (that is $(1+1)$ space time dimensions). 

The results of section \ref{cdl} allow us to interpret this particular family of fixed point models as a boundary theory of a 3D topological model (the Tuarev--Viro model) in section \ref{3Drecursion} and moreover to relate our 2D recursion relations to 3D recursion relations, obtained from the Biedenharn--Elliot identity for these models. This makes the notion of diffeomorphism symmetry obvious, as the 3D recursion relations are the Ward identities of this symmetry for the 3D models.

Section \ref{interpretation} explains the notion of matrix product states and specifies the matrix product state representation of the ground states associated to the fixed point models as well as the Hamiltonians. In quantum gravity language this provides the link between the covariant and canonical description of the models, with the partition function providing the projector on the physical states (i.e. ground states), which are specified by the Hamiltonian (constraints).  Here the problem to understand the symmetry of the ground states motivates the consideration of a notion of finite subgroups in a  quantum group. 

This connects to anyon condensation, which we will discuss in section \ref{condense}. 
 It will explain some of the results we found for the torus partition function, especially for the fixed point models with ground state degeneracy. We finally close with a discussion and outlook in section \ref{discuss}.

\section{Spin net models}\label{spinfoams}

Spin foams \cite{alexreview,eprl,cbook} provide a proposal for a non--perturbative path integral for quantum gravity, based on fundamental building blocks. A key open question \cite{alexreview} to validate this proposal is the continuum limit of the models, which we here understand as the limit including a large number of these building blocks. The complexity of the models made progress difficult, in addition coarse graining and renormalization had to be adapted to a background independent framework \cite{fotini}. In \cite{finite,eckert,bahretal12,bdetal13} a program was started to tackle the problem of the continuum limit first in simpler analogue models.  This  helped to develop tools and techniques and most of all approximations and truncations, that can be tested on increasingly more complicated models. In this way we can hope to build up  to the full models. Indeed, in this work we will consider models which in their algebraic structure (i.e. the structure group) almost reach the full models.

Spin net models have been introduced in \cite{finite,eckert,bdetal13} as analogues to spin foams. They can be interpreted as dimensional reductions of spin foams and for this reason are much easier to investigate, in particular with numerical techniques, as done in \cite{bdetal13}. The results in \cite{bdetal13} motivate us to introduce a further variant of these models in this work, which indeed is closely related to the so called spin net evaluation and the definition of the spin foam dynamics  via these spin net evaluations prescribing intertwiner degrees of freedom \cite{bc,reisenberger}. 

Here we will define this class of intertwiner models and present a large family of triangulation invariant models inside this class.

Spin foams and spin nets are based on some structure group, which for the full models is taken to be $SU(2)\times SU(2)$ or $SL(2,\mathbb{C})$. This however precludes numerical investigations, as the models based on such groups involve infinite summations and potential infinities \cite{riello}. This led to the introduction of finite group models in \cite{finite,eckert,bdetal13}. The use of the group structure simplifies the definition of the so called simplicity constraints, central to the dynamics of spin foams, for the analogue models. A parametrization of possible simplicity constraints can be found in   \cite{bahretal12}. However, there does not exist e.g.  a family of subgroups of the rotation group which would allow to reach the full models in a limiting procedure\footnote{One could consider $SL(2,\mathbb{Q}_p)$ but here a geometrical interpretation is not obvious.}. 

For this reason we introduce in this work intertwiner models based on the quantum group $SU(2)_k$. Here $k\in \mathbb{N}$ denotes the so--called level of the quantum group at root of unity. These quantum groups have a finite number (namely $(k+1)$) of irreducible finite dimensional representations with non--vanishing quantum trace. We will see however that some of the results generalize to the classical group as well as to the quantum groups $SU(2)_q$ with deformation parameter $q$ real. 

The corresponding spin net models will be defined in \cite{toappearBMS}. In the full theory such quantum group models are argued to lead to general relativity with a cosmological constant \cite{noui}. Thus we deal almost with the same algebraic structures as in the full theory, which would involve $SU(2)_k \times SU(2)_k$ and related groups.  The replacement of groups with quantum groups opens up the question of how to describe the simplicity constraints. We will propose a way to construct models with simplicity constraints in \cite{toappearBMS}, based on an idea of Reisenberger \cite{reisenberger} and the fixed points constructed in this work.

Let us shortly explain spin net models. These are versions of vertex models, i.e. models defined on graphs, with weights or amplitudes attached to vertices and depending on labels attached to the adjacent edges.  
Here the edges are decorated with representation labels $\rho$, as well as two magnetic indices, labelling a basis in $V_\rho$ and $V_{\rho^*}$ (the dual representation space) respectively. Thus every edge carries a Hilbert space
\ba\label{intro4}
{\cal H}_e\,=\ \oplus_\rho V_{\rho} \otimes V_{\rho^*}
\ea
where the sum is over all irreducible representations. The contraction of magnetic indices associated to one edge, i.e. between two vertex weights,  and the sum over the representation label $j$ attached to this edge  corresponds then to integrating out all degrees of freedom associated to this edge.

Thus the partition function for a spin net is defined as 
\ba\label{intro5}
Z=\sum_{\rho_e,m_e,n_e}  \prod_v C_v ( \{\rho_e\}_{e\supset v},  \{m_e\}_{e\supset v}, \{n_e\}_{e\supset v})
\ea
where we made the index structure of the vertex weights $C_v$  explicit. 

We considered coarse graining of such models with non--trivial simplicity constraints and for the permutation group of three elements $S_3$ in \cite{bdetal13}.  Under coarse graining the models flow in two ways: the weights change as well as the variables. The coarse graining procedure summarizes a set of edges to effective edges. In terms of the associated Hilbert spaces we just take the tensor product of the 'bare' Hilbert spaces associated to the original edge.  Performing a reduction to a sum over irreducibles again we obtain
\ba
{\cal H}_e \equiv \oplus_{\rho,\rho'} \mu(\rho,\rho') \left(V_\rho \otimes V_{\rho'}\right) \q ,
\ea
with $\mu(\rho,\rho')>1$ in general.

The results of \cite{bdetal13} indicate 
that indeed the models (\ref{intro5}) have the potential to flow to fixed points which carry a notion of simplicity constraints. We will confirm this conjecture in providing an interpretation of the fixed points in terms of imposing simplicity constraints. 

Most of the fixed points found \cite{unpublished} were of a particular factorizing structure. These fixed points are not included in the initial phase space of models decribed by  (\ref{intro4}) but require $\mu(\rho,\rho')=1$ also for pairs $\rho'\neq \rho^*$.  Indeed for these fixed points we find the factorizing structure 
\ba
C_v^{fix} ( \{\rho_e\}_{e\supset v},  \{\rho'_e\}_{e\supset v}, \{m_e\}_{e\supset v}, \{n_e\}_{e\supset v}) \,\, \,=\,\,\,  c_v^{fix} (  \{\rho_e\}_{e\supset v} , \{m_e\}_{e\supset v} ) \,\,\, c_v^{fix} (  \{\rho'_e\}_{e\supset v} , \{n_e\}_{e\supset v} ) 
\ea
This motivates us to introduce models with edge Hilbert spaces 
\ba
{\cal H}_e = \oplus_\rho  V_\rho
\ea
and vertex weights
\ba
c_v(  \{\rho_e\}_{e\supset v} , \{m_e\}_{e\supset v} )
\ea
so that these vertex weights are invariant under the action of the group on the associated tensor product of representation spaces. We will term these models intertwiner models, as the work \cite{bdetal13} led to the conclusion that this choice of intertwiners encode the relevant parameters for spin net (and in this way also for spin foam) dynamics.

The question which arose after the work \cite{bdetal13} was whether the appearance of these fixed points was specific to using the group $S_3$ or a more universal feature. Thus we will investigate the case with quantum group $SU(2)_k$ here. Not only do we find a large family of fixed points, these also generalize to the classical group $SU(2)$, or a real quantum deformation parameter. Moreover we confirm the interpretation of the fixed points carrying a notion of the simplicity constraints. We leave open here the investigation of the stability of these fixed points and hence the question in which sense and whether these fixed points define phases. From the experience with the $S_3$ models one needs to fine-tune first to the phase transition between the equivalent of $BF$ and strong coupling phase to allow a flow to these fixed points. This fine tuning can be understood as imposing a certain weak notion of triangulation invariance (and hence diffeomorphism symmetry \cite{bd12a}) in the sense, that one demands invariance of the partition function under edge (and for spin  foams face) subdivisions \cite{bojowaldperez,bahr,warsaw}.

The appearance of further fixed points opens up new perspectives for spin foams. So far the discussion for spin foam renormalization and phases has been largely confined to $BF$ theory and the (equivalent of) strong coupling fixed point, see for instance the second reference in \cite{fotini}. However there might be more topological theories, which even implement a notion of simplicity constraints. A theory with propagating degrees of freedom might then arise as a perturbation of one of these topological field theories (a similar conjecture has been formulated in \cite{ditt}, however not specifying the topological theory). 
An alternative scenario is to take the continuum limit at a phase transitions between these fixed points.

As mentioned before, a second relation of the intertwiner models to spin foams arises through Reisenberger's construction \cite{reisenberger}: the fixed points define intertwiners for the construction of spin foams that carry simplicity constraints. This notion will be explored in \cite{toappearBMS}.  

\section{Quantum group intertwiner models}\label{models}

Some basic facts on the representation theory of the quantum group $SU(2)_k$ and `diagrammatic calculus' are summarized in appendix \ref{appA}. 

Models with edge Hilbert spaces of the type
\ba\label{t3}
{\cal H}_e \equiv \oplus_{j} V_j \q 
\ea
where $j$ denotes an irreducible finite dimensional representation of $SU(2)_k$
 come up in the description of anyon fusion and splitting \cite{anion}.  Indeed the models we introduce here can be seen as state sum versions of anyon fusion and splitting dynamics. More abstractly we can describe these models as the nesting of intertwiner maps of type $V_1 \otimes V_2 \rightarrow V_3$ and $V_3 \rightarrow V_2\otimes V_1$, where one sums over intertwiner (i.e. representation) labels. (We will often replace indices $j_I$ denoting the representation associated to an edge $I$ with just the index $I$.) 
 In both cases a direction is involved -- in the graphical representation we will take this direction as upwards.  The intertwiner  maps can then be described by two different types of three--valent vertices
\ba\label{d1}
\begin{tikzpicture}
\draw[thick] (-0.3,-0.3) --(0.0,-0.0);
\node[left] at (-0.2,-0.2) {1};
\draw[thick] (0.3,-0.3) --(0.0,-0.0);
\node[right] at (0.2,-0.2) {2};
\draw[thick] (0.0,-0.0) --(0.0,0.4);
\node[right] at (0.0,0.3) {3};
\node at (1.8,0.0) {$ \q=\q{}_q\cc^{j_1j_2j_3}_{m_1m_2m_3}$};
\end{tikzpicture}
\q\q ,\q\q
\begin{tikzpicture}
\draw[thick] (-0.3,0.3) --(0.0,-0.0);
\node[left] at (-0.2,0.2) {2};
\draw[thick] (0.3,0.3) --(0.0,-0.0);
\node[right] at (0.2,0.2) {1};
\draw[thick] (0.0,0.0) --(0.0,-0.4);
\node[left] at (0.0,-0.3) {3};
\node[right] at (0.3,0.0) {$\q=\q{}_{q^{-1}}\cc^{j_1j_2j_3}_{m_1m_2m_3}$};
\end{tikzpicture}
\ea
representing Clebsch Gordan coefficients, which give the components of the maps in the chosen basis. (The $m$ indices label the vectors of the basis in the representation space $V_j$.) Here $q$ denotes a root of unity\footnote{We adopt the conventions of \cite{biedenharn}.} specified by the level $k$ of the quantum group $q=\exp(\frac{2\pi}{(k+2)}i)$.

The intertwiner maps can be combined to give maps between tensor products with more factors. This can be all nicely represented graphically leading to `diagrammatic calculus' \cite{yellowbook,hellmann}. This diagrammatic calculus also includes replacement rules which follow from certain identities for the Clebsch Gordan coefficients (or from the fact that one is considering a certain type of category). A particular important replacement rule is that the following two maps $ \mu_1,\mu_2: V_3 \otimes V_4 \rightarrow V_2 \otimes V_1$ are equal:
\ba\label{tilting0}
\begin{tikzpicture}
\draw[thick] (-0.0,-0.4) --(0,0);
\draw[thick] (0,0) -- (-0.7,0.7);
\draw[thick](0,0)-- (0.2,0.2);
\draw[thick](0.2,0.2)--(0.4,0.4);
\draw[thick](0.4,0.4)--(0.4,0.7);
\draw[thick](0.4,0.4)--(1.2,-0.5);
\node[right] at (0.4,0.6) {1};
\node[left] at (-0.7,0.6) {2};
\node[left] at (0.3,0.3) {6};
\node[left] at (-0.0,-0.25){3};
\node[right] at (1.1,-0.25){4}; 
\node[right] at (1.7,0){$=$};
\draw[thick] (2.5,-0.4)--(3.3,0.4);
\draw[thick](3.3,0.4)--(3.3,0.7);
\draw[thick] (3.3,0.4)--(3.7,0.0);
\draw[thick] (3.7,0.0)--(4.4,0.7);
\draw[thick] (3.7,0.0)--(3.7,-0.4);
\node[right] at (4.4,0.5) {1};
\node[left] at (3.3,0.5) {2};
\node[right] at (3.4,0.3) {6};
\node[right] at (2.7,-0.25){3};
\node[right] at (3.7,-0.25){4}; 
\node[right] at (4.7,0)  {$=:$};
\draw[thick] (6,-0.4)--(6,0.7);
\draw[thick](6,0.1)--(7,0.1);
\draw[thick](7,-0.4)--(7,0.7);
\node[right] at (7,0.5) {1};
\node[left] at (6,0.5) {2};
\node[right] at (6.3,0.3) {6};
\node[left] at (6,-0.25){3};
\node[right] at (7,-0.25){4}; 
\end{tikzpicture}
\ea 
Hence we can think of any diagonal edge also as a horizontal edge.  This allows to put the model also on (three-valent) lattices with horizontal edges. 

To introduce a dynamics we allow the vertices to be dressed with weights $a(j_1,j_2,j_3)$ and $a'(j_1,j_2,j_3)$ which we will indicate graphically by fat vertices
\ba
\begin{tikzpicture}
\draw[thick] (-0.3,-0.3) --(0.0,-0.0);
\node[left] at (-0.2,-0.2) {1};
\draw[thick] (0.3,-0.3) --(0.0,-0.0);
\node at (0,0)  {$\bullet$};
\node[right] at (0.2,-0.2) {2};
\draw[thick] (0.0,-0.0) --(0.0,0.4);
\node[right] at (0.0,0.3) {3};
\node at (2.8,0.0) {$ \q=\q{}_q\cc^{j_1j_2j_3}_{m_1m_2m_3} a(j_1,j_2,j_3)$};
\end{tikzpicture}
\q\q ,\q\q
\begin{tikzpicture}
\draw[thick] (-0.3,0.3) --(0.0,-0.0);
\node[left] at (-0.2,0.2) {2};
\draw[thick] (0.3,0.3) --(0.0,-0.0);
\node at (0,0)  {$\bullet$};
\node[right] at (0.2,0.2) {1};
\draw[thick] (0.0,0.0) --(0.0,-0.4);
\node[left] at (0.0,-0.3) {3};
\node[right] at (0.6,0.0) {$={}_{q^{-1}}\cc^{j_1j_2j_3}_{m_1m_2m_3} a'(j_1,j_2,j_3) \q .$};
\end{tikzpicture}
\ea

From these amplitudes we also require the tilting condition (\ref{tilting0}) to hold, i.e.
\ba\label{tilting1}
a_{362}\,a'_{164}=a'_{623}\,a_{641}
\ea
where $a_{IJK}=a(j_I,j_J,j_K)$.

~\\

\begin{figure}[h!]
\begin{center}
       \includegraphics[scale=0.4]{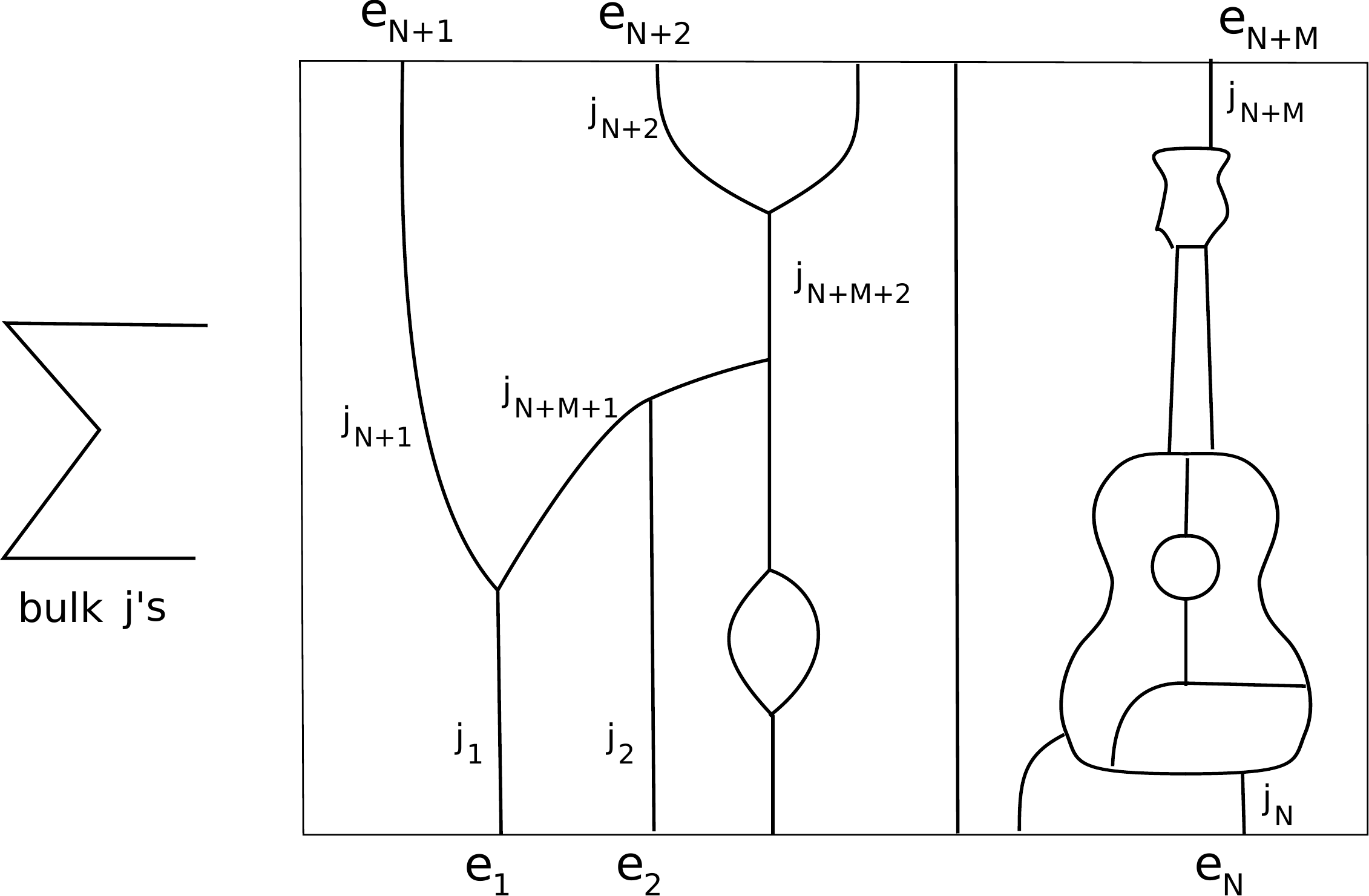}
    \end{center}
       \vspace{-0.5cm}
    \caption{\small \label{itmodel} An intertwiner model defining an intertwining map: ${\cal H}_b\rightarrow {\cal H}_t$. Here all the vertices are `fat vertices'.}
    \vspace{0.5cm}
\end{figure}

We then define an intertwining map between the spaces ${\cal H}_b = \oplus _{j_I} V_{j_1} \otimes \cdots \otimes V_{j_N}$ and ${\cal H}_t=V_{j_{N+1}} \otimes \cdots V_{j_{N+M}}$ by (see figure \ref{itmodel})
\begin{itemize}
\item[(a)] specifying a three--valent graph `in a box'  with edges $e_1 \cdots e_N$ entering at the bottom of the box and edges $e_{N+1} \cdots e_{N+M}$ emerging at the top of the box representing the corresponding tensor product of representation spaces (no edges enter at the vertical sides of the box)
\item[(b)] specifying weights $a(j_1,j_2,j_3)$ and $a'(j_1,j_2,j_3)$ for the vertices
\item[(c)] summing over all representation labels attached to the inner edges, i.e. those not entering at the bottom or emerging at the top.
\end{itemize}

Thus we define a partition function with two boundary components 
\ba\label{defpart}
Z(  \{ j_e,m_e \}_{e \subset \text{bdry}} ) = \sum_{\{{j_e,m_e}\}_{e \in \text{bulk}}} \prod_v  a({j}_{e \supset v} ) \, {}_q{\cal C}^{\{j_e\}_{e \supset v}}_ {\{m_e\}_{e \supset v}} \,\,\,\, \prod_{v'}  a'({j}_{e \supset v'} ) \, {}_{q^{-1}}{\cal C}^{\{j_e\}_{e \supset v'}}_ {\{m_e\}_{e \supset v'}} \q .
\ea


Considering graphs with many edges we can ask how the amplitudes behave under coarse graining: this can be described as taking a given graph in a box and  partitioning either the edges on the top or the bottom into two sets. This gives three sets of edges defining three Hilbert spaces which we associate to three `effective edges' respectively (if one wants to stay with a model based on three--valent vertices). There is also an effective vertex weight given by the intertwiner map defined by the original diagram in the box. 

Now the effective edges rather carry a Hilbert space labelled by a tensor product of representation spaces. This tensor product can be reduced to a sum over representations, which in general might appear with multiplicities larger than one. At this point a truncation is required (alternatively one enlarges the space of models and allows higher multiplicities, i.e. additional multiplicity labels on the edges). Ideally this truncation should pick up the part of the amplitude most relevant for the next coarse graining step. This in particular includes stacking boxes on top of each other, which corresponds to take powers of the corresponding intertwiner maps. Hence one can expect that a truncation determined by the largest singular value  (or the $\chi$ largest singular values)  would lead to satisfying results. 

This  leads to  the tensor network coarse graining algorithm \cite{levin,guwen}.  
The (quantum) group structure allows to introduce symmetry preserving algorithms as described and utilized in \cite{vidalsymm, eckert,  bdetal13}. This symmetry preserving algorithm has several advantages:\footnote{Another advantage is that it takes nicely care of the problem that quantum groups at the root of unity require to factor out trace zero representations. }
besides 
improving the efficiency of the algorithm it allows to extract a lot of information on the structure of the fixed point models. In particular the singular values are attached with representation labels and thus one can follow which representations still appear (are excited) or not by tracking these singular values \cite{bdetal13}.

Most importantly this allows to define easily truncations which stay inside a certain space of models. For the space of models we are considering this would be implemented by keeping only the largest singular value per representation label, so all edges keep carrying a Hilbert space $\oplus_j V_j$. (In a non--symmetry preserving algorithm this  already corresponds to a bond dimension $\chi=\sum_j \text{dim}(j)$ with $\text{dim}(j)$ giving the (classical) dimension of the representation space associated to $j$ and the sum is over all representations that appear, i.e. in $SU(2)_k$ we have $j=0,1/2,\ldots,k/2$ and $\dim(j)=2j+1$.) Larger spaces of models, or less severe truncations, can be obtained by allowing higher multiplicities and keeping the corresponding number of singular values.

\subsection{Fixed points and triangulation invariant models}

We implemented the Levin--Nave algorithm \cite{levin} shortly explained below in its symmetry preserving version, truncating to the largest singular value per representation label. The aim was to identify fixed points of the coarse graining flow. This also allowed us to choose $k$ considerably large (i.e. $k=30, j_{max}=15$ with $SO(3)_k$) and to still obtain a fast (on a laptop) algorithm, which is certainly advantageous if searching `randomly' for fixed points.

These investigations showed a rich structure of fixed points\footnote{Actually most of these fixed points are (small) fixed point cycles where signs still alternate from one coarse graining step to the next. In the course of the work we will see that this is due to the fixed points requiring complex amplitudes. Astonishingly the algorithm still converges to the fixed point cycles even with real amplitudes.}. Although we considered coarse graining on a fixed hexagonal lattice, the fixed points define triangulation invariant models: We defined our models on three--valent graphs, embedded into the plane. Dualization of such graphs (with certain regularity requirements) lead to 2D  triangulations. Triangulations  (of a given topology) can be changed into each other by a sequence of Pachner moves.  In 2D there are three such Pachner moves, which we will call (2--2) (self--inverse) and (3--1) and its inverse (1--3), indicating the number of triangles involved before and after the move. Translating back to the original graphs these numbers indicate the number of vertices involved in this move.

These findings support the strategy advertised in \cite{bahretal09,bahretal12,bd12b,eckert,bdetal13} to look for triangulation invariant models (in the gravitational context tight to diffeomorphism symmetry \cite{bd08,bd12a}) by coarse graining. In particular it seems to be sufficient to consider a regular lattice, which indeed is the only way to actually perform coarse graining in praxis.  

However having established the existence of these fixed points we might ask  whether it is possible to find these also by other means. In the following we will present a strategy to this end. This will allow to consider much more easily more complicated models, which might be not  accessible numerically (for instance with structure group $SU(2) \times SU(2)$).

We will look for triangulation invariant models right away -- these will automatically be fixed points of the coarse graining flow. Hence we ask for the amplitudes to satisfy the (a) tilting condition and to be invariant under (b) the 2--2 move and (c) the two versions of the 3--1 move (as we have two types of three--valent vertices) depicted as follows. 
\ba\label{22pe}
\begin{tikzpicture}
\node[left] at (5.4,0.1) {$\sum_{j_6}$};
\draw[thick] (6,-0.4)--(6,0.7);
\node at (6,0.1) {$\bullet$};
\node at (7,0.1) {$\bullet$};
\draw[thick](6,0.1)--(7,0.1);
\draw[thick](7,-0.4)--(7,0.7);
\node[right] at (7,0.5) {1};
\node[left] at (6,0.5) {2};
\node[right] at (6.3,0.3) {6};
\node[left] at (6,-0.25){3};
\node[right] at (7,-0.25){4}; 
\end{tikzpicture}
\q\q
\begin{tikzpicture}
\node[left] at (-2.3,0.1){$=$};
\node[left] at (-0.8,0.1) {$\sum_{j_5}$};
\draw[thick] (-0.3,-0.3) --(0.0,-0.0);
\node[left] at (-0.2,-0.2) {3};
\draw[thick] (0.3,-0.3) --(0.0,-0.0);
\node at (0,0)  {$\bullet$};
\node[right] at (0.2,-0.2) {4};
\draw[thick] (0.0,-0.0) --(0.0,0.4);
\node[right] at (0.0,0.2) {5};
%
\draw[thick] (-0.3,0.7) --(0.0,0.4);
\node[left] at (-0.2,0.6) {2};
\draw[thick] (0.3,0.7) --(0.0,0.4);
\node at (0,0.4)  {$\bullet$};
\node[right] at (0.2,0.6) {1};
\draw[thick] (0.0,0.4) --(0.0,-0.0);
\end{tikzpicture}
\ea

\ba\label{31p}
\begin{tikzpicture}
\node[left] at (-1,0.5) {$\underset{j_4,j_5,j_6}{\sum}$};
\draw[thick] (-0.6,1.0) --(0.0,0.4);
\node[right] at (0.5,0.9) {1};
\node[right] at (0.25,0.5) {5};
\node[left] at (-0.5,0.9) {2};
\node[left] at (-0.2,0.5) {4};
\node[left] at (0.25,0.9) {6};
\draw[thick] (0.6,1.0) --(0.0,0.4);
\node[left] at (0.0,0.1) {3};
\node at (0,0.4)  {$\bullet$};
\node at (-0.3,0.7) {$\bullet$};
\node at (0.3,0.7) {$\bullet$};
\draw[thick] (0.0,0.4) --(0.0,-0.0);
\draw[thick](-0.3,0.7)--(0.3,0.7);
\node[left] at (1.7,0.5) {$=\; c$};
\draw[thick] (2.2,0.0)--(2.2,0.5);
\node[right] at (2.2,0.1) {3};
\node at (2.2,0.5) {$\bullet$};
\draw[thick] (2.2,0.5)--( 1.7,1.0);
\draw[thick] (2.2,0.5)--( 2.7,1.0);
\node[right] at (1.9,0.9) {2};
\node[right] at (2.6,0.9) {1};
\end{tikzpicture}
\quad,\quad\quad
\begin{tikzpicture}
\node[left] at (-1,0.5) {$\underset{j_4,j_5,j_6}{\sum}$};
\draw[thick] (0.0,0.7)--(-0.7,0.0);
\draw[thick] (0.0,0.7)--(0.7,0.0);
\draw[thick] (0.0,0.7)--(0.0,1);
\draw[thick] (-0.3,0.4) --(0.3,0.4);
\node at (0.0,0.7) {$\bullet$};
\node at (-0.3,0.4) {$\bullet$};
\node at (0.3,0.4) {$\bullet$};
\node[left] at (1.7,0.5) {$= \; c$};
\node at (2.2,0.5) {$\bullet$};
\draw[thick]   (2.2,0.5)-- (2.2,1);
\draw[thick]   (2.2,0.5)--(1.7,0);
\draw[thick]   (2.2,0.5)--(2.7,0);
\node[left] at  (0.0,0.9) {3};
\node[left] at (-0.55,0.1) {1};
\node[left] at (0.55,0.1) {2};
\node[left] at (0.2,0.25) {6};
\node[left] at (-0.3,0.6) {4};
\node[right] at (0.3,0.6) {5};
\node[right] at (1.8,0.1) {1};
\node[right]  at (2.7,0.1) {2};
\node[right] at (2.2,0.9) {3};
%
\end{tikzpicture}
\q\q .
\ea
Here $c$ is a scaling constant, which, if not equal to infinity\footnote{This might only happen for $k\rightarrow \infty$, i.e. $SO(3)$ symmetry.}, can be put to one by rescaling the amplitudes. Note that there are other versions of the 3--1 move (where all three edges are vertical or diagonal) which are however equivalent to the ones in  (\ref{31p})  due to the tilting condition. For the same reason we only need to consider one type of 2--2 move. 

These conditions translate into certain equations that have to hold for the amplitude functions $a,a'$  written out below. The Clebsch Gordon coefficients describing the `bare' intertwiner maps on the left hand side of the graphical equations in (\ref{22pe},\ref{31p}) can be contracted with an intertwiner basis and then expanded into the Clebsch Gordan coefficients appearing on the right hand sides in (\ref{22pe},\ref{31p}) again. This leads to the appearance of $[6j]$ (or $F$--) symbols.  The $[6j]$ symbol is defined graphically in (\ref{def6j}) as the transformation matrix (modulo dimension factors) between two different bases of intertwiner maps. 
\ba\label{def6j}
\begin{tikzpicture}
\draw[thick] (6,-0.4)--(6,0.7);
\draw[thick](6,0.1)--(7,0.1);
\draw[thick](7,-0.4)--(7,0.7);
\node[right] at (7,0.5) {1};
\node[left] at (6,0.5) {2};
\node[right] at (6.3,0.3) {6};
\node[left] at (6,-0.25){3};
\node[right] at (7,-0.25){4}; 
\end{tikzpicture}
\q\q\q
\begin{tikzpicture}
\node[left] at (-4.8,0.1){$=$};
\node[left] at (-0.8,0.1) {$\sum_{j_5}   \Fj{1}{2}{5}{3}{4}{6}  \sqrt{\frac{d_{j_5}}{d_{j_6} }}$};
\draw[thick] (-0.3,-0.3) --(0.0,-0.0);
\node[left] at (-0.2,-0.2) {3};
\draw[thick] (0.3,-0.3) --(0.0,-0.0);
\node[right] at (0.2,-0.2) {4};
\draw[thick] (0.0,-0.0) --(0.0,0.4);
\node[right] at (0.0,0.2) {5};
%
\draw[thick] (-0.3,0.7) --(0.0,0.4);
\node[left] at (-0.2,0.6) {2};
\draw[thick] (0.3,0.7) --(0.0,0.4);
\node[right] at (0.2,0.6) {1};
\draw[thick] (0.0,0.4) --(0.0,-0.0);
\end{tikzpicture}
\ea
It constitutes a transformation between orthonormal (intertwiner) bases, hence
\ba
\sum_{j_6}   \Fj{1}{2}{5}{3}{4}{6}   \Fj{1}{2}{5'}{3}{4}{6} &=& \delta_{j_5 j'_5}  \q .
\ea
The invariance conditions for the amplitudes are then given as
\begin{itemize}
\item[]{\bf Tilting} 
\ba
a_{362}\,a'_{164}=a'_{623}\,a_{641}
\ea
\item[]{\bf 2-2}
\ba\label{22}
\sqrt{\frac{1}{d_6}} \,\, a'_{623}a_{641} &=& \sum _{j_5} \Fj{1}{2}{5}{3}{4}{6} \, \sqrt{\frac{1}{d_5}} \,\,a_{345}a'_{125}  \q .
\ea
\item[]{\bf 3-1} 
\ba\label{31}
c \, a_{123} &=& \sum_{j_4,j_5,j_6}   \Fj{1}{2}{3}{4}{5}{6} (-)^{j_4+j_5-j_3} \frac{1}{\sqrt{d_3d_4}} \,\,a'_{651}a_{624}a_{543} \q , \nn\\
c \, a'_{123} &=& \sum_{j_4,j_5,j_6}   \Fj{1}{2}{3}{4}{5}{6} (-)^{j_4+j_5-j_3} \frac{1}{\sqrt{d_3d_4}} \,\,a_{651}a'_{624}a'_{543}  \q .
\ea
\item[]{\bf Bubble move}\\
The 3--1 move conditions can be replaced by a much more convenient  equation: Given that the 2--2 move equations hold the 3--1 move is equivalent to the bubble move depicted as
\ba
\begin{tikzpicture}
\node at (-1,0.35) {$\underset{j_1,j_2}{\sum}$} ;
\draw[thick] (0,0) --(0,0.3);
\draw[thick] (0,0.3) --(-0.2,0.5);
\draw[thick] (-0.2,0.5)--(-0.0,0.7);
\draw[thick] (0,0.3)--(0.2,0.5);
\draw[thick] (0,0.7)--(0.2,0.5);
\draw[thick] (0,0.7)--(0,1);
\node[left] at (-0.2,0.5) {2};
\node[right] at (0.2,0.5) {1};
\node[right] at (0.0,0.9) {3};
\node[right] at (0.0,0.1) {3};
\node at (0,0.3)  {$\bullet$};
\node at (0,0.7)  {$\bullet$};
\node at (0.9,0.5) {$=$};
\node at (1.9,0.5) {$c \, \Theta(j_3)$};
\draw[thick] (2.5,0)--(2.5,1);
\node[right] at (2.5,0.5){3} ;
\end{tikzpicture}
\ea
Here $\Theta(j)=0$ if the vertex amplitude vanishes if $j$ appears  as an argument, and is equal to one otherwise.
The following gives a graphical proof of this statement:
\ba
\begin{tikzpicture}
\node[left] at (-1,0.5) {$\underset{j_4,j_5,j_6}{\sum}$};
\draw[thick] (0.0,0.7)--(-0.7,0.0);
\draw[thick] (0.0,0.7)--(0.7,0.0);
\draw[thick] (0.0,0.7)--(0.0,1);
\draw[thick] (-0.3,0.4) --(0.3,0.4);
\node at (0.0,0.7) {$\bullet$};
\node at (-0.3,0.4) {$\bullet$};
\node at (0.3,0.4) {$\bullet$};
\node[left] at (1.5,0.5) {$=$};
\node[left] at (2.8,0.5) {$\underset{j_4,j_5}{\sum}$}; 
\draw[thick] (3.5,1)--(3.5,0.7);
\node at (3.5,0.7) {$\bullet$};
\draw[thick] (3.5,0.7)--(3.3,0.5);
\draw[thick] (3.5,0.7)--(3.7,0.5);
\draw[thick] (3.3,0.5)--(3.5,0.3);
\draw[thick] (3.7,0.5)--(3.5,0.3);
\node at (3.5,0.3)  {$\bullet$};
\draw[thick] (3.5,0.3)--(3.5,0.1)  ;
\node at (3.5,0.1)  {$\bullet$};
\draw[thick] (3.5,0.1) --(3.1,-0.3);
\draw[thick] (3.5,0.1) --(3.9,-0.3);
\node[left] at  (3.5,1) {$3$};
\node[left] at  (3.3,0.5) {$4$};
\node[right] at  (3.7,0.5) {$6$};
\node[left] at  (3.1,-0.1) {$2$};
\node[right] at  (3.9,-0.1) {$1$};
%
%
\node[left] at  (0.0,0.9) {3};
\node[left] at (-0.55,0.1) {2};
\node[left] at (0.55,0.1) {1};
\node[left] at (0.2,0.25) {6};
\node[left] at (-0.3,0.6) {4};
\node[right] at (0.3,0.6) {5};
%
%
\node at (4.9,0.5) {$= \q c $};
\node[left] at (6.5,0.9) {$3$};
\node[left] at (6,0.2) {$2$};
\node[right] at (6.8,0.2) {$1$};
\draw[thick] (6.4,1)--(6.4,0.5);
\node at (6.4,0.5) {$\bullet$};
\draw[thick] (6.4,0.5)--(5.9,0.1) ;
\draw[thick] (6.4,0.5)--(6.8,0.1) ;
\end{tikzpicture}
\ea
 To be invariant under the bubble move the amplitudes have to satisfy
\ba\label{bubble}
c \,d_3 \, \Theta(j_3)=\sum_{j_1,j_2} (-)^{j_1+j_2-j_3}  \, \, a'_{123}a_{213} \q .
\ea
\end{itemize}
Finding triangulation invariant models means to find solutions to these equations. 

It will turn out that for constructing fixed point solutions, the 2--2 move conditions are crucial. Indeed, if a model is  invariant under 2--2 moves (or crossing symmetry) then in order to obtain the partition function by reduction of the tri--valent graph to the simplest possible graph, requires only the consideration of bubble moves. Moreover, models with crossing symmetry preserve this property under coarse graining, in particular after incorporating bubbles into effective amplitudes \cite{fendly}.  Thus, models with crossing symmetry can be reduced to a sequence of bubble moves. Such models are always solvable as one can associate a transfer matrix to such a bubble. In our case this transfer matrix is even diagonal due to the (quantum) group symmetry of the models.

The Levin--Nave algorithm implements coarse graining as an alternating sequence of 2--2 moves and 3--1 moves. For the 2--2 moves one contracts first two three-valent tensors and then splits these again to two three--valent tensors (with a different partitioning of the indices) with the help of a singular value decomposition. This is where a truncation needs to be  introduced. The 3--1 move then summarize triples of such three--valent tensors to new effective tensors. The symmetry preserving algorithm operates directly with the amplitudes $a,a'$. The Clebsch--Gordon coefficients contract to $[6j]$ symbols that appear in the same way as in the equations (\ref{22}) and (\ref{31}). 



\section{Recursion relations for fixed point amplitudes} \label{recursion}

Here we will propose a way to construct fixed point solutions analytically.
In the following we restrict to $SO(3)_k$, that is we consider representations $j=0,1,2,\ldots ,j_{max}$ with $j_{max}=\tfrac{k}{2}$ for $k$ even and  $j_{max}=\tfrac{k-1}{2}$ for $k$ odd. As we will later see we can generalize a certain family of fixed points to the classical $SO(3)$ group as well as $SO(3)_q$ with $q$ real.


%
%
%
%
%

\subsection{Permutations and signs}\label{signs}

For the fixed points found numerically the modulus of a given amplitude is invariant under permutation of entries and under replacing $a\rightarrow a'$. We will therefore look for fixed points with such properties.
In this section we will determine the signs (restricting to the case that only signs appear, that is no other complex phases) that might appear in the amplitudes under permutation of the arguments.  

The tilting conditions for a configuration with one zero spin give 
\ba\label{39}
a(j_1,j_2,j_3)\,a'(0,j_2,j_2) &=& a'(j_2,j_3,j_1)\,a(j_2,j_2,0)  \q .
\ea
%
Setting $j_3=0$ in (\ref{39}) we obtain
\be\label{comm}
a'(0,j,j) = a'(j,0,j) \, .
\ee  
Likewise we have 
$
a(0,j,j) = a(j,0,j)
$.

Let us define
\ba
\sigma_j\,=\,\frac{a(j,j,0)}{a'(j,0,j)}\,=\, \frac{a'(j,j,0)}{a(j,0,j)} \q .
\ea
The last equation can be shown from (\ref{39}) with $j_1=0$ (and  (\ref{comm})). 
%
%
%
%
%
%
Thus (\ref{39}) gives
\ba\label{17}
a(j_1,j_2,j_3) \,=\,\sigma_{j_2} \,\, a'(j_2,j_3,j_1) \, ,\q\q
\text{
similarly
}\q\q
a'(j_1,j_2,j_3) \,=\,\sigma_{j_2} \,\, a(j_2,j_3,j_1) \q .
\ea

This allows also to obtain permutations of arguments only involving $a$ or $a'$ amplitudes, i.e. 
\ba\label{18}
&& a(j_1,j_2,j_3)\,\,= \,\, 
 \sigma_{j_3}\sigma_{j_1} a(j_2,j_3,j_1) \,\,=\,\, \sigma_{j_2}\sigma_{j_3} a(j_3,j_1,j_2) \q , \q\q \nn\\
&& a'(j_1,j_2,j_3)\,= \, 
 \sigma_{j_3}\sigma_{j_1} a'(j_2,j_3,j_1) \,=\, \sigma_{j_2}\sigma_{j_3} a'(j_3,j_1,j_2) \q . \q\q
\ea

With (\ref{17}) and (\ref{18}) we reach
\ba\label{44}
a(j_1,j_2,j_3)\,=\,\sigma_{j_1}\sigma_{j_2}\sigma_{j_3} \,a'(j_1,j_2,j_3)  \q . 
\ea

It can be easily checked that the relations (\ref{18}) and  (\ref{44}) imply the tilting condition.

Next we will consider odd permutations of the arguments in the vertex weights. To this end we define the sign function
\ba\label{signdefb}
a'(j_2,j_1,j_3)&=&\sigma(j_1,j_2,j_3)\,a(j_1,j_2,j_3) \q 
\ea
involving an odd permutation of arguments. Together with the even permutations, we will get relations between the signs $\sigma(j_1,j_2,j_3)$ under the action of permuting arguments. 

One can  derive the relations
\ba
\sigma(j_1,j_2,j_3)\,=\, 
 \sigma(j_2,j_3,j_1)\,=\, 
 \sigma(j_3,j_1,j_2)
\ea
and moreover, using (\ref{44})
\ba\label{20}
\sigma(j_1,j_2,j_3)=\sigma(j_2,j_1,j_3) \q .
\ea
Note that the sign function simplifies if any two arguments are equal
\ba\label{21} \sigma(j,j,j')=\sigma_{j'} \q .\ea 

%

\subsection{Using the 2-2 move as recursion relation}

Here we will construct amplitudes that are invariant under some subset of the 2--2 moves: We will first find `initial data' that arise as eigenvectors with eigenvalue $+1$ of a transformation induced by the 2--2 move. Next we will use the 2--2 move directly to determine the remaining amplitudes. In the end one has to check whether the amplitudes satisfy all the 2--2 and bubble moves.

This strategy is especially effective for constructing fixed points with (almost) all amplitudes non--vanishing.


We will adopt the following notation for the diagrams appearing in the 2--2 Pachner move: \\ $A_5(j_1,j_2,j_3,j_4;j_5)$ will denote the `vertical' 2--2 diagram with $j_1,\ldots,j_4$ ordered anti--clockwise at the outer edges and $A_6(j_1,j_2,j_3,j_4;j_6)$ the horizontal one.


\subsubsection{Amplitudes with $j=0$  arguments}

Consider the 2--2 move between $A_6(j,j,0,0;j)$ and $A_5(j,j,0,0;0)$ and between $A_6(0,0,j,j;j)$ and\\ $A_5(0,0,j,j;0)$. Using that the $[6j]$ symbols appearing in this moves are equal to one we have
\ba\label{12}
&& a'(j,j,0) \,a(0,0,0)  \,=\, \frac{1}{\sqrt{d_{j}}} \, a(0,j,j)\,a'(j,j,0) \,=\, \frac{1}{\sqrt{d_{j}}}   a'(j,j,0)\,a(j,0,j) \\
&& a'(0,0,0) \,a(j,j,0)\,=\, \frac{1}{\sqrt{d_{j}}}   a(j,j,0) \, a'(0,j,j)     \,=\, \frac{1}{\sqrt{d_{j}}} \, a'(j,0,j)\,a(j,j,0)   \q .
\ea
Setting $a(0,0,0)=1$ 
 and with the results from section \ref{signs} we therefore have
\ba\label{jzero}
&& a(j,0,j)\,\,=\,\,a(0,j,j)\,\,=\, \,\sigma_j \sigma_0 \,a(j,j,0)\,=\,\,\Theta(j)\sqrt{d_j} \q , \nn\\  
&& a'(j,0,j)\,=\,a'(0,j,j)\,=\,\sigma_j\sigma_0\,a'(j,j,0)\,=\,\sigma_0\,\Theta(j)\sqrt{d_j} \q,
\ea
where $\Theta(j)=1$ or $0$ depending on whether the representation $j$ is excited or not.

Thus we  can fix the constant $c$ in the bubble move: 
\ba\label{bubbleconstant}
c \,=\, \sigma_0\sum_j \Theta(j) \, d_j \q .
\ea

%

\subsubsection{Recursion relations for $a(j,j,1)$}

In the previous step we got all amplitudes involving $j=0$ as one of the arguments. This does however not seem to be sufficient to obtain all other amplitudes by using the 2--2 move directtly. We will rather need  amplitudes with $j=1$ appearing (assuming that these do not vanish). We will find a recursion relation for these amplitudes via an eigenvector condition for a matrix determined by the $[6j]$ symbol.

The general idea is as follows: Consider the 2--2 move equation between the set of diagrams\\ $A_6(j_1,j_2,j_3,j_2;i_6)$ and $A_5(j_1,j_2,j_3,j_2;i_5)$ with $i_6,i_5$ taking all allowed values:
\ba\label{22p}
\begin{tikzpicture}
\node[left] at (5.4,0.1) {$\sum_{j_6}$};
\draw[thick] (6,-0.4)--(6,0.7);
\node at (6,0.1) {$\bullet$};
\node at (7,0.1) {$\bullet$};
\draw[thick](6,0.1)--(7,0.1);
\draw[thick](7,-0.4)--(7,0.7);
\node[right] at (7,0.5) {1};
\node[left] at (6,0.5) {2};
\node[right] at (6.3,0.3) {6};
\node[left] at (6,-0.25){3};
\node[right] at (7,-0.25){2}; 
\end{tikzpicture}
\q\q
\begin{tikzpicture}
\node[left] at (-2.3,0.1){$=$};
\node[left] at (-0.8,0.1) {$\sum_{j_5}$};
\draw[thick] (-0.3,-0.3) --(0.0,-0.0);
\node[left] at (-0.2,-0.2) {3};
\draw[thick] (0.3,-0.3) --(0.0,-0.0);
\node at (0,0)  {$\bullet$};
\node[right] at (0.2,-0.2) {2};
\draw[thick] (0.0,-0.0) --(0.0,0.4);
\node[right] at (0.0,0.2) {5};
%
\draw[thick] (-0.3,0.7) --(0.0,0.4);
\node[left] at (-0.2,0.6) {2};
\draw[thick] (0.3,0.7) --(0.0,0.4);
\node at (0,0.4)  {$\bullet$};
\node[right] at (0.2,0.6) {1};
\draw[thick] (0.0,0.4) --(0.0,-0.0);
\end{tikzpicture}
\q\q\q\q\q\q\q\q\;
\ea
\ba
\frac{1}{\sqrt{d_{i_6}}}\,a'(i_6,j_2,j_3)\,  a(i_5,j_2,j_1) &=& \sum_{i'} \Fj{j_1}{j_2}{i_5}{j_3}{j_2}{i_6}  \frac{1}{\sqrt{d_{i_5}}} \,  a(j_3,j_2,i_5)a'(j_1,j_2,i_5) \q .
\ea
Using the tilting condition for the amplitudes appearing on the RHS of this equation, we can replace these by $a(i_5,j_1,j_2)\,a'(i_5,j_3,j_2)$. These are almost equal to the amplitudes on the LHS, however we still have to exchange certain arguments. We therefore use the sign $\sigma(j_1,j_2,j_3)$ defined in (\ref{signdefb}) to write
\ba\label{signdef}
a'(j_2,j_1,j_3)&=&\sigma(j_1,j_2,j_3)\,a(j_1,j_2,j_3) \q .
\ea
In this way we obtain an eigenvalue condition
\ba
a(j_3,i_6,j_2)\,a'(j_1,i_6,j_2) &=& \sum_{i_5} \Fj{j_1}{j_2}{i_5}{j_3}{j_2}{i_6} \sqrt{ \frac{  d_{i_6}}{d_{i_5}}} \sigma(i_5,j_1,j_2) \sigma(j_3,i_5,j_2) \,a(j_3,i_5,j_2)\,a'(j_1,i_5,j_2)\q\q\nn\\
\ea
for the vector $v_i=a(j_3,i,j_2)\,a'(j_1,i,j_2)$  and involving the matrix
\ba\label{matrix1}
\tilde M_{i_6i_5} &\,:=\,&  \Fj{j_1}{j_2}{i_5}{j_3}{j_2}{i_6} \sqrt{ \frac{  d_{i_6}}{d_{i_5}}} \sigma(i_5,j_1,j_2) \sigma(j_3,i_5,j_2)  \nn\\
&\,=:\,& M_{i_6i_5}\sqrt{ \frac{  d_{i_6}}{d_{i_5}}} \sigma(i_5,j_1,j_2) \sigma(j_3,i_5,j_2)  \q .
\ea
We are searching for an eigenvector with eigenvalue $\lambda=+1$. As we will see this does not turn out to be as strong a condition as one would expect. However one can find recursive families of equations that would fix all amplitudes (at least in the case that these are all non vanishing) depending on one initial value.  The aim is to find instances where this matrix is small, but not just one--dimensional (as this would typically lead to tautological equations).

This strategy can be applied to i.e. $A_6(j,l,j+2l-1,l,i_6)$ and $A_5(j,l,j+2l-1,l,i_5)$. For generic choices of $j,l$ the intertwiner space is two--dimensional: $i_5,i_6$ can both take the values $j+l-1,j+l$.

For $l=1$ the two--dimensional vector in question is given by $\tilde v_x=a'(j,j+x,1) a(j+1,j+x,1)$ with $x=0,1$. We need to restrict to $1\leq j\leq \frac{k-3}{2}$ to obtain a two--dimensional intertwiner space. Thus we might hope to fix the ratio ( in case that $a(j+1,j,1)$ is not vanishing)
\ba\label{fixratio}
\sigma(j+1,j,1) \frac{a'(j,j,1)}{a(j+1,j+1,1)}
\ea
which would fix  amplitudes (modulo sign) of the form $a(j,j,1)$. 

The matrix (\ref{matrix1}) has three parts: Firstly the matrix $M_{i_6i_5}$ which due to the symmetries of the $[6j]$ symbols is symmetric. It is also a real matrix and furthermore due to the orthogonality of the $[6j]$ symbol we have
\ba\label{Mcond}
\sum_{i_6} M_{i_6i_5} M_{i_6 i'_5} = \delta_{i_5 i'_5} \q . 
\ea
Hence all eigenvalues have to be either equal to $+1$ or equal to $-1$.  If $M$ is not equal to $\pm {\mathbb I}$ it has to have vanishing trace and hence one eigenvalue $+1$ and one eigenvalue $-1$. 

The factors $\sqrt{ \frac{  d_{i_6}}{d_{i_5}}}$ just redefine the eigenvectors of a given matrix but do not change the eigenvalues. The signs  $\sigma(i_5,j,1)\sigma(j+1,i_5,1)$ might change the eigenvalues, however given that there is a finite set of possibilities one can easily check which choices lead to eigenvalues $+1$ and which do not.  Moreover, using  (\ref{21}) we have that for both values of the intertwiner $i_5=j,j+1$
\ba
\sigma(i_5,j,1)\sigma(j+1,i_5,1) \,=\, \sigma_1\sigma(j+1,j,1)
\ea
that is we can exclude a relative sign.

The $[6j]$ symbols defining the components of $M$ can be explicitly evaluated (see appendix \ref{appA})  and it can be indeed verified that its trace vanishes.  Due to the condition (\ref{Mcond}) and the vanishing trace the matrix is of the form
\ba\label{33}
M=
\begin{pmatrix} 
-a &\sqrt{1-a^2} \\
\sqrt{1-a^2} & a
\end{pmatrix}
\ea
with $a=\frac{[2]}{[2j+2]}< 1$. (This holds for $1\leq j\leq  \frac{k-3}{2}$.)  
The $+1$ eigenvector is given by $v_+=( \sqrt{(1-a)},\sqrt{(1+a)})$.




Thus the eigenvectors for $\tilde M$ (obtained by rescaling the eigenvectors of $M$ by the dimension factors) are given by
\ba
\tilde v_-=
\begin{pmatrix}
\sqrt{[2j+1] ([2j+2]+[2])} \\
-\sqrt{[2j+3] ([2j+2]-[2])}
\end{pmatrix}   \q,\q\q\q
\tilde v_+=
\begin{pmatrix}
\sqrt{[2j+1] ([2j+2]-[2]) }\\
\sqrt{[2j+3] ([2j+2] +[2])}
\end{pmatrix}   \q.
\ea

For $\sigma_1\sigma(j+1,j,1)=+1$ we reach the conclusion (using the relations in section \ref{signs})
\ba\label{rec1}
a(j+1,j+1,1) &=& 
a(j,j,1) \frac{ \sqrt{[2j+3] ([2j+2]+[2])}}  { \sqrt{[2j+1] ([2j+2]-[2])}   } \, \nn\\
&\underset{(\ref{qind2})}{=}& 
a(j,j,1) \frac{ \sqrt{[2j+3] [2j] [j+2]^2}}  { \sqrt{[2j+1] [2j+4][j]^2}   }  \nn\\
&\underset{q=1}{=}&a(j,j,1)       \sqrt{\frac{2j+3}{2j+1}}     \sqrt{\frac{(j+2)}{j}}
\ea
and for $\sigma_1\sigma(j+1,j,1)=-1$
\ba
a(j+1,j+1,1) &=& 
a(j,j,1) \frac{ \sqrt{[2j+3] ([2j+2]-[2])}}  { \sqrt{[2j+1] ([2j+2]+[2])}   }\nn\\
&\underset{(\ref{qind2})}{=}& 
a(j,j,1) \frac{ \sqrt{[2j+3] [2j+4] [j]^2}}  { \sqrt{[2j+1] [2j][j+2]^2}   }  \nn\\
&\underset{q=1}{=}&
a(j,j,1)  \sqrt{\frac{2j+3}{2j+1}}  \sqrt{\frac{j}{ (j+2)}}  \q .
\ea

Note that this fixes $\sigma_2=\sigma(2,1,1)$.

\subsubsection{Recursion relations for $a(j,j+1,1)$}

This fixes the amplitudes of the form $a(j,j,1)$. 
To fix also amplitudes non--diagonal in the first two entries consider the 2--2 move transformation between
$A_6(j,1,j,1,i_6)$ and $A_5(j,1,j,1,i_5)$.  The same line of arguments as before leads us to seek for   eigenvectors 
\ba\label{28}
\tilde v_{i_5}=a(j,i_5,1)\, a'(j,i_5,1)
\ea
with eigenvalue $+1$ of the matrix
\ba\label{mat31}
\tilde M_{i_6i_5}&=&\Fj{j}{1}{i_5}{j}{1}{i_6} \sqrt{ \frac{  d_{i_6}}{d_{i_5}}} \sigma(i_5,j,1) \sigma(j,i_5,1) \,:=M_{i_6i_5} \sqrt{ \frac{  d_{i_6}}{d_{i_5}}} \sigma(i_5,j,1) \sigma(j,i_5,1) 
\ea
with $i_5,i_6$ taking the three values $j-1,j,j+1$. According to (\ref{20}) we have $\sigma(i_5,j,1) \sigma(j,i_5,1) =1$ so neither a global nor a relative sign for the rows arises.

A three--dimensional intertwiner space is obtained for $1\leq j\leq \tfrac{k}{2}-1$. As we know already the amplitudes for $i_5=j$ we can hope to fix all the remaining amplitudes which include a spin 1 representation as one argument.

The matrix $M$ can be readily computed. Due to the identities (\ref{Mcond}) this matrix can be parametrized by just two (continuous) parameters: $M=\pm(\mathbb{I}-2 v_{3}v_{3}^{t})$ where $v_{3}$ is a normalized eigenvector and we assume that $M$ is not proportional to the identity.  Indeed one finds that  $\text{Tr}(M)=+1$, hence we have eigenvalues $\lambda_1=\lambda_2=+1,\,\lambda_3=-1$ and $v_{3}$ has eigenvalue $-1$.   Explicitly, the matrix $M$ is given by 
\ba
M&=&
\begin{pmatrix}
a & -\sqrt{(1-a)(a+b)}& \sqrt{(1-a)(1-b)}   \\
-\sqrt{(1-a)(a+b)} &1-a-b& \sqrt{(1-b)(a+b)}\\
 \sqrt{(1-a)(1-b)}  & \sqrt{(1-b)(a+b)}& b
\end{pmatrix} ,\ea
with
\ba\label{30}
a&=&\frac{[2]}{[2j][2j+1]} < 1\q ,\q\q b\,=\,\frac{[2]}{[2j+1][2j+2]}\leq 1 \q .
\ea

The two eigenvectors for eigenvalue $+1$ are given as
\ba\label{32}
v_1=(\sqrt{1-b} ,0,\sqrt{1-a} )\q,\q\q v_2=(-\sqrt{a+b}, \sqrt{1-a},0)
\ea
These are orthogonal to 
\ba\label{32b}
v_{3}=\tfrac{1}{\sqrt{2}}(\sqrt{1-a},\sqrt{a+b},-\sqrt{1-b}) \q .
\ea

Thus  we obtain the recursion (using  $a'(j_1,j_2,j_3)=\sigma_{j_1}\sigma_{j_2}\sigma_{j_3} a(j_1,j_2,j_3)$)
\ba\label{35}
 \sigma_{j+1}\,a(j,j+1,1)^2&\!\!=\!\!& \sqrt{\frac{[2j+3]}{[2j-1]}} \sqrt{\frac{1-a}{1-b}}\,\,\sigma_{j-1} \,a(j-1,j,1)^2  \nn\\&& 
  +  \sqrt{\frac{[2j+3]}{[2j+1]}}   \sqrt{\frac{a+b}{1-b}} \,\, \sigma_{j}\, a(j,j,1)^2  \nn\\
  &\underset{(\ref{qind1},\ref{qind2})}{=}&     \frac{[2j+2]}{[2j]}      \,\, \sigma_{j-1}\, a(j-1,j,1)^2    
  + \frac{[2]}{[2j]} \,\, \sigma_j\, a(j,j,1)^2  \q .
\ea
with $a$ and $b$ defined in (\ref{30}). In the classical limit $q\rightarrow 1$
\ba
 \sigma_{j+1}\,a(j,j+1,1)^2 &\underset{q=1}{=}& \frac{j+1}{j} \,\,\sigma_{j-1} \,a(j-1,j,1)^2 \,\,+\,\,  \frac{1}{j} \,\, \sigma_j\, a(j,j,1)^2  \q .
\ea
As $a(0,1,1)^2=d_1$, the initial data are given by $a(1,1,1)^2$ and $\sigma_0,\sigma_1$. 




Finally we can use the bubble move (with the outer edges carrying $j_3=1)$ to fix the initial datum $a(1,1,1)$ for the recursion relations.

\subsection{Recursion relations for the remaining amplitudes}

\begin{figure}[bt]
\begin{center}
       \includegraphics[scale=0.27]{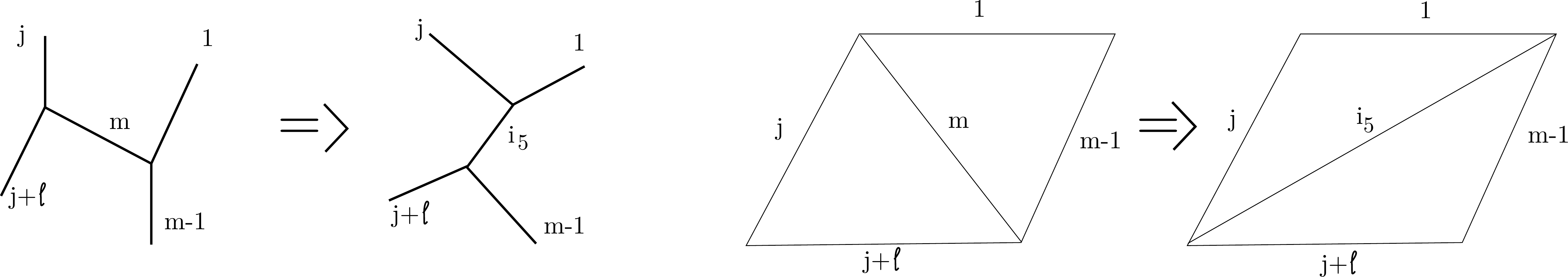}
    \end{center}
    \caption{\small \label{2dpic} The 2--2 move relations (\ref{36}) and their geometrical interpretation as re--gluing two triangles.}
\end{figure}

For amplitudes where all arguments are $j\geq 2$ we can use the 2--2 moves directly to obtain recursion relations. To this end consider the diagram $A_6(1,j,j+l,m-1;m)$, see figure (\ref{2dpic})  in which the amplitudes
\ba\label{36}
&&a'(m,j,j+l)\,a(m,m-1,1)\,=\,a(j+l,m,j)\,a'(1,m,m-1) \nn\\
&=&\sum_{i_5=\max(j-1,j-m+l+1)}^{ 
j+1
} \Fj{1}{j}{i_5}{j+l}{m-1}{m} \sqrt{\frac{d_m}{d_{i_5}}}\,\, a'(1,j,i_5) \,a(j+l,m-1,i_5) \q . 
\ea
appear. 
Knowing all amplitudes in which spins up to $(m-1)$ appear as arguments, there will be only one potentially unknown amplitude in (\ref{36}), namely $a'(m,j,j+l)$. This will give  all amplitudes in which one of the arguments is equal to $m$.

We can restrict to the following range of spins, either due to coupling rules, or because we would otherwise obtain an equation only involving already  known amplitudes:
\ba\label{47}
2\leq m \leq \tfrac{k}{3} ,\q\q m\leq j \leq \tfrac{k}{2}-1 ,  \q\q 0\leq l \leq m ,\q\q m+2j+l\leq k \q .
\ea
The sum (\ref{36}) includes maximally three terms, for $l =m-1$ two terms and collapses to one term for $l=m$.  In this case the coefficients also simplify. The $[6j]$ symbols and these special cases are given in appendix \ref{appC}.

To determine the sign function  $\sigma(j,m,j+l)$ (assuming sign functions are known for all cases where one of the arguments is smaller $m$) one can for instance consider the same diagram but with the second and third outer edge exchanged, i.e. $A_6(1,j+l,j,m-1;m)$. This gives the following equation 
\ba\label{36b}
&&a'(m,j+l,j)\,a(m,m-1,1)\,=\,a(j,m,j+l)\,a'(1,m,m-1) \nn\\
&=&\sum_{i_5=j+l-1}^{\min(j+m-1,j+l+1
)} \Fj{1}{j+l}{i_5}{j}{m-1}{m} \sqrt{\frac{d_m}{d_{i_5}}}\,\, a'(1,j+l,i_5) \,a(j,m-1,i_5) \q . 
\ea
The number of summands is the same as for the previous sum (\ref{36}), i.e. one for $l=m$, two for $l=m-1$ and three otherwise. The relevant $[6j]$ symbols are displayed in appendix \ref{appC}.

~\\

This gives in principle all the amplitudes, depending on a choice of $a(1,1,1)$ and some sign functions. For the construction of these amplitudes we only used a specific set of the 2--2 Pachner moves, hence one has to check the bubble moves for $j_3\geq 1$ and all other 2--2 moves. This will in particular fix the value for $a(1,1,1)$.

Furthermore one should be aware that `branchings' can arise, if certain amplitudes are set to zero. We will discuss such examples in section \ref{solving}.

\section{Solutions to the recursion relations}\label{solving}

Here we are going to solve the recursion relations for the amplitudes $a(j,j,1)$ and $a(j,j+1,1)$. We will choose all sign functions to be positive. This will indeed lead to a large family of fixed points, however in almost all cases certain amplitudes will be imaginary. With our choice of signs this applies to amplitudes whose arguments sum to an odd number. (Examining the bubble move condition one sees that the imaginary nature is necessary to obtain only positive contributions to the bubble move equation.) 

Let us start with recursion relation (\ref{rec1}), which for positive permutation signs is
\ba
a(j+1,j+1,1) &=& 
a(j,j,1) \frac{ \sqrt{[2j+3] [2j] [j+2]^2}}  { \sqrt{[2j+1] [2j+4][j]^2}   }  \q .
\ea
The solution can be readily found to be
\ba\label{recsol1}
a(j,j,1)&=&a(1,1,1) \sqrt{ \frac{[4]}{[2][3]} [2j+1] \frac{[j]^2[j+1]^2}{[2j][2j+2]     }       } \q .
\ea
Next consider the recursion relations for $a(j,j+1,1)^2$
\ba\label{53}
a(j,j+1,1)^2&\!\!=\!\!&    \frac{[2j+2]}{[2j]}     a(j-1,j,1)^2    
  + \frac{[2]}{[2j]} \ a(j,j,1)^2  \q .
\ea
Let us introduce the coefficients $X^0(j)$ and $X^1(j)$ by
\ba
a(j,j+1,1)^2&=& X^0(j)+X^1(j) a(1,1,1)^2 \q .
\ea
The recursion relation (\ref{53}) can then be easily solved for $X^0(j)$ (with $X^0(0)=a(0,1,1)^2=d_1=[3])$ giving
\ba\label{recsol2}
X^0(j)&=&\frac{[3]}{[2]} [2j+2] \q .
\ea
Using this solution for $X^0(j)$ we arrive at the following equation for $X^1(j)$ 
\ba
X^1(j)=\frac{[2j+2]}{[2j]} X^1(j-1)  \,\, +\,\, [2j+2]\left(  \frac{[4]}{[3]}[2j+1] \frac{[j]^2[j+1]^2}{[2j]^2[2j+2]^2}\right)
\ea
with initial value $X^1(0)=0$. This is a recursion relation which is sourced by an inhomogeneous term. It is solved by the sum
\ba
X^1(j)&=&[2j+2]\frac{[4]}{[3]} \sum_{m=1}^j [2m+1]\frac{[m]^2[m+1]^2}{[2m]^2[2m+2]^2} \,\,=\,\, [2j+2]\frac{[4]}{[3]} \frac{[j][j+1]^2[j+2]}{[2]^2[2j+2]^2}\q .
\ea
The last equation, i.e. the evaluation of the sum, is proven in appendix \ref{appb}.

This gives the explicit solutions to the recursion relations. We are left with determining $a(1,1,1)$. Here, it turns out that we can choose $a(1,1,1)$ such that only representations $j\leq J$ appear. Amplitudes with any argument having representation $j>J$ are vanishing. (We have also to set $a(0,j,j)$ (and all permutations) for $j>J$ to zero.) One might be worried that this violates the recursion relations  (\ref{rec1}), however remember that these only hold for $j$ such that $a(j,j+1,1)$ is not vanishing.  Thus we actually require $a(J,J+1,1)=0$ and this will determine $a(1,1,1)^2$:
\ba\label{recsol3}
a(1,1,1)^2 &=& - \frac{X^0(J)}{X^1(J)} \,\,=\,\, -\frac{[3]^2[2]}{[4]} \frac{[2J+2]^2}{[J][J+1]^2[J+2]} \,\,\underset{q=1}{=}\,\,-\frac{18}{J^2+2J}
\ea
so that
\ba\label{recsol4}
a(j,j+1,1)^2 &=& X^0(j)\left(1-  \frac{X^0(J)}{X^1(J)}  \frac{X^1(j)}{X^0(j)}  \right) \q .
\ea
We see also the need for imaginary amplitudes. An exception is provided by the cases with even level $k$ and $J$ equal to the maximal representation $J=j_{max}=\frac{k}{2}$.  In this case $a(1,1,1)=0$ (as $X^0(j_{max})=0$ for even level) and hence also $a(j,j,1)=0$ for all $j$. For odd level we have $j_{max}=\frac{k-1}{2}$ and $a(1,1,1)$ will be non--vanishing for $J=j_{max}$, but decreasing with the level growing.  In general all amplitudes whose arguments sum to an odd number are either imaginary or vanish. We can use any sign for the square root of $a(j,j+1,1)^2$ in (\ref{recsol4}).

The surprising fact is that this choice for $a(1,1,1)$ leads also to the bubble condition being satisfied, which we can show explicitly for $j=1$. 

To this end we have to show that
\ba\label{ind1}
\sum_{j=0}^J a(j,j,0)^2 &=& -\frac{1}{d_1}\sum_{j=1}^J a(j,j,1)^2  + \frac{2}{d_1}\sum_{j=0}^{J-1}a(j,j+1,1)^2 \q ,
\ea
where $d_1=[3]$ and $a(j,j,0)^2=[2j+1]$. Putting in the solutions for $a(j,j,1)$ and $a(j,j+1,1)$ there are two summations we can perform right away (\cite{biedenharn}, p.58)
\ba\label{ind2}
\sum_{j=0}^{J-1} [2j+2]  &=& [J+1][J] \q ,\nn\\
\sum_{j=0}^J [2j+1] &=& [J+1]^2 \q .
\ea
Using the q-number identity 
\ba
[2][J+1]^2 -2 [J][J+1]\,=\, [2J+2]
\ea
which follows from (\ref{qind1},\ref{qind2}) we can rewrite (\ref{ind1}) into
\ba\label{ind3}
\frac{[J][J+1]^2[J+2]  }{[2J+2]} \,=\, [2] \sum_{j=1}^J [2j+1] \frac{[j]^2[j+1]^2}{[2j][2j+2]} - 2\sum^{J-1}_{j=0} \frac{[j][j+1]^2[j+2]  }{[2j+2]}  \q .
\ea
We prove this equation by induction. After checking the case $J=1$ the induction step $J\rightarrow J+1$ amounts to the statement
\ba
[J+2][J+3][2J+2]\,+\,[J][J+1][2J+4]\,-\,  [2][2J+3][J+1][J+2] \,=\,0\, .\q
\ea
This equation can be shown to hold by applying repeatedly the q--number identities (\ref{qind1},\ref{qind2}).

Thus the bubble move condition holds for the outer edges carrying representations $j=1$. Amplitudes with all arguments $j\geq 2$ are determined by the recursions (\ref{36}).  

In section \ref{cdl} we will provide an alternative construction of these fixed point amplitudes which will show that all 2--2 move and bubble move conditions are satisfied.

We have found a family of fixed points indexed by (a) the level $k$ and (b) the maximal representation $J$. These fixed points generalize to the classical group $SO(3)$ and to quantum groups $SO(3)_q$ with $q$ real, for which we can allow arbitrary maximal representation $J$. (Indeed the solutions to the recursion relations simplify considerably and the bubble move condition can now be shown by explicit summation.) For $J\rightarrow \infty$ we will obtain an infinite constant in the bubble (and 3--1 moves), indicating the same kind of non--compact gauge symmetry as for the 3D Ponzano Regge model based on $SU(2)$.

We also see here nicely the principle of triangulation invariance (related to a discrete notion of diffeomorphism symmetry as argued in \cite{bahretal09,bd12a}) at work.  The triangulation invariant models are basically determined by the amplitude $a(1,1,1)$ for the smallest (non--degenerate) triangle.  Amplitudes for larger triangles are obtained by gluing (e.g. through 2--2 moves) these basic triangles together. The amplitudes $a(0,j,j)$ are fixed according to whether the representation $j$ as argument leads to vanishing amplitudes or not.

Also for fixed points where $a(1,1,1)=0$ we expect that all amplitudes are determined from the amplitudes associated to the  `smallest' triangles. Thus the method of the recursion relations seems to allow the construction of all fixed points for this class of models.

\section{Fixed point models:  Examples}\label{examples}



%
%
%
%

\subsection{All representations excited and $k$ even }

The coarse graining algorithm starting with  initial data where all amplitudes are set to one leads to a fixed point in which all representations are excited. However, as mentioned in section \ref{solving}  the amplitudes $a(j,j,1)$ are vanishing. The non--vanishing amplitudes  are all positive. 

%
To obtain the amplitudes with all entries $j\geq 2$ we can use the 2--2 move relations (\ref{36}) with $m=2,3,\ldots,\frac{k}{3}$.  
Using these relations one can show (iteratively in $m$ and by setting $l=m-(2n+1)$ in (\ref{36}) with $n\geq 0$) that amplitudes $a(j_1,j_2,j_3)$ with $j_1+j_2+j_3$ odd are vanishing. 

For $k$ even, we will have $j_{max}=\frac{k}{2}$ with quantum dimension one. In general $j$ and $j_{mir}=j_{max}-j$ will have the same quantum dimension. Indeed the amplitudes are (in a certain sense, as triangle conditions have to be satisfied) invariant under this reflection symmetry, for instance $a(j,j+1,1)=a(j_{mir},j_{mir}-1,1)$.


%

\subsection{All representations excited and $k$ odd}

\subsubsection{Fibonacci: $k=3$}

For $k=3$, which gives the so called Fibonacci fusion category $SO(3)_{3}$, we just have the two representations $j=0,1$ and the only non--trivial coupling rule is $1\otimes 1=0\oplus 1$. Hence none of the recursion relations applies and we can fix the only non--trivial amplitude $a(1,1,1)$ by the bubble move. This leads to the condition $a(1,1,1)^2 \sigma_1=-\sigma_0$.  With $a(1,1,1)=\pm 1$ and $\sigma_1=-\sigma_0$ we obtain a real solution that satisfies all Pachner moves. This also applies to the choice  $a(1,1,1)=i$ and $\sigma_1=+\sigma_0$.

\subsubsection{$k$ odd and $k\geq 5$}

Here a certain subset of amplitudes has to be imaginary. This can be either the subset for which the sum of the arguments is odd (as in section \ref{solving}). The recursion relations (\ref{36}) indeed just give (purely) imaginary amplitudes for this subset. In this case all sign functions are $+1$ which makes this choice the most convenient one. Alternatively one can also choose the complementary set of amplitudes to be imaginary, this will then require sign functions $\sigma_j=-\sigma_0$  (and $\sigma(j,j+1,1)=\sigma_1)$.

%
%
Also, as in the $k$ even case, all contributions to the bubble move will be positive (or all negative for $\sigma_0=-1$). Indeed, the amplitudes for $k$ odd and $k$ even approach each other for $k$ large, in particular as the imaginary amplitudes are decreasing (at fixed $j$'s) with the level $k$ growing. 

A interesting point to make is that although we did not allow complex amplitudes in the coarse graining algorithm we found these fixed points also numerically. However instead of true fixed points, we obtained (small) fixed points cycles, in which only the signs of the amplitudes changed periodically.


\subsection{ Only $j=0,1$ excited}

%
As the simplest example for a fixed point in which only a certain number of representations are excited we can consider the case that only $j=0,1$ appear. 
The amplitude $a(1,1,1)=i \sqrt{[3]([3]-1)}$. (However also a real choice of amplitudes is possible if one chooses $\sig_1=-\sig_0$.) We will later see that these fixed points lead to (quantum deformations of)  generalized AKLT states \cite{aklt} and hence these fixed points can be understood to describe the Haldane phase \cite{guwen,wenclass}.

\subsection{ Further examples}

We constructed explicitly the families of fixed points where representations $j=0,1\ldots,J$ appear and in particular $j=1$. With the coarse graining algorithm one can find more examples, in which $j=1$ is not excited. A simple example, which exist for all even $k$, is with only $j=0$ and $j_{max}$ are excited. In other examples (which might exist only for specific levels) there are only $j=0,j_{max}, j_{max}/2$ excited  (for $j_{max}$ an even number and even level) or $j=0,j_{max},(j_{max}-1)/2, (j_{max}+1)/2$ (for $j_{max}$ an odd integer and even level). These cases in which only very few excitations appear can be checked explicitly and conditions on the $[6j]$ symbols (that is on the level) can be derived. There are however also examples in which for instance all the even $j$ are excited.   

A systematic investigations  and classification of this more complicated fixed point structure will appear elsewhere \cite{bw}. 
 Here we will   just describe a  set of examples, which will be relevant for making to anyon  condensation in section \ref{condense}.

\subsubsection{Only $j=0,j_{max}$ excited}\label{Z2}

For even level $k$ we have that $j_{max}=k/2$ has quantum dimension equal to one. We also have the coupling rule $j_{max}\otimes j_{max}\equiv 0$. Thus we can consider the case where the only non--vanishing amplitudes are given by
\ba
a(0,0,0)=a'(0,0,0)=1 \q , \q\q a(j_{max},j_{max},0)=a'(j_{max},j_{max},0)=1
\ea
and the amplitudes obtained by permutation of the arguments. For $k=2$ this example coincides with the $J=2$ model. 

One can check that all 2--2 moves and the bubble move are satisfied for this example.

\subsubsection{Only $j=0,j_{max}/2,j_{max}$ excited}\label{symmm}

Here we will consider the level $k$ to be a multiple of $4$, that is $k=4 l$, so that $j_{max}=2l$. We then try to find fixed point amplitudes, such that only $j=0,l,2l$ are excited. For the case $k=4$ such an example is provided by the $J=2$ model. 

As $j=2l$ has quantum dimension equal to one, we will assume that the amplitudes involving $j=2l$ are equal to the ones, obtained by replacing $j=2l$ by $j=0$, as long as these are allowed by the coupling rules. 

This leaves as only non--trivial amplitude $a(l,l,l)$. The bubble move with $j=l$ at the outer edges gives
\ba\label{sund1}
(-)^l\,a(l,l,l)a'(l,l,l) &=& [2l+1]([2l+1]-2)  \q .
\ea
For $l=1$ and hence $k=4$ the RHS of (\ref{sund1}) vanishes, which is consistent with the $J=2$ model. 

This fixes (the absolute values of) all amplitudes. We have to check the 2--2 moves. For instance the move involving $A_5(l,l,l,2l;l)$ requires that
\ba
 \Fj{l}{l}{l}{l}{2l}{l}&=&+1 \q .
\ea
This is the case for $k=8$ and $k=16$ but not for $k=12$ (where it gives $-1$).

Thus we stick to even $l$ and fix the sign of the amplitudes
\ba
a(l,l,l)=a'(l,l,l)&=& \sqrt{ [2l+1]([2l+1]-2)} 
\ea
(The other non--vanishing amplitudes are given by (\ref{jzero}) with all sign functions being equal to one.)

A rather non--trivial condition arises from the 2--2 move with $A_5(l,l,l,l;l)$. This move is only satisfied if
\ba
 \Fj{l}{l}{l}{l}{l}{l}&=& \frac{[2l+1]-4}{[2l+1]-2} \q .
\ea
This indeed holds for $k=8$ and $k=16$, but not for $k=24$ and not for $k=32$.  

One can check that all other 2--2 moves are satisfied for $k=8$ and $k=16$.

\subsubsection{Only two representations excited}\label{tworep}

Here we consider the case that only $j=0$ and another representation $j=j_1$ lead to non--vanishing amplitudes. Moreover, we choose $j_1$ such that $j_1 \otimes j_1$ includes $j_1$ as a summand.

From the bubble move equation we obtain the condition
\ba\label{oct0}
(-)^{j_1} a(j_1,j_1,j_1) a'(j_1,j_1,j_1) &=& [2j_1+1] ([2j_1+1]-1) \q .
\ea
Here we see that $[2j_1+1]=1$ leads to a solution with $a(j_1,j_1,j_1)=0$, which is the one described in section \ref{Z2}.  Let us therefore assume that $[2j_1+1]>1$. The 2--2 move with all outer edges carrying the representation $j_1$ leads to the condition
\ba\label{oct1}
 \Fj{j_1}{j_1}{j_1}{j_1}{j_1}{j_1}&=& \frac{[2j_1+1]-2}{[2j_1+1]-1} \q .
\ea
Using the formula (\ref{6jb}) for the $[6j]$ symbol  one can show that this is satisfied for $j_1=1$ for all $k\geq 3$. Indeed this leads to the family in section \ref{solving} with $J=1$.

There are however other pairs $(k,j_1)$, for which (\ref{oct1}) is satisfied, for instance $(k=6,j_1=2)$ and $(k=10,j_1=3)$.

For $(k=6,j_1=2)$ we therefore choose the amplitudes
\ba\label{oct2}
&&a(0,0,0)=a'(0,0,0)=1\q,\q\q a(j_1,j_1,0)=a'(j_1,j_1,0)=\sqrt{[2j_1+1]} \nn\\ &&a(j_1,j_1,j_1)=a'(j_1,j_1,j_1)=\sqrt{[2j_1+1]([2j_1+1]-1)} \q .
\ea
All other non--vanishing amplitudes are obtained by permutations of (\ref{oct2}).

For $(k=10,j_1=3)$, because of the sign in (\ref{oct0}), we change the last amplitudes in (\ref{oct2}) to
\ba\label{oct3}
a(j_1,j_1,j_1)=a'(j_1,j_1,j_1)=i\sqrt{[2j_1+1]([2j_1+1]-1)} \q .
\ea
For both cases one can check that all the 2--2 moves and the bubble moves are satisfied.
\\
~\\

The method of using first the bubble move to fix a maximal set of amplitudes and then to check the 2--2 moves can be generalized and allows to establish the existence (or non--existence) of certain types of fixed points, which implement certain effective coupling rules. In particular 2--2 moves with  a diagram $A_5(j_1,j_2,j_3,j_4,j_5)$ where $j_5$ is a `forbidden' representation due to the effective coupling rules, will rule out many possibilities. However the method gets naturally very involved, the more allowed representations appear.


\section{Partition function for the torus}\label{torus}

We introduced the intertwiner models as partition functions associated to a box -- with edges entering the bottom of the box and emerging from the top of the box. If we associate a triangulation to the three--valent graph inside this box we will obtain a triangulation of a disc, whose boundary is partitioned into two components. 

One might ask whether it is possible to associate a partition function to other two--dimensional manifolds, also to manifolds without boundary. We will extend here the definition to the cylinder and to the torus. The partition function of the torus gives the degeneracy of the ground states associated to the fixed points model for periodic boundary conditions, as will be explained further below. This will give important hints on the structure of the fixed point model, i.e. whether it represents a symmetry broken phase or not. 

\begin{figure}[bt]
\begin{center}
       \includegraphics[scale=0.15]{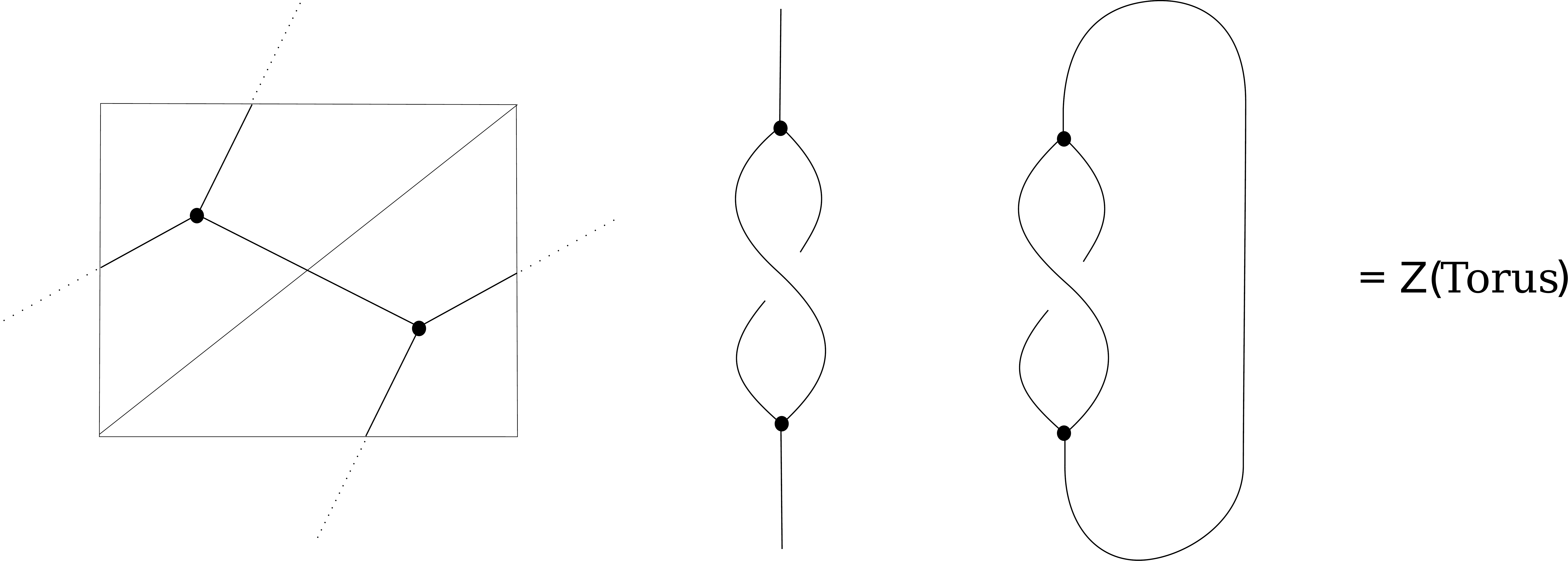}
    \end{center}
    \caption{\small \label{tor1} The two left panels show a simple triangulation of a cylinder and its dual graph. The right panel shows how to identify the bottom and top boundary of the cylinder to obtain a torus. The corresponding graph (with fat vertices) can be evaluated within diagrammatical calculus and defines the partition function of the torus.}
\end{figure}

A minimal triangulation of the cylinder is given by two triangles, as shown in figure \ref{tor1}.

Obviously we cannot draw the dual triangulation onto the plane without crossing the strands. We cannot ignore the crossing -- as it exchanges two factors in the tensor product of representation spaces and the core of the quantum group is to render this non--commutative. Indeed ignoring the crossing leads to a map between bottom and top boundary Hilbert space that is not an intertwiner. 

However, the quantum group provides us with a notion of crossing and (the inverse) anti--crossing
\ba
\begin{tikzpicture}
\node[right] at (-2,0.1) {$R\,\,\,=$};
\draw[thick] (-0.3,-0.3) --(0.3,0.3);
\node[left] at (-0.2,-0.2) {1};
\draw[thick] (0.3,-0.3) --(0.1,-0.1);
\node[right] at (0.2,-0.2) {2};
\draw[thick] (-0.1,0.1) --(-0.3,0.3);
\node at (5.5,0.0) {
\begin{tikzpicture}
\node[left] at (-0.8,0.3) {$=\q\sum_{j} [2j+1] q^{-\tfrac{1}{2}(j_1(1_1+1)+j_2(j_2+1)-j(j+1))}$};
\draw[thick] (-0.3,-0.3) --(0.0,-0.0);
\draw[thick] (0.3,-0.3) --(0.0,-0.0);
\node[right] at (0.2,-0.2) {2};
\draw[thick] (0.0,-0.0) --(0.0,0.4);
\node[right] at (0.0,0.2) {$j$};
%
\draw[thick] (-0.3,0.7) --(0.0,0.4);
\node[left] at (-0.2,0.6) {2};
\draw[thick] (0.3,0.7) --(0.0,0.4);
\node[right] at (0.2,0.6) {1};
\draw[thick] (0.0,0.4) --(0.0,-0.0);
\node[left] at (-0.2,-0.2) {1};
\end{tikzpicture} \q ,
};
\end{tikzpicture}
\ea
\ba
\begin{tikzpicture}
\node[right] at (-2,0.1) {$R^{-1}=$};
\draw[thick] (-0.3,-0.3) --(-0.1,-0.1);
\node[left] at (-0.2,-0.2) {1};
\draw[thick] (0.3,-0.3) --(-0.3,0.3);
\node[right] at (0.2,-0.2) {2};
\draw[thick] (0.1,0.1) --(0.3,0.3);
\node at (5.5,0.0) {
\begin{tikzpicture}
\node[left] at (-0.8,0.3) {$=\q\sum_{j} [2j+1] q^{\tfrac{1}{2}(j_1(1_1+1)+j_2(j_2+1)-j(j+1))}$};
\draw[thick] (-0.3,-0.3) --(0.0,-0.0);
\draw[thick] (0.3,-0.3) --(0.0,-0.0);
\node[right] at (0.2,-0.2) {2};
\draw[thick] (0.0,-0.0) --(0.0,0.4);
\node[right] at (0.0,0.2) {$j$};
%
\draw[thick] (-0.3,0.7) --(0.0,0.4);
\node[left] at (-0.2,0.6) {2};
\draw[thick] (0.3,0.7) --(0.0,0.4);
\node[right] at (0.2,0.6) {1};
\draw[thick] (0.0,0.4) --(0.0,-0.0);
\node[left] at (-0.2,-0.2) {1};
\end{tikzpicture} \q .
};
\end{tikzpicture}
\ea

We can therefore associate the dual graph in figure \ref{tor1}  to the simplest triangulation of a cylinder. For the torus we have to identify the bottom with the top or in other words perform a trace. Here we have to take first  the quantum trace in a given representation, which can be obtained as a combination of unit and co--unit (or cap and cup), see appendix \ref{appA}. Secondly we have to sum over the representations. This results in the evaluation of the diagram in figure \ref{tor1}, which gives
\ba\label{torus1}
Z(\text{Torus})&=& \sum_{j_1,j_2,j_3} a(j_1,j_2,j_3)a'(j_1,j_2,j_3) (-)^{2j_1+2j_2} q^{\tfrac{1}{2} (j_1(j_1+1) +j_2(j_2+1) - j_3(j_3+1))}
\ea 

Let us discuss the consistency of this definition. Going back to the case of the cylinder, one would expect that gluing the partition functions of two cylinders (i.e. summing over the  representations associated to the glued edges) one again obtains the partition function for the cylinder. In other words the partition function for the cylinder, which is rather a map from the bottom to the top boundary Hilbert space, should be a projector.

Having just one boundary edge, makes the understanding of this projector difficult, as the maps have to be intertwiners. Intertwiners between irreducible representation spaces are just multiples of the identity or equal to the zero map.

\begin{figure}[bt]
\begin{center}
       \includegraphics[scale=0.15]{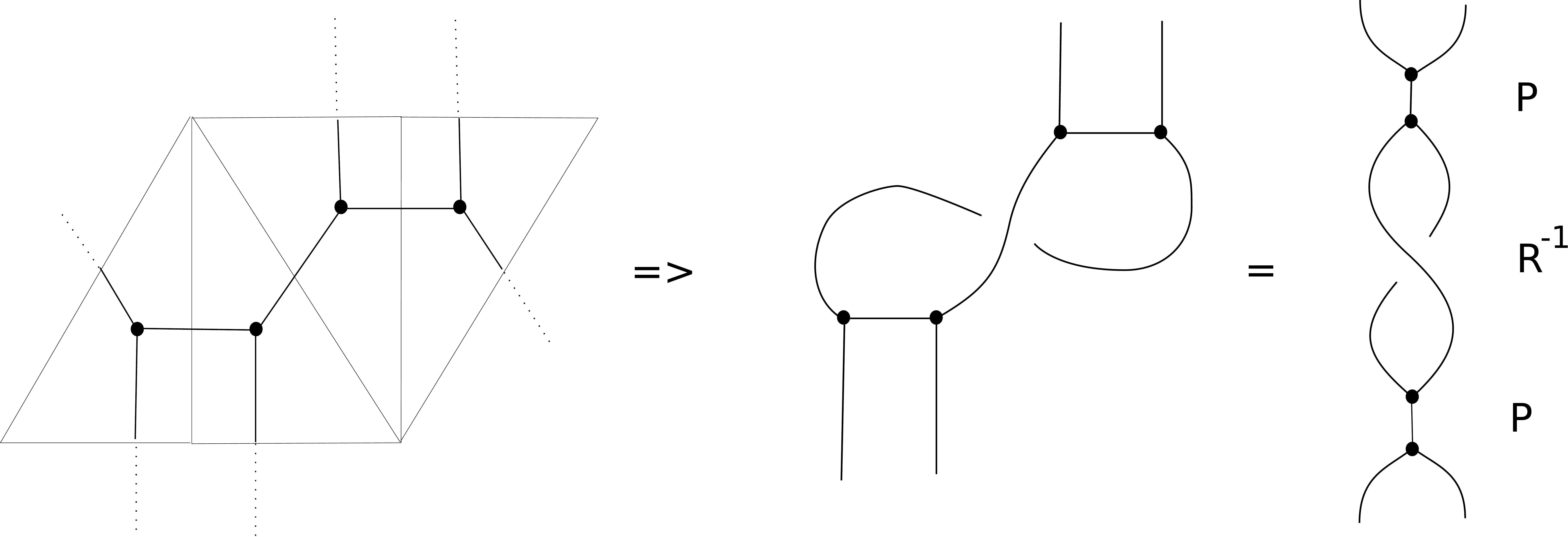}
    \end{center}
\caption{ A refined triangulation of the cylinder and the associated dual graph. We apply the 2--2 move to reach the graph on the right \label{tor2}.}
\end{figure}

\begin{figure}
\begin{center}
\includegraphics[width=0.5\textwidth]{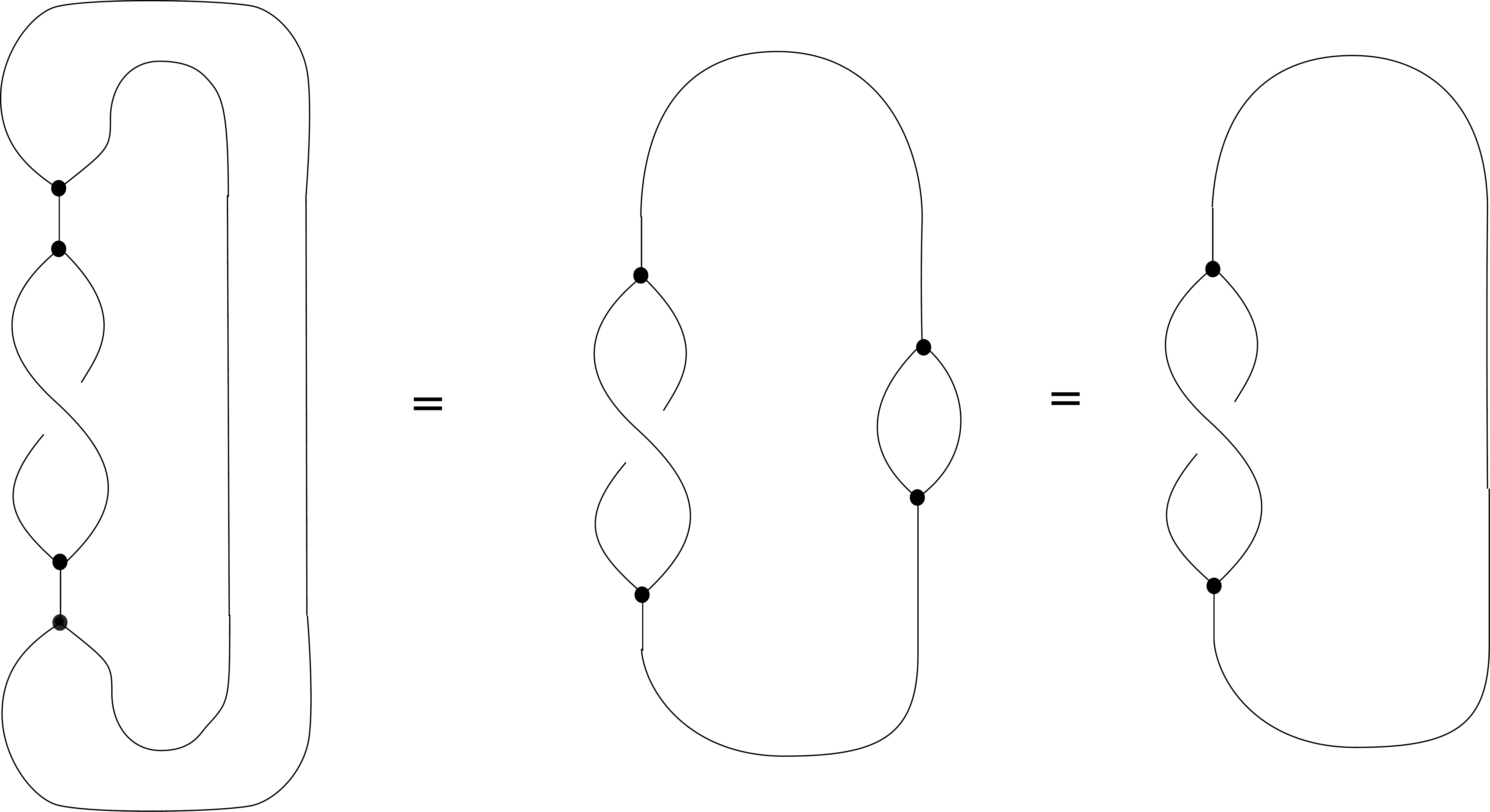}\hspace{2.5cm}
\includegraphics[width=0.15\textwidth]{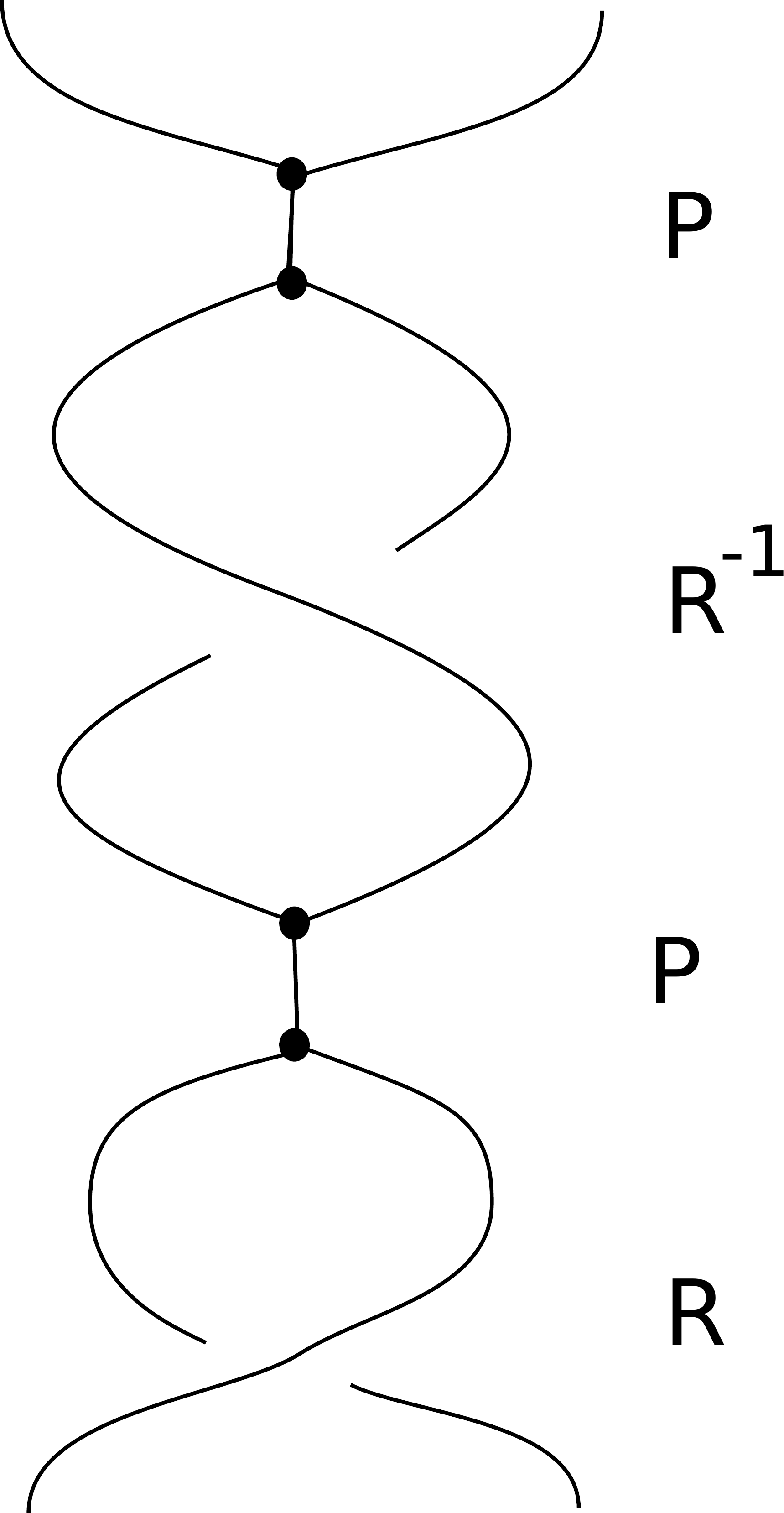}
\caption{ The  diagrammatic equations on the left show how to evaluate the graph associated to the torus, obtained by closing the cylinder in figure \ref{tor2}.  The figure on the right shows a choice of dual graph for the cylinder, giving a projector map.
  \label{tor3}}
\end{center}
\end{figure}

Let us therefore consider the slightly more complicated triangulation of the cylinder in figure \ref{tor2}. This can be nicely understood as a combination of three maps $ P\circ R^{-1}\circ P$, where \ba
\begin{tikzpicture}
\node[left] at (-0.5,0.3){$P=$};
\draw[thick] (-0.3,-0.3) --(0.0,-0.0);
\draw[thick] (0.3,-0.3) --(0.0,-0.0);
\node at (0,0)  {$\bullet$};
\draw[thick] (0.0,-0.0) --(0.0,0.4);
%
\draw[thick] (-0.3,0.7) --(0.0,0.4);
\draw[thick] (0.3,0.7) --(0.0,0.4);
\node at (0,0.4)  {$\bullet$};
\draw[thick] (0.0,0.4) --(0.0,-0.0);
\end{tikzpicture} \q .
\ea
$P$ itself is indeed a projector, as follows from the bubble move.  For this amplitudes have to be normalized such that $c=1$, which we will assume in this section.   Also we can compute the partition function of the torus by taking the quantum trace over the tensor product of representation spaces, as shown in figure \ref{tor3}. Using diagrammatical calculus (and the tilting condition for the amplitudes) this agrees with the previous definition in figure \ref{tor1}, as can be seen from figure \ref{tor3}. Just taking the quantum trace of $P$ will give a different result equal to $\sum_j\Theta(j) d_j (-1)^{2j}$. Thus the braiding is important to impose a two site projector $P$ not only on sites  ordered as $(1,2)$ but also on the ordering $(2,1)$.


However $P\circ R^{-1}\circ P$ is in general not a projector.  Squaring $P\circ R^{-1}\circ P$ might be understood as twisting the cylinder twice. Thus, if we consider doubling the height of the cylinder, we rather should use $ R \circ  P\circ R^{-1}\circ P$. This is indeed a projector, and thus can be taking to give the partition function of the cylinder with two boundary edges. 
Taking the quantum trace we get in our examples the same result for the torus partition function as previously.

%

We established the cylinder as a projector. The range of this projector can be understand as the span of ground states of a Hamiltonian, which itself can be written as a sum of projectors. For the two edge cylinder we have (see also the first reference in \cite{vidalsymm})
\ba
H=(\mathbb{I}-P)+(\mathbb{I}- R\circ P\circ R^{-1}) \q .
\ea
Thus taking the (quantum) trace of the cylinder we obtain the ground state degeneracy (generalized to a system with quantum group symmetry) of this Hamiltonian.

Finally, let us give the results for the torus partition functions. For the models parametrized with a maximal spin $J$ an explicit numerical evaluation of (\ref{torus1}) gives $Z(\text{Torus})=1$. Thus we can assume that these have no ground state degeneracy and are in a symmetry unbroken phase. Indeed, in the next section \ref{cdl}, we will show that the amplitudes for this models can be reconstructed as a so called Corner Double Line structure, which is only possible for systems with unique ground states \cite{guwen,schuch}.

For the models introduced in section \ref{Z2} the torus partition function (\ref{torus1}) can be evaluated explicitly with the result
\ba
Z(\text{Torus})\,=\, \frac{3}{2}+\frac{1}{2} \exp(\frac{2\pi i \,k}{4})  \q .
\ea
As $k$ is even for these models we obtain $Z=2$ for $k$ a multiple of $4$ and $Z=1$ otherwise. This agrees with $Z=1$ for the $k=2,J=1$ model.

Thus for $k$ a multiple of $4$, we can assume that the ground states are degenerate and the fixed point describes a `symmetry broken' phase. 
This broken symmetry can be either some subset of the quantum group symmetry or some additional symmetry of the model. In the latter case, the fixed point will in general not be stable under perturbations, which violate this symmetry (as these will lift the ground state degeneracy), i.e. one would have to exclude such perturbations in order to obtain a stable phase. It would be therefore interesting to clarify this symmetry.

For the models discussed in section \ref{symmm}, with three excited representations, and for $k=8,16$ we also obtain $Z(\text{Torus})=2$, indicating degenerate ground states.

Finally, for the models in section \ref{tworep}, with two representations excited, we find  for the case $(k=6,j_1=2)$, that $Z(\text{Torus})=1$. But for the case $(k=10,j_1=3)$ we find $Z(\text{Torus})=d_{j=0}+d_{j=3} \approx 4.73$. 

We will explain the significance of these results for the theory of anyon condensation in section \ref{condense}.

We can in general expect that phase transitions occur between phases with differing ground state degeneracy. We have found several different cases of ground state degeneracies. However, phase transitions also occur between phases with the same ground state degeneracy. One mechanism for this to happen is symmetry protected topological order, as discussed in \cite{guwen,wenclass}. Thus, as we will find in the next sections, the $J$ even and the $J$ odd models can be understood to be in two different phases with respect to the $SO(3)_k$ symmetry. This is a generalization of the two different $SO(3)$ phases that arise in the classification of phases with (proper) groups \cite{wenclass,schuch}.

\section{Structure of  the fixed point amplitudes}\label{cdl}

Previously we constructed a family of fixed points via the recursion relation which exemplifies the geometric interpretation of the models. It turns out that this fixed point family is of a particular form, known as Corner Double Line (CDL) structure \cite{guwen,levinslides}. 

For models described by a CDL structure we replace every edge by two strands or lines. That is edges are now ribbons. Again we can understand a single strand as describing a map  $V\rightarrow V$. (One could associate $V$ to the left strand and $V^*$ to the right strand but we will work with  associating $V \otimes V$  to the strands.)

\begin{figure}[t]
\begin{center}
       \includegraphics[scale=0.35]{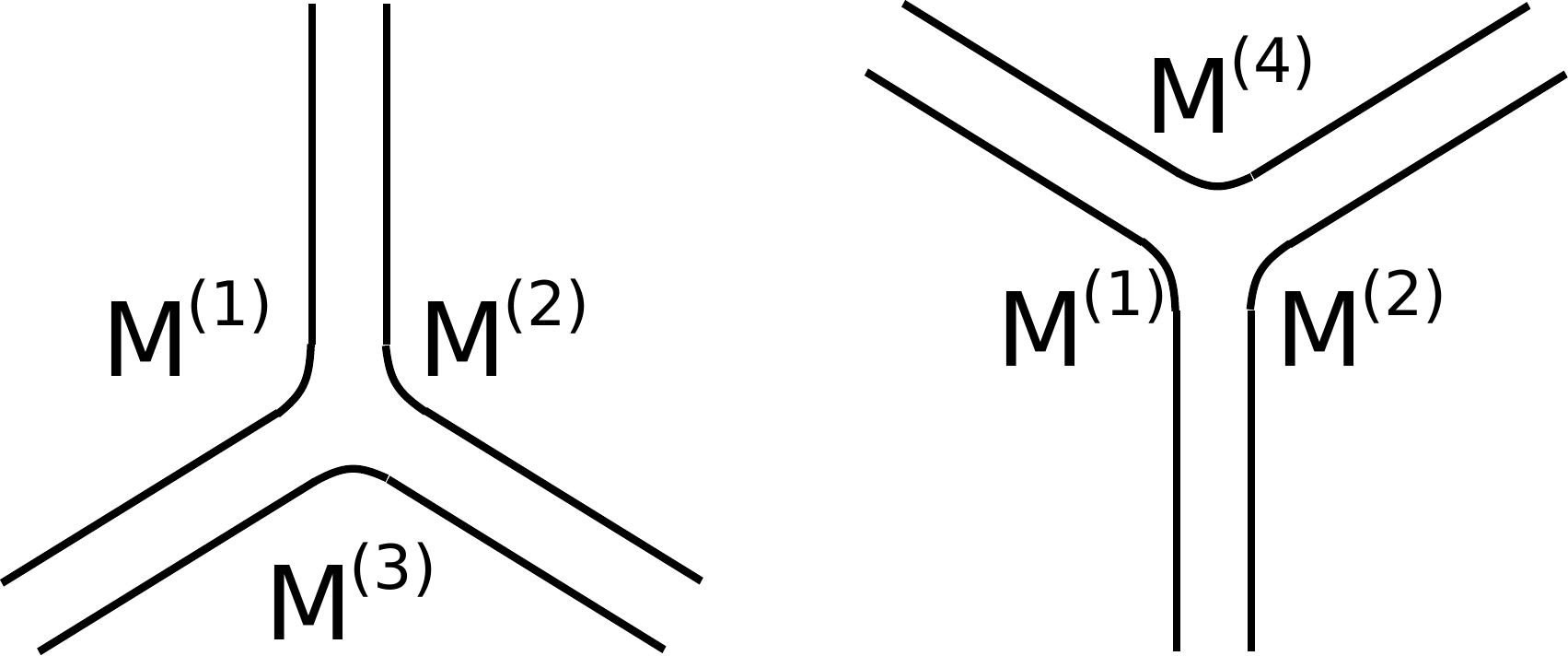}
    \end{center}
    \caption{\small \label{fig1} The matrices $M^{(i)}$ are attached to the corners of the double line vertices.}
\end{figure}

The vertices are depicted in figure \ref{fig1} and are parametrized by four matrices $M^{(i)}$, $i=1,\ldots 4$, attached to the `corners' of the (bloated) vertices, and describing maps between the appropriate vector spaces: $M^{(i)},i=1,2$ are maps from $V$ to $V$, $M^{(3)}$ is a map $V\otimes V\rightarrow \mathbb{C}$ and  $M^{(4)}$ acts as $ \mathbb{C} \rightarrow V\otimes V$.

For the  tilting, 2--2 move  and bubble move conditions to be satisfied, the following requirements are sufficient.\footnote{If one considers coarse graining on a regular lattice, the fixed point conditions might be less restrictive than the full invariance under these moves. In this case the matrices are allowed to be more general. For instance for the square lattice the $M$ can be arbitrary, leading to the problem of non--isolated fixed points \cite{guwen}.}  $M^{(1)}$ and $M^{(2)}$, have to be projectors. Furthermore $M^{(1)}$ and $M^{(2)}$ have to stabilize the maps $M^{(3)}$ and $M^{(4)}$ in the following sense:
\ba\label{cdl1}
M^{(3)} &=& M^{(3)} \circ (M^{(1)} \otimes \mathbb{I}) \,=\, M^{(3)} \circ (\mathbb{I} \otimes M^{(2)} ) \nn\\
M^{(4)} &=& (M^{(1)} \otimes \mathbb{I})  \circ M^{(4)}  \,=\, (\mathbb{I} \otimes M^{(2)} ) \circ M^{(4)}  \q .
\ea

\begin{figure}[h!]
\begin{center}
       \includegraphics[scale=0.2]{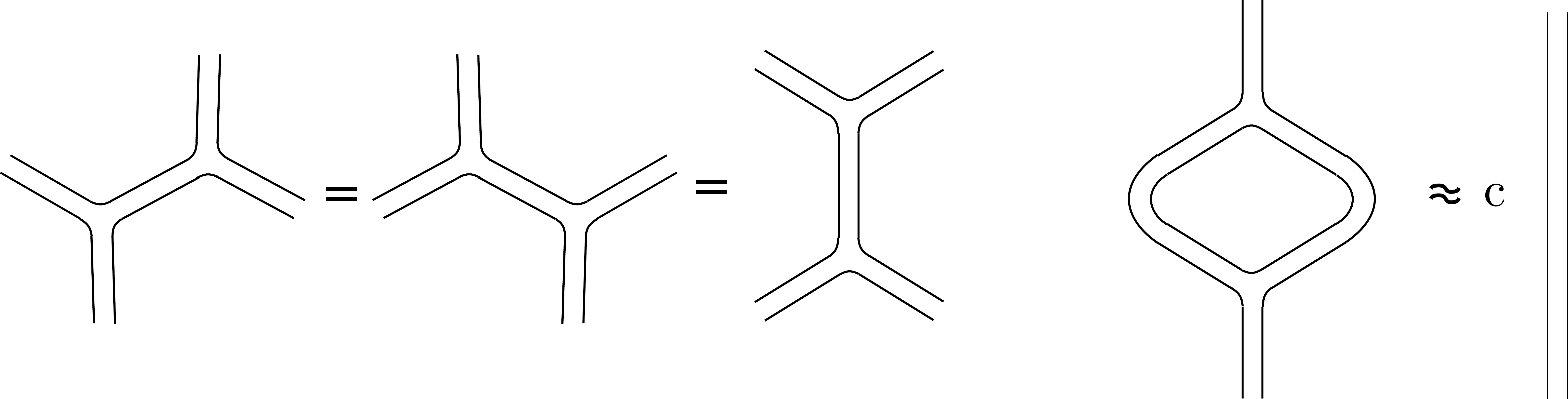}
    \end{center}
    \caption{\small \label{fig2} The 2--2 and bubble move conditions for  CDL structure. The bubble move gives a map that is proportional to the tensor product of projectors $M^{(1)}$ and $M^{(2)}$. However the projectors  can be absorbed into a neighbouring vertex.}
\end{figure}

Then the right hand side of the bubble move (see figure \ref{fig2}) is not given  by a multiple of the identity map but by multiples of the projector maps $M^{(1)}$ and $M^{(2)}$.  Due to the invariance  properties the maps can be absorbed into one of the neighbouring vertices.  Equivalently we can understand the strands as representing directly the projectors. 
The constant $c$ appearing in the bubble or 3--1 move (\ref{31p}) is given by
\ba
c&=& M^{(3)} \circ M^{(4)}    \q ,
\ea
which is a linear map: $\mathbb{C}\rightarrow \mathbb{C}$ and hence just given by a complex number.

We can then re--construct the models described by a maximal representation $J\leq \tfrac{1}{2}k$ (where $J$ is integer) as follows. We take as vector space $V$ the representation space $V_{J/2}$. The maps $M^{(1)}, M^{(2)}$ are given by the identity maps and the maps $M^{(3)},M^{(4)}$ are given by the co--unit and unit map respectively (see appendix \ref{appA}).  
In this description we have vector spaces $V_{J/2} \otimes V_{J/2}$ associated to one edge. We can easily change to edge vector spaces $\oplus_{j} V_j$ by inserting a projector\footnote{For quantum groups at root of unity these project out the trace zero parts in the tensor product $V_{J/2} \otimes V_{J/2}$, see appendix \ref{appA}. However for $J\leq k/2$ the trace zero part is empty.}
\ba\label{proje}
\begin{tikzpicture}
\node[left] at (-0.7,0.2){$ \Pi^{\frac{J}{2} \frac{J}{2}} \,=\,\,\sum_{j =0}^{J} (-)^{J-j} \,[2j+1]$};
\node[right] at (0.2,0.6) {$\tfrac{J}{2}$};
\node[left] at (-0.2,0.6) {$\tfrac{J}{2}$};
\node[left] at (-0.2,-0.3) {$\tfrac{J}{2}$};
\node[right] at (0.2,-0.3) {$\tfrac{J}{2}$};
\draw[thick] (-0.3,-0.3) --(0.0,-0.0);
\draw[thick] (0.3,-0.3) --(0.0,-0.0);
\node[right] at (0,0.2) {$\small j$};
\draw[thick] (0.0,-0.0) --(0.0,0.4);
%
\draw[thick] (-0.3,0.7) --(0.0,0.4);
\draw[thick] (0.3,0.7) --(0.0,0.4);
\draw[thick] (0.0,0.4) --(0.0,-0.0);
\end{tikzpicture} 
\ea
at each of the two ends of each edge. 
\begin{figure}[bt]
\begin{center}
       \includegraphics[scale=0.15]{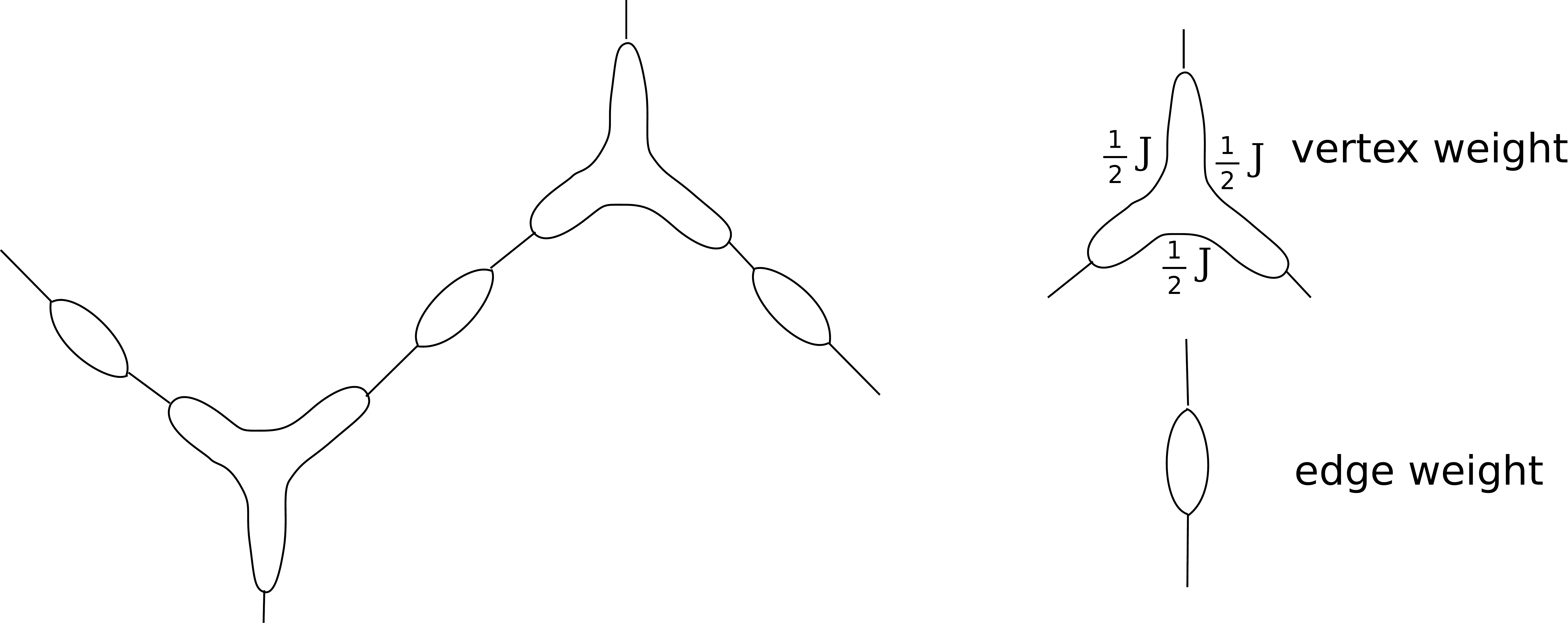}
    \end{center}
    \caption{\small \label{wurst} After inserting two projectors (\ref{proje}) on each edge of the CDL fixed points the model can be expressed in terms of vertex and edge weights. }
\end{figure}
Figure \ref{wurst}  then shows that we obtain a model with vertex and edge weights. The edge weights can be absorbed into the vertex weights  by taking square roots (this leads to imaginary amplitudes, alternatively one might leave the edge weights, or associate the sign of an edge weight always to the upper or always to the lower vertex).  This leads to vertex amplitudes 
\ba\label{cdlvertex1}
a_{CDL}(j_1,j_2,j_3)\,=\, a'_{CDL}(j_1,j_2,j_3) &=& \sqrt{(-)^{J-j_1}} \sqrt{(-)^{J-j_2}}  \sqrt{(-)^{J-j_3}} (-)^{2J-j_1-j_2} \times \nn\\&& \q\q\q\q\q\q \frac{[2j_1+1]^{1/2} [2j_2+1]^{1/2}}{[J+1]^{1/2} }  \Fj{j_1}{j_2}{j_3}{\tfrac{J}{2}}{\tfrac{J}{2}}{\tfrac{J}{2}} \q .
\ea
The amplitudes are invariant under permutations of the arguments. It can be shown that the amplitudes $a_{CDL}(j,j,1)$ and $a_{CDL}(j,j+1,1)$ coincide modulo signs with the amplitudes (\ref{recsol1}) and (\ref{recsol4}) obtained by solving the recursion relations. The difference in signs is due to choices for the $\pm$ branches for the square roots, which have to be made both for (\ref{cdlvertex1}) and for the solutions of the recursion relations. (In general we can multiply  $j_I$ dependent signs $\text{sign}(j_1)\text{sign}(j_2)\text{sign}(j_3)$  to the vertex amplitudes without changing the (bulk) partition function. The signs cancel out, as an edge connects two vertices.)

\begin{figure}[bt]
\begin{center}
       \includegraphics[scale=0.2]{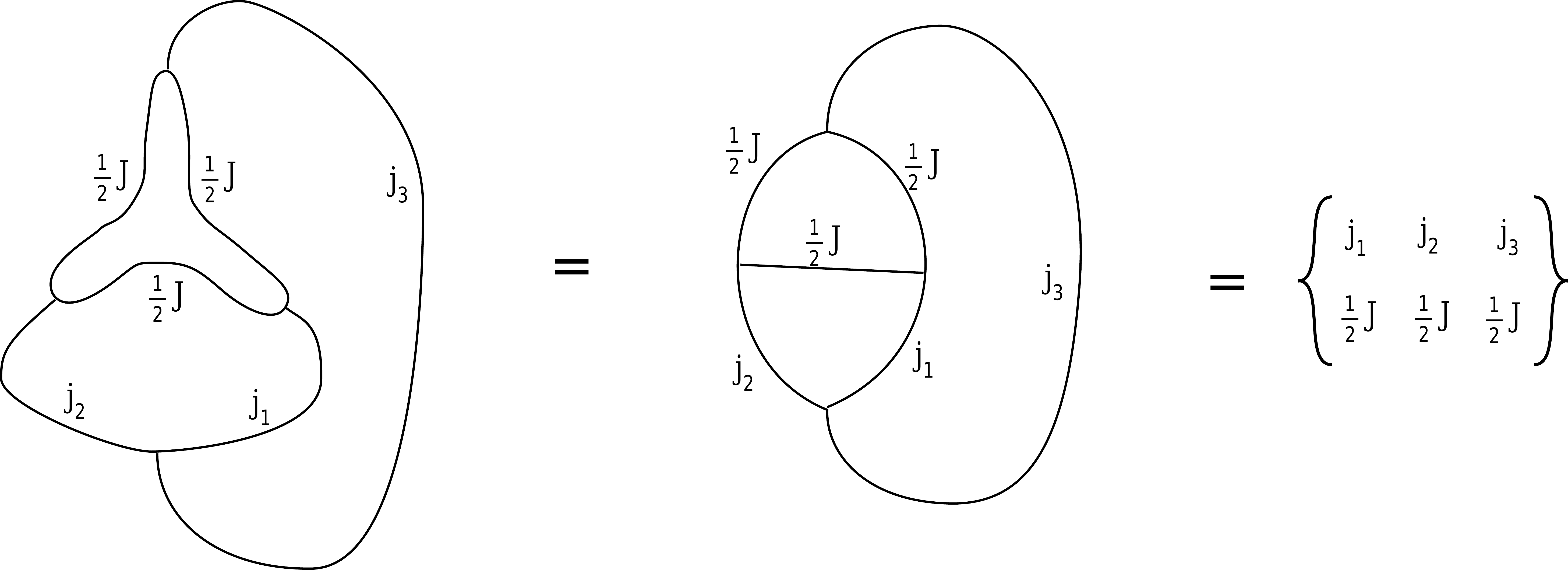}
    \end{center}
    \caption{\small \label{cdlweights} This figure shows how to arrive at the $[6j]$ symbol for the vertex weights.  The vertex weight in figure \ref{wurst} represents an intertwiner and hence can be expressed in an intertwiner basis. This leads to the $\{6j\}$ symbol shown here. See appendix \ref{appA} for the definition of the $\{6j\}$ and $[6j]$ symbols. }
\end{figure}


In section \ref{3Drecursion} we will prove that the amplitudes also satisfy the recursion relations (\ref{36}), which are used to construct the remaining amplitudes. Indeed this shows that the 2D recursion relations (\ref{36}) can be interpreted to arise as projections of 3D recursion relations, that can be derived from the Biedenharn Elliot (aka pentagon) identity \cite{cranebarrett}.

Variants of such models can be easily constructed. For instance we can introduce twists of the ribbons (i.e. braidings). Having a twist and and anti--twist on each edge we again will not change the (bulk) partition function. This is an example of a `weak gauge transformation' for vertex models, which changes the weights but not the partition function by inserting a representation of unity $\mathbb{I}=UU^\dagger$ for each edge. 

Furthermore we can generalize the models by allowing different representations $J$ for the strands of the ribbons in a vertex. This leads to certain gluing rules for the vertices that can be encoded in a colouring (given by the representation $J$). For the triangulation it implies that vertices are colored  and triangles can be only glued so that shared vertices agree in their colour. Partition function with boundaries, which define intertwining operators from one boundary to the other, do not depend (modulo constants) on the choice of the representations for any strands forming closed loops.

Also  choosing the vector space $V$ as a tensor product or direct sum of representation spaces we will obtain fixed point models with edge Hilbert spaces $\oplus_j \mu(j)V_j$ with $\mu(j) \geq 1$.  In this case the fixed point models can be interpreted to be built from fundamental building blocks carrying more boundary data then in the $\mu(j)=1$ case. As is explained in \cite{bd12b} this allows to incorporate non--local couplings into the model, and in this way move away from topological models. (To obtain a field theory with propagating degrees of freedom in the limit we need however to allow infinite multiplicity.)

In section \ref{examples} we discussed further fixed point models and determined the associated partition function on the torus.  We argued that the resulting number gives the ground state degeneracy  of the Hamiltonian projector associated to the model. We found examples where this ground state degeneracy is larger than one. Such models cannot arise (purely) from CDL tensors. CDL tensors rather lead to a unique ground state \cite{schuch,guwen}. (In section \ref{interpretation} we will discuss the ground states in more detail.  However from the graphical representation of the ground states one can see that a CDL tensor leads to a ground state built from tensoring the unit maps $M^{(4)}$. Thus the ground states is given by a unique map from $\otimes^N \mathbb{C}\equiv \mathbb{C}$ to the corresponding boundary Hilbert space.)

However, a general classification of fixed points that can arise (satisfying certain stability requirements) is available \cite{schuch,wenclass}. This suggests that the fixed points with ground state degeneracy can be understood as direct sums of such CDL tensors. The direct sum structure might however only arise after a variable transformation (or field redefinition) has been performed. This will in general induce a mixing of the $j$-- representations.

A simple example -- where we ignore the structure in the magnetic indices -- is given by fixed points where only $j=0$ and $j=j_{max}$ are excited, with coupling rule $j_{max}\otimes j_{max} \equiv 0$. The amplitudes coincide with the Ising model in the symmetry broken phase. Hence  let us identify the two representations with two basic vectors $v_0$ and $v_1$. We introduce a new basis $v_{\pm}=\frac{1}{\sqrt{2}}(v_0 \pm v_1)$. The amplitudes in the new bases decouple completely between the $+$ and $-$ sector and hence these can be understood to give the two ground states.\footnote{One might wonder how the examples with exactly the same amplitudes but with no ground state degeneracy arise. This might be due to ignoring the magnetic indices in this discussion, the fact that trace zero parts have to be projected out, as well as the behaviour under braiding.}

\section{The 2D recursion relations as projection of 3D recursion relations} \label{3Drecursion}

We have seen that the amplitudes for the fixed points described by a maximal spin $J$ are given by $\{6j\}$ symbols in (\ref{cdlvertex1}). This allows to readily  obtain a geometric interpretation of the models, at least in the limit of large level $k$ and maximal spin $J$. The asymptotic behaviour of the $\{6j\}$ symbols is known  \cite{asympt1} to be given by the (cosine of the) 3D Regge action \cite{regge}, a simplicial discretization of the gravitational action.  
This persists to the quantum group case \cite{6jc}, for which one obtains the Regge action with a (positive for $q$ a root of unity) cosmological constant. For this the Regge action involving curved tetrahedra is appropriate \cite{bd09curved}.  Here we give the asymptotics for the classical $\{6j\}$ symbol which is
\ba\label{asump}
\Tj{j_1}{j_2}{j_3}{j_4}{j_5}{j_6} &\underset{\lambda \rightarrow \infty}{\rightarrow} &   \frac{1}{\sqrt{12\pi V(l_I)}}
\cos \left(S_{Regge}(l_I)+\frac{\pi}{4}\right)       \left(1+O(\frac{1}{\lambda} )\right) \q .
\ea
$S_{Regge}$ is a specific function \cite{regge} of the six lengths $l_I$ of a tetrahedron associated to the $\{6j\}$ symbol. The lenghts are related to the spins by $l_I=j_I+\frac{1}{2}$. For this limit we have to assume a homogeneous scaling of the spins $j \sim \lambda$. This allows to arrive at a classical action principle for our models, where also the issue of the sum over orientations (leading to the cosine in \ref{asump})  \cite{orient}  can be studied in a simpler context.

\begin{figure}[bt]
\begin{center}
       \includegraphics[scale=0.3]{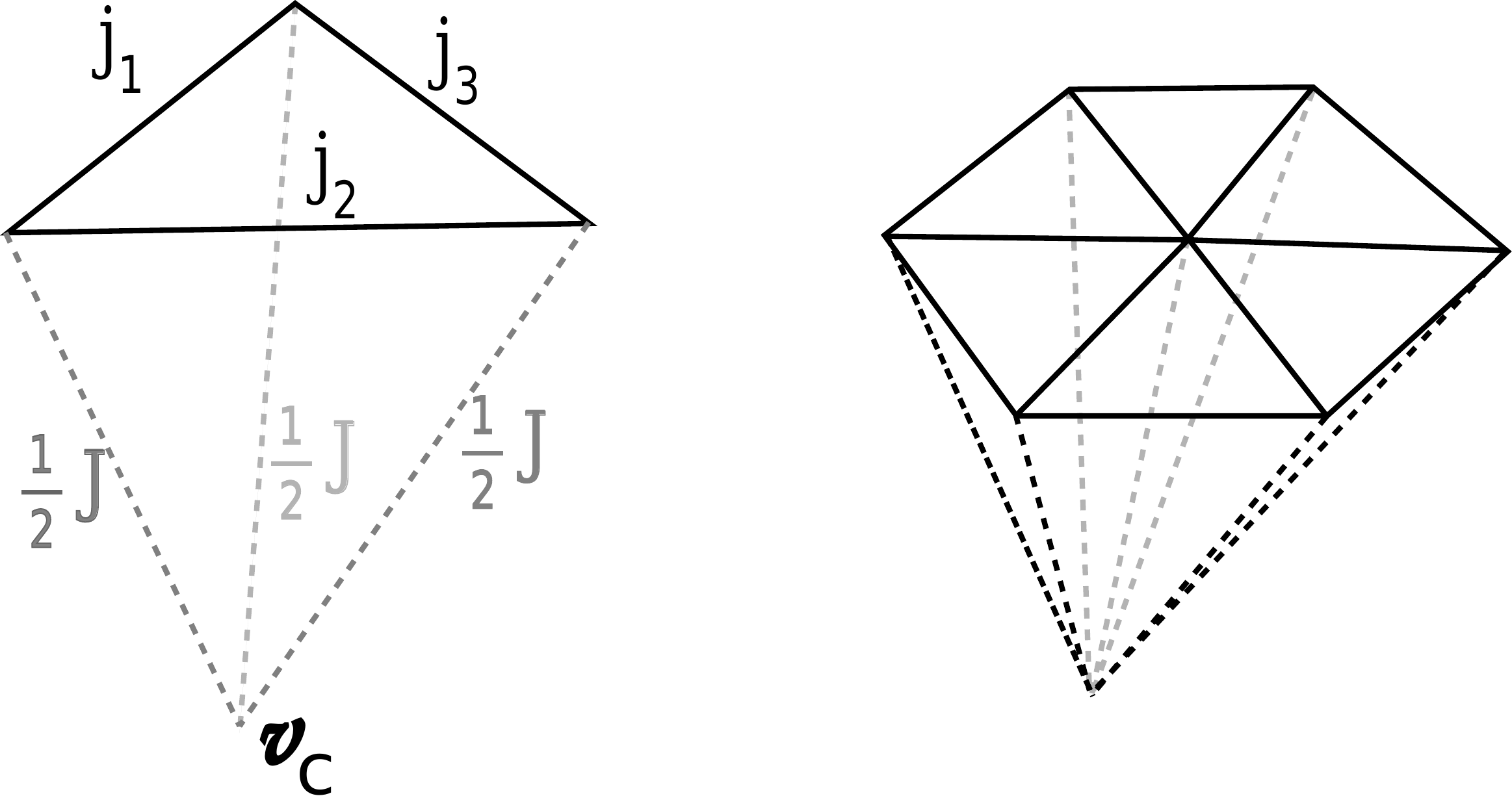}
    \end{center}
    \caption{\small \label{3dpic} The CDL fixed point amplitudes lead to $\{6j\}$ symbols. These can be associated to tetrahedra. In this way the 2D triangulation of a disc arises as a top of a triangulation of a 3D cone.}
\end{figure}

Thus the fixed point models parametrized by a maximal representation $J$ arise as  boundary of a specific 3D triangulation, see figure \ref{3dpic}. All tetrahedra $\tau$ in this triangulation have three edges of length $J/2 \approx J/2+1/2$, starting from a vertex $v_\tau$. The other three edges form a triangle and carry the dynamical degrees of freedom. The tetrahedra are glued along the $J/2$ edges, so that all vertices $v_\tau$ are identified to one vertex. That is we obtain the triangulation of a cone, with a 2D disc at the top of the cone. The fixed point models are associated to this disc. 

There is  an interesting connection to 3D topological models here, as the Ponzano--Regge \cite{ponzanoregge} or Tuarev--Viro \cite{tuarev} models are built from $\{6j\}$ symbols associated to tetrahedra. 

For these models the Biedenharn--Elliott identity is essential, and is responsible for the invariance of the models under 3--2 Pachner moves. We can obtain the 2--2 move invariance for our amplitudes also from the Biedenharn--Elliott identity. We will apply this to an `infinitesimal version' of the Biedenharn--Elliott identity, which allows to derive the Hamiltonian constraints for the Ponzano--Regge model \cite{cranebarrett,bonzomfreidel}, and will lead to the recursion relations (\ref{36}) in our case. This makes the geometric interpretation of these recursion relations, as implementing a notion of vertex translations for the triangulation  obvious, as the same interpretation holds in the 3D case \cite{bd08}.

\begin{figure}[bt]
\begin{center}
       \includegraphics[scale=0.4]{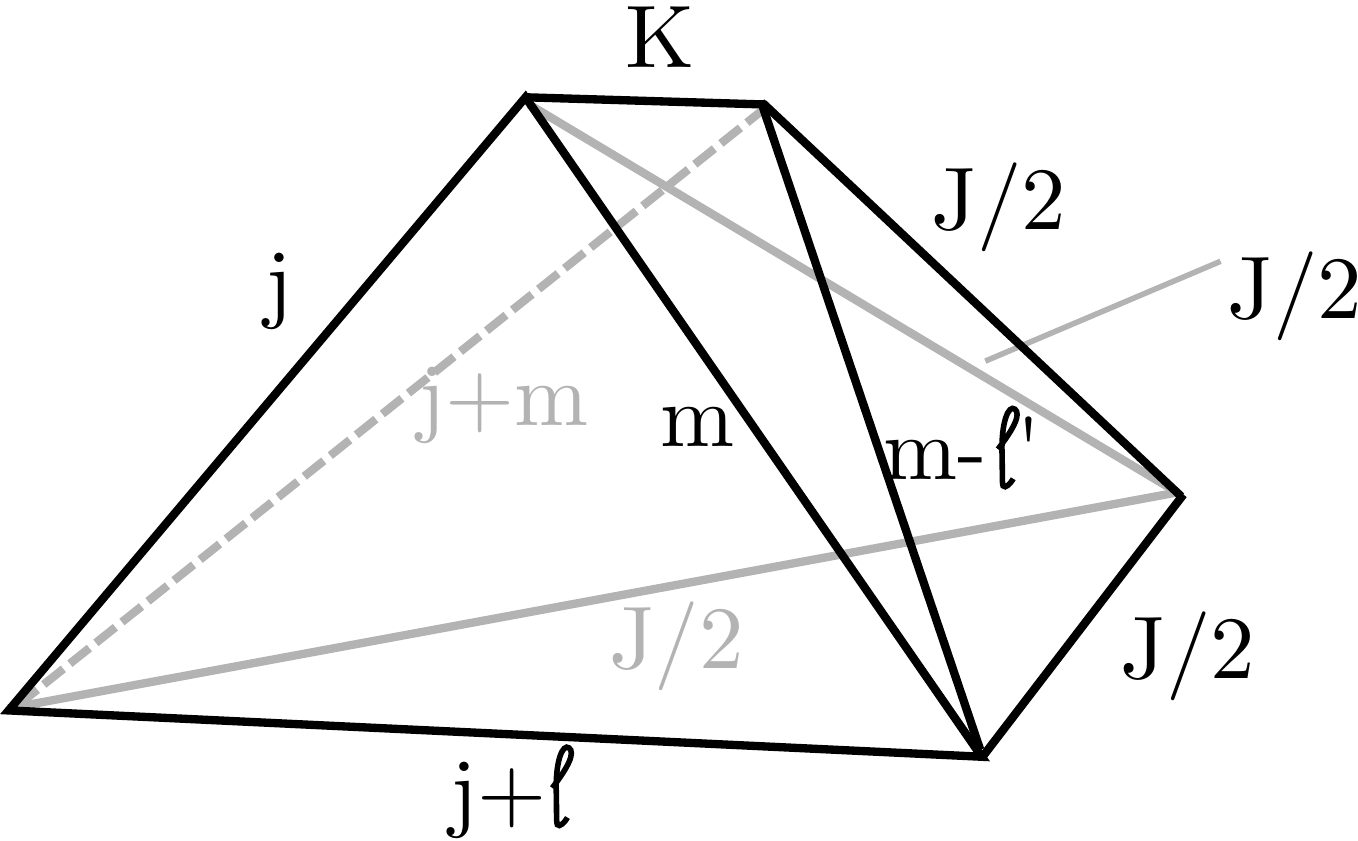}
    \end{center}
    \caption{\small \label{figBE} The 3D complex for the Biedenharn--Elliot identity.}
\end{figure}

The Biedenharn--Elliott identity involves two $\{6j\}$ symbols on the one side and three on the other side of the equation
\ba
&&\Tj{k_1}{k_2}{k_3}{j_1}{j_2}{j_3}\Tj{k_1}{k_2}{k_3}{l_1}{l_2}{l_3} \,= \nn\\ &&\q\q\q
\sum_n(-)^{n+\sum k_I+\sum l_I +\sum j_I} [2n+1] 
\Tj{k_1}{j_2}{j_3}{n}{l_3}{l_2}\Tj{j_1}{k_2}{j_3}{l_3}{n}{l_1}  \Tj{j_1}{j_2}{k_3}{l_2}{l_1}{n}   \q ,
\ea
describing a complex of two tetrahedra on the left hand side and the gluing of three tetrahedra around one edge, labelled by $n$ on the right hand side.
The work \cite{cranebarrett} relates this equation to the Hamiltonian constraints for 3D gravity. Following \cite{cranebarrett} we apply the Biedenharn--Elliott identity to the complex depicted in figure \ref{figBE}. 

This results into
\ba\label{be1}
&&\Tj{K}{m}{m-l'}{J/2}{J/2}{J/2}\Tj{j}{j+l}{m}{J/2}{J/2}{J/2} \,= \nn\\ &&
\sum_{n=-K}^K \sigma\,\, [2(j+n)+1] 
\Tj{K}{j}{j+n}{j+l}{m-l'}{m}\Tj{j+n}{j+l}{m-l'}{J/2}{J/2}{J/2}  \Tj{j+n}{j+l}{m-l'}{J/2}{J/2}{J/2}   \q \q\q
\ea
with $\sigma=(-)^{2J+3j+l-l'+2m+n+K}$.

Using the amplitudes for the fixed point models
\ba
a^{(')}(j_1,j_2,j_3) =  \sqrt{(-)^{J-j_1}}  \sqrt{(-)^{J-j_2}} \sqrt{(-)^{J-j_3}} (-)^J  \, \sqrt{[2j_1+1]  [2j_2+1]  [2j_3+1]}   \Tj{j_1}{j_2}{j_3}{J/2}{J/2}{J/2}
\ea
we arrive at the equation
\ba
&&a'(m,j,j+l)a(m,m-l',K)\,=\, \nn\\ &&\q\q\sum_{n=-K}^K\Fj{K}{j}{j+n}{j+l}{m-l'}{m} \sqrt{ \frac{[2m+1]}{[2(j+n)+1]} } a'(K,j,j+n) \, a(j+l,m-l',j+n) \q .
\ea
This is a generalization of  the previous recursion relation (\ref{36}), for which we need to choose $K=1,l'=1$.

An interesting question is whether also the other fixed points (or fixed points which describe non--local couplings) can be constructed from this 3D model -- possibly by introducing some summation of degrees of freedom in the 3D bulk. A further question is whether 3D invariant models can be built in a similar way from 4D topological theories. See also \cite{chenetal} for topological models arising as boundaries of higher dimensional models in a different context.

\section{Matrix product states and parent Hamiltonians}\label{interpretation}

We constructed  families of partition functions  invariant under change of triangulation. 
In section \ref{models} we defined the models not for arbitrary 2D manifolds but only for the disc, whose boundary is partitioned into two parts. One part corresponds to the lower boundary of the `box'  discussed in relation with graphical calculus in section \ref{models} and the other part to the upper boundary.  The edges $e_1,\ldots,e_N$ entering the bottom of the box and the edges $e_{N+1},\ldots, e_{N+M}$ emerging from the top of the box correspond to  Hilbert spaces
\ba
{\cal H}_b= \oplus_{j_1,\ldots ,j_N} V_{j_1} \otimes \cdots \otimes V_{j_N}  \q \text{and} \q {\cal H}_t= \oplus_{j_{N+1},\ldots ,j_{N+M}} V_{j_{N+1}} \otimes \cdots \otimes V_{j_{N+M}} \q .
\ea
As previously discussed the intertwiner models define $SU(2)_k$ intertwining maps between these Hilbert spaces. These maps can be also seen as transfer operators of the system. (Here we have a generalized notion of transfer operators as in principle the Hilbert spaces can differ in dimension.)

For $N=M$ the triangulation invariance of the models implies that these transfer operators are actually projectors. (We need to rescale all amplitudes by $\sqrt{c}^{-1}$, where $c$ is the constant appearing in the bubble move (\ref{bubble}), so that one obtains a true fixed point under coarse graining.) These projectors specify physical states as elements of their image \cite{proj,nouip,finite} and allow the definition of a physical inner product. In the language of Dirac quantization for constraint systems the projector specifies a rigging map \cite{algquant}. 

This implies that there are constraints which are implemented by the projector. As we will argue below these constraints can be interpreted as simplicity constraints\footnote{Due to the triangulation independence there is also an alternative interpretation as Hamiltonian and diffeomorphism constraints, as we will discuss in section \ref{discuss}.}, in the sense that a certain set of representations is forbidden.

Concretely, fixing the Hilbert spaces ${\cal H}_t= {\cal H}_b$ (i.e. fixing $N$ and $M=N$), we can consider the following bulk triangulation interpolating between the corresponding boundaries:
\ba \label{proj1}
\begin{tikzpicture}
\draw[thick] (0,0)--(2,-2);
\node at (0.5,-0.5)  {$\bullet$};
\draw[thick] (0.5,-0.5)--(1,0);
\node at (1,-1)  {$\bullet$};
\draw[thick] (1,-1)--(2,0);
\node at (2,-2)  {$\bullet$};
\node at (2,-1) {$\cdots $};
\draw[thick] (2,-2)--(4,0);
\draw[thick] (2,-2)--(2.5,-2.5);
\draw[thick] (2.5,-2.5)--(2,-3);
\node at (2,-3)  {$\bullet$};
\draw[thick] (2.5,-2.5) --(0,-5);
\draw[thick]   (2,-3) --(4,-5);
\node at  (1,-4)  {$\bullet$};
\node at (2.2,-4) {$\cdots$};
\draw[thick]  (1,-4) --(2,-5);
\node at (0.5,-4.5)  {$\bullet$};
\draw[thick] (0.5,-4.5)--(1,-5);
\end{tikzpicture}
\ea

One sums over the indices attached to all bulk edges, i.e. all edges which connect fat vertices. It is easy to see that this map defines a projector by putting two diagrams (\ref{proj1}) on top of each other (summing over all indices associated to the  glued edges) . One then applies the bubble move repeatedly from left to right and arrives back at the original diagram.

So what states does the projector leave invariant?  The rank of the projector can be maximally as large as the smallest possible horizontal cut through the diagram (\ref{proj1}). This is just given by a cut through the edge in the middle of the diagram (\ref{proj1}), i.e. the rank is equal to or smaller than $\text{dim}( \oplus_{\Theta(j)=1} V_j)$. 

We can find $\text{dim}( \oplus_{\Theta(j)=1} V_j)$ states that are left invariant by the projector. These states are defined by applying the following map to the basis $|j,m\rangle$ in $ \oplus_{\Theta(j)=1} V_j$: 

\ba \label{proj2}
\begin{tikzpicture}
\draw[thick] (0,0)--(2,-2);
\node at (0.5,-0.5)  {$\bullet$};
\draw[thick] (0.5,-0.5)--(1,0);
\node at (1,-1)  {$\bullet$};
\draw[thick] (1,-1)--(2,0);
\node at (2,-2)  {$\bullet$};
\node at (2,-1) {$\cdots $};
\draw[thick] (2,-2)--(4,0);
\draw[thick] (2,-2)--(2.5,-2.5);
\node[right] at (2.5,-2.5) {$|j,m\rangle$};
\end{tikzpicture}
\ea

Again, putting the projector (\ref{proj1}) on top of the map in (\ref{proj2}) shows that the  $\text{dim}( \oplus_{\Theta(j)=1} V_j)$ states are left invariant. 
%
Hence we identified the image of the projector and the physical Hilbert space. A similar (generalized) projector can be obtained for the $M\neq N$ -- in this case the co--kernel and the image are given by states spanned by (\ref{proj2}) with $M$ and $N$ upper edges respectively.

The diagram (\ref{proj1}) defines a global projector. One can also show that the states (\ref{proj2}) are invariant under the following local projectors acting on any two neighbouring sites
\ba\label{proj3}
\begin{tikzpicture}
\draw[thick] (-0.3,-0.3) --(0.0,-0.0);
\draw[thick] (0.3,-0.3) --(0.0,-0.0);
\node at (0,0)  {$\bullet$};
\draw[thick] (0.0,-0.0) --(0.0,0.4);
%
\draw[thick] (-0.3,0.7) --(0.0,0.4);
\draw[thick] (0.3,0.7) --(0.0,0.4);
\node at (0,0.4)  {$\bullet$};
\draw[thick] (0.0,0.4) --(0.0,-0.0);
\end{tikzpicture}
\ea
where again we sum over the spin associated to the middle edge. 
We can stack these together to built projectors of the form (\ref{proj1}):
\ba
\begin{tikzpicture}
\draw[thick] (-0.3,-0.3) --(0.0,-0.0);
\draw[thick] (0.3,-0.3) --(0.0,-0.0);
\node at (0,0)  {$\bullet$};
\draw[thick] (0.0,-0.0) --(0.0,0.4);
%
\draw[thick] (-0.4,1.2) --(0.0,0.4);
\draw[thick] (0.3,0.7) --(0.0,0.4);
\node at (0,0.4)  {$\bullet$};
\draw[thick] (0.0,0.4) --(0.0,-0.0);
\node at (0.3,0.7)  {$\bullet$};
\draw[thick] (0.3,0.7)--(1,-0.3);
\draw[thick] (0.3,0.7)--(0.3,1);
\node at (0.3,1) {$\bullet$};
\draw[thick] (0.3,1)--(0.1,1.2);
\draw[thick] (0.3,1)--(0.5,1.2);
\node at (1.8,0.7)  {$=$};
\draw[thick] (2.5,1.2)--(  3.25  ,0.45);
\draw[thick]  (3.25  ,0.45)--(2.5,-0.3);
\draw[thick] (2.7,-0.1) --(2.9,-0.3);
\node at (2.7,-0.1) {$\bullet$};
\draw[thick] (2.9,0.1) --(3.3,-0.3);
\node at (2.9,0.1) {$\bullet$};
\draw[thick]  (2.7,1.0)--(2.9,1.2);
\node at (2.7,1.0) {$\bullet$};
\node at (2.9,0.8) {$\bullet$};
\draw[thick]  (2.9,0.8)--(3.3,1.2);
\end{tikzpicture}
\q .
\ea

The interpretation of these projectors is that they project out any states in $\otimes_{I=1}^N( \oplus_{j} V_j)$, that in any coupling scheme would lead to the appearance  of representations $j$ with  $\Theta(j)=0$. 

The implementation of simplicity constraints for spin foams \cite{alexreview} leads to a set of representations that is allowed and representations that are forbidden. In this sense the fixed points implement simplicity constraints. In future work we will consider $SU(2)_k \times SU(2)_k$ and $SU(2)\times SU(2)$ models, which appear for the full spin foam models,  and investigate the fixed point structure there.


States in the form (\ref{proj2}) are known as matrix product states\footnote{The state (\ref{proj2}) represents a MPS with two free ends, if the `physical index' on the  top left is taken as a `virtual index'. 
} (MPS) \cite{mps}.   These provide ansaetze for ground state wave functions of Hamiltonians. 
In these kind of problems the Hamiltonian is given and one uses the MPS as a variational ansatz (where the amplitudes are varied) for the ground state(s). On the other hand one can also start from a given MPS and ask which Hamiltonian has this MPS as a ground state. This leads to the notion of parent Hamiltonians associated to an MPS. This relationship between MPS and parent Hamiltonian can be rendered one--to one by formulating certain conditions for the parent Hamiltonian and the MPS \cite{schuch}. One of the conditions is that the parent Hamiltonian is a sum over two--site local projectors.

We constructed such a Hamiltonian with (\ref{proj3}). Lets denote  the maps in (\ref{proj3}) by $P_I$ where $I$ denotes the site index say of the left vertex. As $P_I$ are projectors, so are $(\mathbb{I}-P_I)$. Moreover these are positive maps. Hence the states (\ref{proj2}) are ground states to 
\ba\label{100}
H=\sum_{I=1}^{N-1} (\mathbb{I}-P_I)  \q .
\ea
An alternative view point is to take the $ (\mathbb{I}-P_I)$  as a set of constraint operators and to interpret the system as one with gauge symmetries. Indeed these constraints should be related to the recursion relations used to construct the fixed points, see also \cite{cranebarrett, bonzomfreidel,recursion}. As argued in \cite{bahretal09,hoehn,bd12a} in the context of gravitational systems, we can interpret these gauge symmetries as diffeomorphisms. In particular here we have a geometrical interpretation of the representation labels as geometric length variables. In future work we will explore this analogy in more detail and investigate the (Virasoro) algebra of the constraints. 

 This provides then a realization of the path integral (partition function) (\ref{proj1}) providing the projector onto the physical states, see for instance \cite{nouip,finite} for a realization for 3D topological field theories. The reason why we do not need to integrate over infinitely many `time steps' in the path integral is rooted in the triangulation invariance (connected to diffeomorphism symmetry).

These remarks are very general and apply to any triangulation invariant  2D model\footnote{These kind of arguments can also be generalized to higher dimensions, the definition of ground states and parent Hamiltonians built from local projectors is of course more involved. See \cite{hoehn} for a canonical (classical) implementation of  Pachner moves, which could be used to build such local projectors.},  with the same graphical representation of ground states and parent Hamiltonians.

More concretely we can interpret our fixed point family parametrized by a maximal representation $J$ as providing a large generalization of  AKLT states \cite{aklt}. The AKLT Hamiltonian is defined on the $N$--fold tensor product of the $j=1$ representation space. (This can be generalized to other representations, see for instance \cite{iran}. For a $q$--deformed version of the AKLT system see \cite{qaklt}.)  The Hamiltonians project out the spin $j=2$ part from the tensor product of any two neighbouring $j=1$ spins. Indeed the valence bond construction of the AKLT states pairs two spin $j=1/2$ spins to a spin $j=1$ site. This picture corresponds to the CDL construction in section \ref{cdl}, where for $J=1$ the single lines in the Double Lines denote $J/2=1/2$ representation spaces.

For $q=1$ we obtain the standard version of the AKLT states, with (proper) symmetry group $SO(3)$. The phases described by MPS with on-site symmetry described by a group can be classified \cite{wenclass,schuch}. For on-site $SO(3)$ symmetry and systems with non--degenerate ground states (as is the case for the $J$ family of fixed points), there are two classes of states, which are specified by the two elements in the second cohomology group $H^2(SO(3),U(1))$ and giving rise to projective representations of $SO(3)$. The half spin representation of $SU(2)$ correspond to the choice of the non-trivial cocycle. 

These two classes of states define two different phases, where the phase corresponding to the choice of non--trivial cocycle is an example of symmetry protected topological order \cite{wenclass}. Phase transitions occur between these two different phases \cite{guwen}.

Here the AKLT states with odd maximal representation $J$ provide the non--trivial class of examples. This is due to the appearance of half spin representations for the virtual bonds. Thus we can expect phase transitions to occur between phases with $J$ even and $J$ odd. Although this classificaion only applies to the case of proper groups, it is quite likely that this behaviour generalizes to quantum groups.

Note that the definition of (equivalence classes of) phases in \cite{wenclass, schuch} depends, among other details, on the specification of the symmetry (group). Thus, specifying additional symmetries might change the classification.

Such a classification is also available for systems with ground state degeneracy \cite{schuch}, but again only for the case where the symmetry is described in terms of groups. Thus it would be good to understand the ground state degeneracy for the examples with a non--trivial torus partition function. This calls for a generalization of the classification \cite{schuch} to the case of symmetries described by quantum groups or representation categories. In particular the notion of subgroups needs to be generalized to quantum groups. In the next section we will point out a relation to the theory of anyon condensation, whose formulation in terms of category theory provides such a generalization \cite{kirillov,bais1,kong}.

\section{Relation to anyon condensation}\label{condense}

Here we will point out an interpretation of the fixed points as describing the condensation of anyons into an effective vacuum \cite{bais1,kong,bais2}.  This also explains of how to derive a fusion category from the $SU(2)_k$ categories, see the recent \cite{bais2}. For this, i.e. the computation of the new (effective) $[6j]$ symbols, one needs so-called Vertex Lifting Coefficients \cite{bais2}. It turns out, that the fixed point models give these Vertex Lifting Coefficients for the new effective vacuum.

Anyons are characterized by charges, in this case by the representations $j$. The neutral charge is given by the representation label $j=0$. The anyon dynamics is described by fusion and splitting of charges into new charges, as given in our model with bare vertices.  Additionally anyons can braid around each other. 

For anyon condensation one imagines that by some physical process, several anyon types condense to a new vacuum. This new vacuum corresponds to a (neutral) representation in a new fusion category, that can be understood to arise as a quotient from the original category.

As we will argue in the following, we provide here the physical process, namely the Hamiltonian (\ref{100}), that makes a number of anyon charges to condense into a new vacuum.

 There are several consistency conditions on the effective vacuum, for the construction of a derived fusion category to make sense. To explain this, let us denote the new vacuum representation by a dashed line, representing a direct sum of all the representation spaces included into the effective vacuum. This vacuum can also fuse with itself, which then is given as a combination of all the fusions between the representations making up the new vacuum:
 \ba
 \begin{tikzpicture}
\draw[dashed] (-0.3,-0.3) --(0.0,-0.0);
\draw[dashed] (0.3,-0.3) --(0.0,-0.0);
\draw[dashed] (0.0,-0.0) --(0.0,0.4);
\node at (3,0.1) {
\begin{tikzpicture}
\node at (-2.7,0) {$ \q=\q \sum_{j_1,j_2,j_3} a(j_1,j_2,j_3)$};
\draw[thick] (-0.3,-0.3) --(0.0,-0.0);
\node[left] at (-0.2,-0.2) {$j_1$};
\draw[thick] (0.3,-0.3) --(0.0,-0.0);
\node[right] at (0.2,-0.2) {$j_2$};
\draw[thick] (0.0,-0.0) --(0.0,0.4);
\node[right] at (0.0,0.3) {$j_3$};
\end{tikzpicture}};
\end{tikzpicture}
 \ea
and similarly for the splitting. Here we used the same coefficients $a(j_1,j_2,j_3)$ ( and $a'(j_1,j_2,j_3)$ for the splitting vertex) as for our (fixed point) models. 

For the new vacuum to be consistent one should be free to re--connect vacuum lines, which gives (part of) the 2--2 move condition (\ref{22pe}):
\ba\label{vacuumre}
\begin{tikzpicture}
\draw[dashed] (6,-0.4)--(6,0.7);
\draw[dashed](6,0.1)--(7,0.1);
\draw[dashed](7,-0.4)--(7,0.7);
\end{tikzpicture}
\q\q
\begin{tikzpicture}
\node at (-1.5,0.1){$=$};
\draw[dashed] (-0.3,-0.3) --(0.0,-0.0);
\draw[dashed] (0.3,-0.3) --(0.0,-0.0);
\draw[dashed] (0.0,-0.0) --(0.0,0.4);
%
\draw[dashed] (-0.3,0.7) --(0.0,0.4);
\draw[dashed] (0.3,0.7) --(0.0,0.4);
\draw[dashed] (0.0,0.4) --(0.0,-0.0);
\end{tikzpicture}
\ea
This identifies an associative algebra, which turns out to be a Frobenius algebra\footnote{The definition of a  Frobenius algebra $A$  includes two morphisms $\eta:  \mathbb{C}\simeq V_0 \rightarrow A$ and $\epsilon: A \rightarrow  V_0 \simeq\mathbb{C}$. Diagrammatically these are just given as two--valent vertices, where one edge represents the kinematical vacuum $V_0$ and the other the algebra $A$. As  $j=0$ edges can be omitted, these maps can be just represented by the now one-valent vertices. These signify to project the algebra $A$ to the $V_0$ component.}. As is well known 2D (lattice) topological field theories are classified by Frobenius algebras \cite{ltqft}. 

Note that there is a subtle difference between the condition (\ref{vacuumre}) and the 2--2 move conditions (\ref{22pe}). One can understand (\ref{vacuumre}) as involving summations of the condensed charges only. Whereas in the 2--2 move conditions (\ref{22}) one might also have charges $j_6$ appearing that are not condensed. This gives the (non--trivial) condition that the RHS of (\ref{22}) should be vanishing, whereas a priori, there is no such condition in (\ref{vacuumre}).

For a condensation to take place one usually needs bosonic particles, see \cite{bais1,bais2} for the subtleties to define bosons. This is a condition we did not implement and as we will see is also not satisfied for most of our fixed points. However if one does not consider braiding properties, such a condition is not needed. 

The bosonic condition implies in particular that the braiding of the vacuum with itself is trivial and the corresponding Frobenius algebra commutative:
\ba
\begin{tikzpicture}
\draw[dashed] (-0.3,-0.3) --(0.3,0.3);
\draw[dashed] (0.3,-0.3) --(0.1,-0.1);
\draw[dashed] (-0.1,0.1) --(-0.3,0.3);
\draw[dashed] (-0.3,-0.3) --(0.0,-0.6);
\draw[dashed] (0.3,-0.3) --(0.0,-0.6);
\draw[dashed] (0.0,-0.6) --(0.0,-0.9);
\node at (2,-0.3) {
\begin{tikzpicture}
\node[left] at (-0.8,0.3) {$=$};
\draw[dashed] (0.0,-0.4) --(0.0,0.4);
%
\draw[dashed] (-0.3,0.7) --(0.0,0.4);
\draw[dashed] (0.3,0.7) --(0.0,0.4);
\draw[dashed] (0.0,0.4) --(0.0,-0.0);
\end{tikzpicture} \q .
};
\end{tikzpicture}
\ea
This in particular implies that in the notion of section \ref{torus}, $R\circ P \circ R^{-1}=P$. Hence the torus partition function, given as the quantum trace of $R\circ P \circ R^{-1}=P$, should be equal to the quantum trace of $P$. The latter just equates to $\sum_j \Theta(j) d_j$. 

In section \ref{examples} we have described several examples for fixed point (families). These examples fit nicely into the classification \cite{kirillov}.

The work \cite{kirillov} gives  a classification of the commutative associative algebras that can be obtained as subalgebras of the fusion category $SU(2)_k$. (See \cite{ostrik} for the case with general quantum deformation parameter $q$.) Such algebras are classified by  Dynkin diagrams, as are finite subgroups of $SU(2)$ (in this case the Dynkin diagrams are actually affine) via the McKay correspondence. We find almost all the cases discussed in \cite{kirillov} (see Table 1 in this reference): The diagram $A_n$ corresponds to the trivial fixed points, where only $j=0$ appears. Algebras with diagram $D_{2l+2}$ arise from level $k=4l$ and correspond to the fixed points in section \ref{Z2}. Diagram $E_6$ corresponds to the $(k=10,j_1=3)$ case in section \ref{tworep}. Indeed all these examples give a consistent result for the partition function of the torus.

As is discussed in \cite{kirillov} the diagram $E_{7}$ cannot be realized. We however find a corresponding fixed point in section \ref{symmm} with $k=16$. The partition function for the torus shows that indeed in this case the vacuum has non--trivial self--braiding. Indeed \cite{kirillov} conjectures that for $E_7$ one needs to consider a non--commutative Frobenius algebra.    
The case we did not encounter here\footnote{Additionally we do not have the one family $T_n$ discussed in \cite{kirillov}, as these require half spin representations. In any case this family does not lead to a commutative Frobenius algebra.}, is $k=28$, where representations $j=0,5,9,14$ condense into a vacuum, giving the diagram $E_8$. 

As we have a large set of fixed point models with non--trivial braiding, leading to non--commutative algebras, it will be interesting to extend this classification also to these cases. Indeed this allows to characterize all possible fixed points in this this class of models, as will be shown in  \cite{bw}. Whereas the braiding properties are essential for anyons, there might be interesting applications where one is not interested in the braiding aspects and hence a non--commutative algebra will suffice. Furthermore this allows to interpret the CDL fixed points in the framework of anyon condensation.  Here a Cooper pair of charges $J/2\otimes J/2$ make up the vacuum condensate.

We so far discussed only the new effective vacuum. \cite{kirillov,bais1,bais2} explains how to construct the other effective charges and the full `effective' fusion categories. In particular the vertex lifting coefficients are needed to find the new $[6j]$ symbols for this new fusion category. \cite{bais2} provides consistency conditions in order to find also the  coefficients involving other charges than the vacuum. It will be interesting to see, whether a systematic derivation via recursion relations, as we provided here for the vacuum vertices, is possible.

This relation to anyon condensation is very interesting also for quantum gravity for the following reason. The notion of the kinematical vacuum for loop quantum gravity is based on the trivial representation $j=0$ in a way very similar to anyon dynamics. In particular any state can be refined by adding $j=0$ lines, without changing the physical meaning of this state. This allows to embed any given state into the continuum theory via  the concept of cylindrical consistency, see for instance \cite{al,bd12b} and references therein.  

One would however expect a second notion of a `physical' vacuum, that includes geometric excitations connected to representations $j\neq 0$. Similarly as for anyons, such a geometric vacuum is envisaged to condense from a number of non--trivial representations, for instance all representations satisfying the conditions set by the so--called simplicity constraints \cite{alexreview}. Such a new vacuum should also lead to a new way to refine given states by adding lines representing the new vacuum. Such added vacuum lines should not alter the physics encoded by the state. Indeed this leads to  consistency conditions  \cite{bais2} for the vertices involving  charges, not condensed into the new vacuum. An algebraic formulation in terms of tensor categories has been given in \cite{kirillov}, leading to a classification of finite subgroups in $U_q(sl(2,\mathbb{C})$.

\begin{figure}[bt]
\begin{center}
       \includegraphics[scale=0.3]{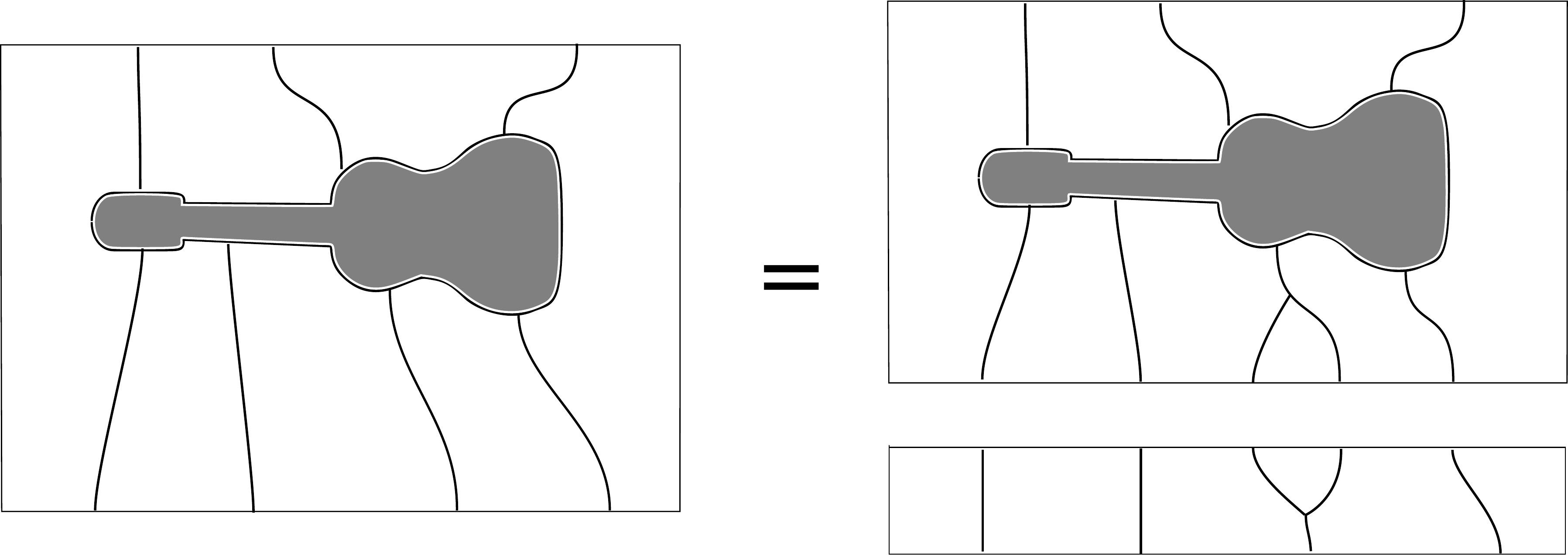}
    \end{center}
    \caption{\small \label{cc} Consider an intertwining map $\iota_{bt}: {\cal H}_b \rightarrow {\cal H}_t$ defined by a fixed point partition function. Note that the choice of interpolating graph, here represented by a guitar shaped box, does not matter due to the fixed point conditions. Applying the refinement map (\ref{refine}): $\rho_{bb'}: {\cal H}_b\rightarrow {\cal H}_{b'}$ on some factor in ${\cal H}_b$ and then using the definition by the same fixed point partition function for the intertwiner map between the refined states and ${\cal H}_t$ denoted by $\iota_{b't}: {\cal H}_{b'} \rightarrow {\cal H}_t$ does lead to the same intertwining map, i.e. $\iota_{bt}= \iota_{b't} \circ \rho_{bb'}$. As can be seen from the picture this follows easily from the bubble move invariance (vertices in this picture are `fat vertices').}
\end{figure}

This is similar to the notion of dynamical cylindrical consistency as discussed in \cite{bd12b}. One can check that, due to the triangulation invariance, the following refinement map (here acting on the bottom Hilbert space ${\cal H}_b$)
\ba\label{refine}
\begin{tikzpicture}
\draw[dashed] (-0.3,0.3) --(0.0,-0.0);
\draw[dashed] (0.3,0.3) --(0.0,-0.0);
\draw[dashed] (0.0,0.0) --(0.0,-0.4);
\node[right] at (0.4,0.0) {
\begin{tikzpicture}
\node[right] at (-0.8,0.0) {$=$};
\draw[thick] (-0.3,0.3) --(0.0,-0.0);
\draw[thick] (0.3,0.3) --(0.0,-0.0);
\node at (-0.2,-0.0)  {$\bullet$};
\draw[thick] (0.0,0.0) --(0.0,-0.4);
\end{tikzpicture}
};
\end{tikzpicture}
\ea
makes the partition function (\ref{defpart}) for the corresponding fixed point model cylindrically consistent, as defined in \cite{bd12b}. Figure \ref{cc}  gives a pictorial explanation. The effective fusion category and its effective charges will provide a basis for cylindrically consistent observables in the theory, a notion that we will explore elsewhere.

\section{Discussion and Outlook}\label{discuss}

We introduced a class of 2D models describing intertwiner degrees of freedom and identified families of triangulation invariant models inside this class. To this end we presented  techniques for finding triangulation invariant models and derived a set of recursion relations that allows the construction of such 2D models. Here the invariance under 2--2 Pachner moves, also known as crossing symmetry turned out to be essential. For a certain family of fixed points we showed how these models arise from the 3D Turaev--Viro model. 

The fixed point models admit different interpretations, depending on the context one is interested in. Taken as intertwiners for spin foam models, the fixed points implement simplicity constraints in the sense that a certain set of representations is not allowed to appear. This notion will be further investigated in \cite{toappearBMS}. The family of models with maximal spin $J$ lead to generalized AKLT states and admit a generalization to the $q=1$ and  $q \in \mathbb{R}$ case. Finally we can interpret the models as providing a means for anyon condensation, where the representations leading to non--vanishing amplitudes, condense into a new effective vacuum. 

A question we have only touched upon, is the classical (large spin $j$) limit. One would expect that a classical interpretation might be only possible for families that admit a $q \rightarrow 1$ limit, which also allows for arbitrary large spin $j$'s to appear. For the family with maximal spin $J$ we found such a classical limit, provided by the asymptotics of the $\{6j\}$ symbol. This gives a relation of the amplitudes to the (cosine of the) Regge action for specific flat ($q=1$) or curved ($q\neq 1)$ tetrahedra. An alternative means is provided by the Fourier transform, for possible generalization to the quantum group case, see \cite{biedenharn}. It gives the partition functions as  integrals over variables on the sphere $S^2=SU(2)/U(1)$, attached to the edges of the dual graph.  This might lead to an interpretation of the models in terms of an action principle, which simplifies  for instance the consideration of perturbations around the fixed points.  Such perturbations can describe propagating degrees of freedom. That would be in particular interesting for defining a dynamics for the effective charges on top of the effective vacuum.

We encountered different classifications, one via the relation of topological lattice field theories to the MPS framework and another one via the notion of anyon condensation. The classification \cite{wenclass,schuch} for the MPS framework requires so far classical groups, here it would be interesting to extend this to quantum groups, for instance via the categorical framework, developed for anyon condensation \cite{kirillov,bais1,bais2,kong}. Also, as we have examples where the condensed vacuum shows non--trivial self--braiding, corresponding to non--commutative Frobenius algebras. These cases can also be classified \cite{bw}.
This will enlighten the cases with ground state degeneracy and allow to understand the kind of  symmetry responsible for this degeneracy. Let us mention \cite{ltqft}, which shows that 2D topological lattice field theories are classified by non--commutative Frobenius algebras, but that the continuum limit requires the restriction to the centre of this algebra. This would then provide another means to classify phases described by MPS. As we will discuss below, it would be also interesting to develop a classification of phases, taking diffeomorphism symmetry into account. Such a symmetry would typically be more non--locally realized than the on--site symmetries discussed so far in  \cite{wenclass,schuch}.

The description of anyon condensation in categorical language leads to the problem of classifying module categories over $SU(2)_k$ representations, which turn out to be described by the A--D--E Dynkin diagrams \cite{kirillov}. Interestingly such diagrams also classify conformal field theories \cite{capelli,kirillov}. This link could be even made more concrete: integrable restricted  solid--on--solid (RSOS)  models can again be classified by Dynkin diagrams and lead in the continuum theory to conformal field theories \cite{pasquier}. Such models can be expressed as intertwiner models \cite{witten}. Indeed also the golden chain model (an anyon fusion model with $k=3$)  can be transformed into an RSOS model \cite{anion}.


The investigation of the models introduced here was motivated by the question of how many fixed points we can expect in spin net models \cite{toappearBMS}, which themselves are analogue models of spin foams. We see that there is a rich fixed point structure for these models, which opens up interesting prospects for the continuum limit of spin foams. We also argued that we can expect phase transitions between phases. First of all between all phases with differing ground state degeneracy, secondly between the even $J$ and odd $J$ members of the family with maximal representation $J$. Another question, not touched upon in this work, is whether one flows to these models, starting from a given fixed initial space of models. This will answer whether these fixed points lead to phases or whether some fine tuning (as actually observed in \cite{bdetal13}) is necessary.  Related to this we will investigate the stability of the fixed points, i.e. a linarization of the coarse graining flow equations.

For the full spin foam models, we have to investigate the structure group $SU(2)\times SU(2)$ or the corresponding quantum groups. One choice of simplicity constraints for these models is parametrized by the Barbero--Immirzi parameter $\gamma$ \cite{eprl,alexreview}. Here the mentioned relation of fixed points to spin foam intertwiners might give constraints on this parameter: the question will be whether we can find a fixed point family for arbitrary Barbero--Immirzi parameter or not.


Finally, as the original motivation for the models has been a gravitational theory, and the models indeed allow for a geometric interpretation, let us discuss the role of diffeomorphism symmetry, which is crucial for general relativity \cite{bd08,ditt}.

The models and the recursion relations illustrate nicely an important point for the construction of diffeomorphism invariant models. This  is that a single vertex amplitude needs to encode both microscopic as well as macroscopic physics, as the boundary variables (here representation labels) can be chosen to be small or large \cite{bd12a}. Such a requirement leads to stringent conditions -- which are exactly the recursion relations. These can be interpreted as obtaining the amplitude for a `larger' vertex (or dual triangle) from gluing smaller vertices (or dual triangles). On this note we remark that the family of models with maximal spin $J$ is uniquely determined by the amplitude $a(1,1,1)$ for the smallest possible non--degenerate triangle. This is similar to 1D discretizations, where the continuum limit is determined by a few couplings describing the most singular terms in the short distance limit \cite{steinhaus}. Similarly we expect that the other models are uniquely determined by the amplitudes for the smallest non--degenerate triangles.

We considered here triangulation invariant models with local couplings only. These necessarily lead to topological theories. For instance figure \ref{proj1} shows that the maximal number of propagating degrees of freedom is given by the number of variables associated to one edge, irrespective of the size (number of edges) of initial or final Hilbert space. (See also \cite{phil13}, for how to define the number of propagating degrees of freedom in theories where the dimension of the Hilbert spaces changes during (discrete) time evolution.) For theories with  propagating field degrees of freedom we need to allow non--local couplings, as naturally produced by coarse graining, see for instance \cite{he}. Alternatively, as discussed in \cite{bd12b}, one can introduce more complicated building blocks, allowing for more boundary data. This is very natural if working with tensor network methods. The requirement of triangulation invariance is then replaced by a notion based on cylindrical consistency and allows a precise notion of discretization independence, and hence of the continuum limit, of the theory, see \cite{bd12b,loops}. In future work we will investigate the question whether we can derive recursion relations for transition amplitudes from these requirements.

As is argued in \cite{bd08,bahretal09,steinhaus} discretizations of diffeomorphism invariant theories typically lead to a violation of diffeomorphism symmetry. One way to regain these symmetries is via coarse graining \cite{bahretal09,bd12a}. Indeed the fixed points realize diffeomorphism symmetry, in the sense of being triangulation invariant. This has been argued to be equivalent to diffeomorphism symmetry, for instance in \cite{steinhaus,bojowaldperez}. The corresponding constraint equations are provided by the Hamiltonian constraints (\ref{proj3},\ref{100}) with (\ref{proj2}) realizing the projector on physical states.

As shown in \cite{steinhaus} triangulation invariance can be seen as equivalent to diffeomorphism symmetry in particular because diffeomorphism symmetries in discrete models are realized as vertex translations. To explain this we need a model with a geometric interpretation, i.e. in our case we can interpret the representations $j$ to encode the lengths of the edges in the dual triangulation. Shifting the vertex in such a triangulation will induce a change of the edge length. If vertex translations are realized as symmetry, this shift, and hence change of representation labels, will not change the weights in the partition functions. 

This notion is well established for the 3D Ponzano Regge model  \cite{cranebarrett,bonzomfreidel,louapre} and leads to recursion relations \cite{bonzomfreidel} similar to ours, as well as the notion of (infinitesimal) Hamiltonian and diffeomorphism constraints. The difference to the constraints (\ref{proj3}) is that these are projectors and hence should be seen as exponentiated constraints. For the infinitesimal  constraints one can consider the algebra, which should lead to a representation of a discrete hypersurface deformation algebra \cite{vbbd}. In 2D this algebra is equivalent to a Virasoro algebra (with vanishing central charge). It would be interesting to derive these constraints and the corresponding algebra. This could then give another way to classify the fixed points, as each fixed point should lead to a representation of this hypersurface deformation algebra. Here the relation of the maximal spin $J$ models to the Tuarev--Viro and Ponzano--Regge models is particular helpful, as the 2D recursion relations and algebra should follow from the 3D algebra (that is actually not available yet in the quantum case, but derived for the classical theory of flat and curved tetrahedra in \cite{vbbd}). We expect that a $q$--deformed version \cite{biedenharn} of the $U(N)$ formalism developed in \cite{etera} will provide the Hamiltonian operators.  

The hypersurface deformation algebras correspond to  Virasoro algebras with vanishing central charge. Virasoro algebras with non--vanishing charges describe  theories with propagating degrees of freedom, hence it would be useful to understand these algebras and to compare with possible discretizations of the case with non--vanishing charges. This might allow a generalization of the recursion relations to theories with propagating degrees of freedom.

A further aspect to investigate in these models is  whether violations of diffeomorphism symmetry generate irrelevant or relevant couplings. Indeed one would rather expect the latter. As diffeomorphism symmetry is generically violated by discretizations, one cannot start with an initial phase space of diffeomorphism invariant models. These are rather expected to appear as fixed points of the coarse graining flows -- i.e. either leading to topological theories, or in case one tunes to a phase transition, a theory with propagating degrees of freedom. Fine tunings might then be interpretable as avoiding diffeomorphism violating couplings. This holds in particular for the fine tuning of face weights for spin foams -- here invariance under subdivision of faces (which fixes the face weights) can be understood as a weak form of discretization independence \cite{bojowaldperez} and hence diffeomorphism symmetry.  In this sense the definition of phase, which depends on the chosen symmetries \cite{wenclass}, is not clear cut, as we are lacking a phase space of diffeomorphism invariant initial lattice models. The models introduced here might provide an ideal set of examples to study these issues. A possible general line of attack is provided by \cite{fendly}, finding different behaviour for perturbations violating either 2--2 or 3--1 move invariance.

Another prospect is to generalize the technique of deriving recursion relations from the requirement of triangulation invariance to higher dimensions. In some sense such a framework is provided by category theory, where the requirements of triangulation invariance (i.e. the Pachner moves) are encoded in certain properties (i.e. associativity) for the category.  Indeed this leads to the construction of Tuarev--Viro models. However, using directly the recursion relations might allow more flexibility, as the category framework in 4D leads to the notion of 2--categories, which from a physical viewpoint appear as quite rigid \cite{girelli}. 
\appendix

\section{$SU(2)_k$ basics}\label{appA}

A thorough introduction to the quantum deformation of the rotation group can be found in \cite{biedenharn,yellowbook}. Our notation and conventions will follow \cite{biedenharn}. Here we will just give the definitions for the basic objects needed for our work.

Firstly, we need to introduce quantum numbers, denoted by square brackets $[n]$ where $q\in \mathbb{R}/\{0\}$ or $q$ a root of unity
\ba
[n]&:=& \frac{q^{\frac{n}{2}}-q^{-\frac{n}{2}}  }{q^{\frac{1}{2}}-q^{-\frac{1}{2}}  }    \q .
\ea
We employ the following identities \cite{biedenharn} for the $q$--numbers
\ba\label{qind1}
[a][b-c]+[b][c-a]+[c][a-b]&=& 0 
\ea
and
\ba\label{qind2}
 [a]-[b] &=& \frac{[a+b][\tfrac{1}{2}(a-b)]}{[\tfrac{1}{2}(a+b)]}\q .
\ea

 The finite dimensional representations of $SU(2)_q$ are, as for the classical case, labelled by half integers $j$ and can be defined on $(2j+1)$ dimensional representation spaces $V_j$.    The so--called quantum dimensions are given by
\ba
d_j=[2j+1]  \q .
\ea

For $q$ a root of unity, $q=\exp(\frac{2\pi}{(k+2)}i)$ the quantum numbers become periodic 
\ba
[n]&=&\frac{\sin(\frac{2\pi i n}{2k+4})}{\sin(\frac{2\pi i}{2k+4})}  \q ,
\ea
having zeros at $n=0$ and $n=k+2$. Thus $j=\frac{k}{2}$ with $d_{k/2}=1$ is the last representation with a strictly positive quantum dimension. Representations $j=0,\frac{1}{2},\ldots,\frac{k}{2}$ are called admissible, representations $j>\frac{k}{2}$ are of so called quantum trace zero. The number $k\in\mathbb{N}$ is referred to as level.

We will need furthermore invariant tensors between representation spaces, where the basic (unique) intertwiner is between triplets of representation spaces. This leads to the Clebsch Gordan tensors, which we denote diagrammatically as
\ba\label{d1app}
\begin{tikzpicture}
\draw[thick] (-0.3,-0.3) --(0.0,-0.0);
\node[left] at (-0.2,-0.2) {1};
\draw[thick] (0.3,-0.3) --(0.0,-0.0);
\node[right] at (0.2,-0.2) {2};
\draw[thick] (0.0,-0.0) --(0.0,0.4);
\node[right] at (0.0,0.3) {3};
\node at (1.8,0.0) {$ \q=\q{}_q\cc^{j_1j_2j_3}_{m_1m_2m_3}$};
\end{tikzpicture}
\q\q ,\q\q
\begin{tikzpicture}
\draw[thick] (-0.3,0.3) --(0.0,-0.0);
\node[left] at (-0.2,0.2) {2};
\draw[thick] (0.3,0.3) --(0.0,-0.0);
\node[right] at (0.2,0.2) {1};
\draw[thick] (0.0,0.0) --(0.0,-0.4);
\node[left] at (0.0,-0.3) {3};
\node[right] at (0.3,0.0) {$\q=\q{}_{q^{-1}}\cc^{j_1j_2j_3}_{m_1m_2m_3}$};
\end{tikzpicture}
\ea
We use rescaled Clebsch Gordans with ${}_q{\cal C}^{j_1j_2j_3}_{m_1m_2m_3}= (d_{j_3})^{-1/2}  \,{}_qC^{j_1j_2j_3}_{m_1m_2m_3}$ for the diagrammatical calculus. The Clebsch--Gordan's satisfy the following orthogonality relation
\ba\label{orthogonal}
\sum_{m_1,m_2} {}_qC^{j_1j_2j_3}_{m_1m_2m_3} \,\, {}_qC^{j_1j_2j'_3}_{m_1m_2m'_3}&=& \delta_{m_3m'_3}\delta_{j_3j'_3} \q .
\ea

The coupling coefficients are only non--vanishing if the following conditions hold
\ba
j_I+j_K &\geq& j_L \q\text{for all permutations} \; \{J,K,L\} \;\text{of}\; \{1,2,3\} \q ,\nn\\
j_1+j_2+j_3 &=& 0\!\mod 1 \q , \nn\\
j_1+j_2+j_3 &\leq& k \q .
\ea
Note the last condition, that is special to the quantum deformed case at root of unity. This condition signifies that  $V_1\otimes V_2$ can include trace zero parts. These trace zero parts can be modded out \cite{yellowbook}. Some equations (for instance the one defining the $[6j]$ symbol) will however hold only modulo such trace zero parts. In particular we have the completeness relation
\ba\label{complete}
\sum_{m_3,\, j_3 \, \text{admiss.}} {}_qC^{j_1j_2j_3}_{m_1m_2m_3} \,\, {}_qC^{j_1j_2j_3}_{m'_1m'_2m_3}&=&\Pi^{j_1 j_2}_{m_1m_2\,,m'_1m'_2}
\ea
where $\Pi^{j_1 j_2}_{m_1m_2\,,m'_1m'_2}$ projects onto the complement of the trace zero part in $V_{j_1}\otimes V_{j_2}$

The Clebsch Gordan coefficients satisfy a number identities under permutation of arguments \cite{biedenharn}, in particular ${}_qC^{j1j2j3}_{m_1m_2 m_3}=(-)^{j_1+j_2-j_3}{}_{\bar q}C^{j_2j_1j_3}_{m_2m_1m_3}$. With these identities we can rewrite (\ref{complete}) into the diagrammatic equation
\ba\label{complete2}
\begin{tikzpicture}
\node[left] at (-0.7,0.2){$ \Pi^{j_1j_2} \,=\,\,\sum_{j \, \text{admissable}}(-)^{j_1+j_2-j} \,[2j+1]$};
\node[right] at (0.2,0.6) {$j_2$};
\node[left] at (-0.2,0.6) {$j_1$};
\node[left] at (-0.2,-0.3) {$j_1$};
\node[right] at (0.2,-0.3) {$j_2$};
\draw[thick] (-0.3,-0.3) --(0.0,-0.0);
\draw[thick] (0.3,-0.3) --(0.0,-0.0);
\node[right] at (0,0.2) {$ j$};
\draw[thick] (0.0,-0.0) --(0.0,0.4);
%
\draw[thick] (-0.3,0.7) --(0.0,0.4);
\draw[thick] (0.3,0.7) --(0.0,0.4);
\draw[thick] (0.0,0.4) --(0.0,-0.0);
\end{tikzpicture} \q .
\ea
and similarly for (\ref{orthogonal}):
\ba
\begin{tikzpicture}
\draw[thick] (0,0) --(0,0.3);
\draw[thick] (0,0.3) --(-0.2,0.5);
\draw[thick] (-0.2,0.5)--(-0.0,0.7);
\draw[thick] (0,0.3)--(0.2,0.5);
\draw[thick] (0,0.7)--(0.2,0.5);
\draw[thick] (0,0.7)--(0,1);
\node[left] at (-0.2,0.5) {2};
\node[right] at (0.2,0.5) {1};
\node[right] at (0.0,0.9) {3};
\node[right] at (0.0,0.1) {3};
\node at (4,0.5) {$=\, \Theta(j_1,j_2,j_3)\, \frac{1}{[2j_3+1]} (-)^{j_1+j_2-j_3}$};
\draw[thick] (7,0)--(7,1);
\node[right] at (7.2,0.5){3} ;
\end{tikzpicture}
\ea
where $\Theta(j_1,j_2,j_3)=1$ if the coupling $(j_1,j_2,j_3)$ is allowed and vanishing otherwise.

We have two further structures, unit or cup: $\mathbb{C}\rightarrow V_j\otimes V_j$  and co--unit or cap: $ V_j\otimes V_j  \rightarrow\mathbb{C}$ , that allow to obtain closed diagrams. These are given as
\ba
\begin{tikzpicture}
\draw[thick] (-0.4,0.4)--(0,0);
\draw[thick] (0.4,0.4)--(0,0);
\node[left] at (-0.4,0.4) {$m$};
\node[right] at (0.4,0.4) {$m'$};
\node[left] at (0.2,-0.2) {$j$};
\node[right] at (1.2,0.2) {$=(-)^{j+m} q^{ \frac{m}{2} } \delta_{m,-m'}$};
\end{tikzpicture}
\q,\q\q
\begin{tikzpicture}
\draw[thick] (-0.4,-0.4)--(0,0);
\draw[thick] (0.4,-0.4)--(0,0);
\node[left] at (-0.4,-0.4) {$m$};
\node[right] at (0.4,-0.4) {$m'$};
\node[left] at (0.2,0.2) {$j$};
\node[right] at (1.2,-0.2) {$=(-)^{j-m} q^{ \frac{m}{2} } \delta_{m,-m'}$};  \q ,
\end{tikzpicture}
\ea 
where we made the magnetic indices (labelling the basis elements in $V_j$) explicit.
Cap and cup are inverse to each other, we have for instance
\ba
\begin{tikzpicture}
\draw[thick] (-0.4,0.4)--(0,0);
\draw[thick] (0.4,0.4)--(0,0);
\draw[thick] (0.4,0.4)--(0.8,0);
\node at (1,0.2) {$=$};
\draw[thick] (1.4,0)--(1.4,0.4);
\end{tikzpicture} \q .
\ea

\begin{figure}[bt]
\begin{center}
       \includegraphics[scale=0.25]{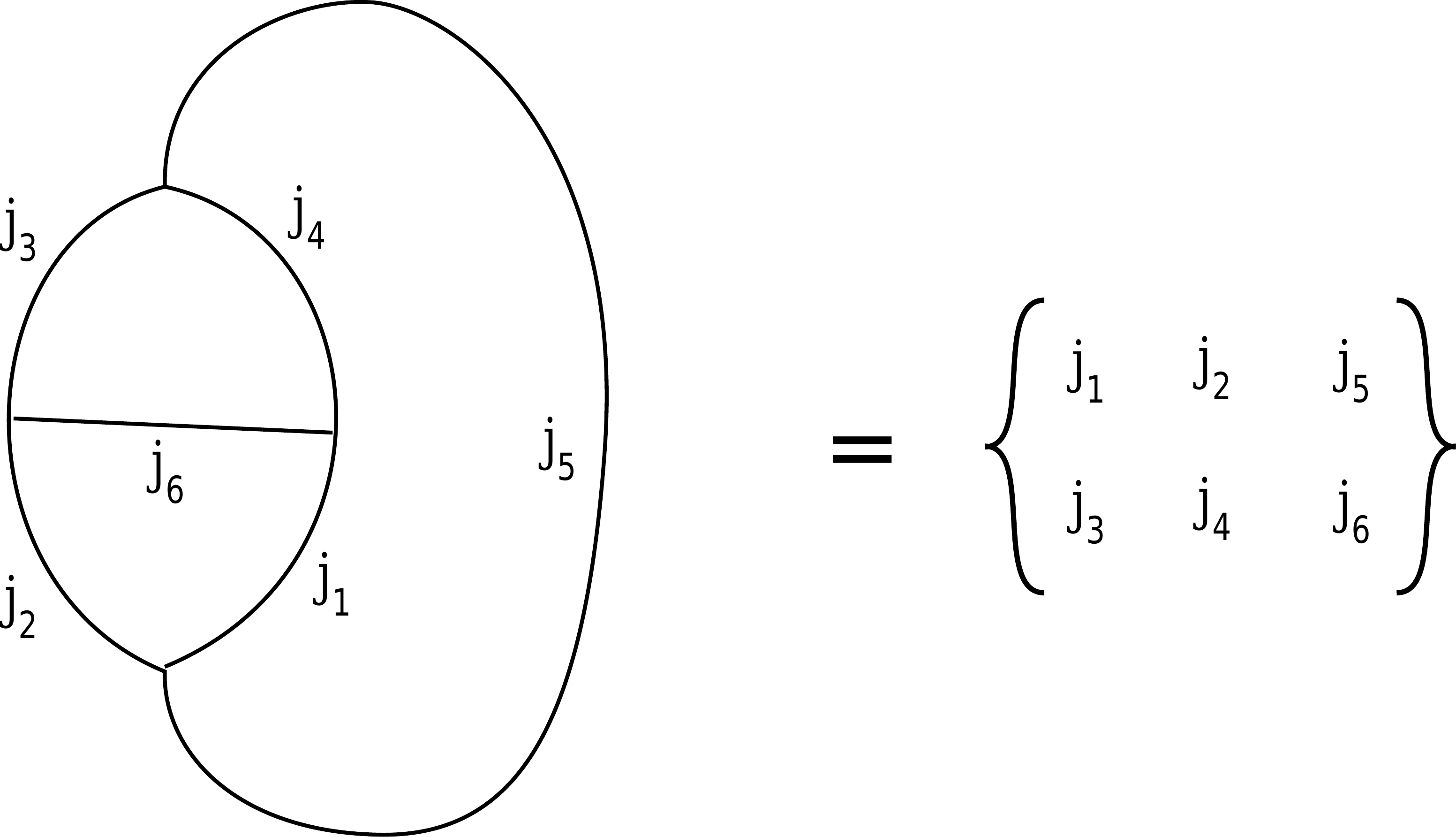}
    \end{center}
    \caption{\small \label{fig6j} The $\{6j\}$ symbol in diagrammatical calculus.}
\end{figure}

Cap and cup also allow to define the quantum trace. We will take the quantum trace to the right of the part of the diagram describing the intertwining map, see for instance the diagram on the left in figure \ref{tor3}. A particular example for a closed diagram is the $\{6j\}$ symbol in figure \ref{fig6j}, 
where \cite{biedenharn}
\ba\label{6jb}
\Tj{j_1}{j_2}{j_5}{j_3}{j_4}{j_6}&=&\Fj{j_1}{j_2}{j_5}{j_3}{j_4}{j_6}  (-)^{j_1+j_2+j_3+j_4} \frac{1}{\sqrt{[2j_5+1][2j_6+1]}}\nn\\
&=&
\Delta(j_1j_2j_3)\Delta(j_3j_4j_5)\Delta(j_1j_4j_6)\Delta(j_3j_2j_6)\nn\\
&&\sum_n  \frac{(-)^n[n+1]!}{ [n-j_1-j_2-j_5]! \,[n-j_1-j_4-j_6]! \,[n-j_2-j_3-j_6]! \,[n-j_3-j_4-j_5]! } \times \nn\\
&&\q\q \frac{1}{[j_1+j_2+j_3+j_4-n]!\,[j_1+j_3+j_5+j_6-n]!\, [j_2+j_4+j_5+j_6-n]!} \q .\q\q
\ea
The summation is over integers with
\ba
&&
\max(j_1+j_2+j_5,j_1+j_4+j_6,j_2+j_3+j_6,j_3+j_4+j_5)\,\leq n\leq\nn\\ &&\q\q\q\q\q\q \min(j_1+j_2+j_3+j_4,j_1+j_3+j_5+j_6,j_2+j_4+j_5+j_6 ) \q ,
\ea
and $\Delta(j_1j_2j_3)$ is given as
\ba
\Delta(j_1j_2j_3)&=& \left(       \frac{[j_1+j_2-j_3]! \, [j_1-j_2+j_3]! \, [-j_1+j_2+j_3]! }{[j_1+j_2+j_3+1]! }         \right)^{\frac{1}{2}}
\ea
with $[n]!=\prod_{m=1}^n [m]$.  
The $\{6j\}$ symbol is invariant under the symmetry group of the tetrahedron.

\section{Proof of summation identity}\label{appb}

Here we will prove the evaluation of the sum
\ba
\sum^j_{m=1} [2m+1]\frac{[m]^2[m+1]^2}{[2m]^2[2m+2]^2} &=& \frac{[j][j+1]^2[j+2]}{[2j+2]^2[2]^2} \q .
\ea
After verifying the induction hypothesis for $j=1$ the induction step $(j-1)\rightarrow j$ amounts to the claim
\ba
[2j]^2[j+1][j+2]-[2j+2]^2[j-1][j]&=&[2]^2[2j+1][j][j+1] \q .
\ea
After applying
\ba\label{auxapp}
\frac{[2j][j+2]}{[j]}\,\,=\,\,[2j+2]+[2] \q,\q\q \frac{[2j+2]}{[j+1]} \,\,=\,\, [j+2]-[j]  \q,\q\q [2][2j+1]\,\,=\,\,[2j+2]+[2j]
\ea
we arrive at
\ba
[2j+2] \,\,( [2j]- [2]+[j][j-1] - [j+2][j-1]) &=&0
\ea
which can be shown to hold with the help of the same identities (\ref{auxapp}).

\section{Further 2--2 move relations}

Here we will shortly discuss the 2--2 move between
$A_6(j,l,j+2l-1,l,i_6)$ and $A_5(j,l,j+2l-1,l,i_5)$. This move might be needed if some or all amplitudes with $j=1$ arguments are vanishing. (However it will in general not lead to closed recursion relations.)

For $1\leq j\leq \tfrac{k+1}{2}-2l$ we have a two--dimensional intertwiner space: $i_5,i_6$ can both take the values $j+l-1,j+l$. The matrix $\tilde M$ defined in (\ref{matrix1}) is given by
\ba\label{a49}
\tilde M_{i_6i_5} &\,=\,&  \Fj{j}{l}{i_5}{j+2l-1}{l}{i_6} \sqrt{ \frac{  d_{i_6}}{d_{i_5}}} \sigma(i_5,j,l) \sigma(j+2l-1,i_5,l)  \nn\\
&\,=\,& M_{i_6i_5}\sqrt{ \frac{  d_{i_6}}{d_{i_5}}} \sigma(i_5,j,l) \sigma(j+2l-1,i_5,l)   \q .
\ea

The vector on which this $\tilde M$ is acting is given by
\ba
\tilde v= \begin{pmatrix}
a(j+2l-1,j+l-1,l) \, a'(j,j+l-1,l) \\
a(j+2l-1,j+l,l)\,a'(j,j+l,l)  \q .
\end{pmatrix}
\ea

The matrix $M$ is of the same form as (\ref{33}) with the parameter $a$ given by
\ba
a&=& \frac{[2l]}{[2j+2l] }\q .
\ea
Thus $M=\mathbb{I}-2vv^t$ where $v_-=\frac{1}{\sqrt{2}}(\sqrt{(1+a)},-\sqrt{(1-a)})$ is the (normalized) eigenvector with eigenvalue $-1$. 

The signs in (\ref{a49}) can lead to a multiplication of the rows of the matrix with $(-1)$. However  a relative sign between rows leads only to eigenvalues $+1$ if $a=1$. But $a=1$ only occurs for $j= \tfrac{k+2}{2}-2l$, which is outside the range of allowed values for $j$, as $j\leq \tfrac{k+1}{2}-2l.$
Thus we can conclude that for $1\leq j \leq  \tfrac{k+1}{2}-2l$ we have (if all related amplitudes are non--vanishing)
\ba
\sigma(j+l-1,j,l)\,\sigma(j+l-1,j+2l-1,l)\,=\,\sigma(j+l,j,l)\,\sigma(j+l,j+2l-1,l) \q .
\ea
Note that for $j=1$ we obtain $\sigma(j+l-1,j,l)\,\sigma(j+l-1,j+2l-1,l)=\sigma_1 \sigma_{2l}$. We have to distinguish again two cases $\sigma_1 \sigma_{2l}=+1$ and $\sigma_1 \sigma_{2l}=-1$. For $\sigma_1 \sigma_{2l}=+1$ we obtain
\ba
a(j+2l-1,j+l,l)\,a'(j,j+l,l) &\,=\,& \sqrt{ \frac{[2j+2l+1]([2j+2l]+[2l])}{[2j+2l-1]([2j+2l]-[2l])}} \times \nn\\&& \q\q a(j+2l-1,j+l-1,l)\,a'(j,j+l-1,l) \nn\\
&\,=\,& \sqrt{ \frac{[2j+2l+1]  \,[2j][j+2l]^2  }{[2j+2l-1]  [2j+4l][j]^2   }} \times \nn\\&& \q\q a(j+2l-1,j+l-1,l)\,a'(j,j+l-1,l)
\ea
and for $\sigma_1 \sigma_{2l}=-1$
\ba
a(j+2l-1,j+l,l)\,a'(j,j+l,l) &\,=\,&- \sqrt{ \frac{[2j+2l+1]([2j+2l]-[2l])}{[2j+2l-1]([2j+2l]+[2l])}} \times \nn\\&& \q\q a(j+2l-1,j+l-1,l)\,a'(j,j+l-1,l)\nn\\
&\,=\,&- \sqrt{ \frac{[2j+2l+1] \, [2j+4l][j]^2   }{[2j+2l-1] \,[2j][j+2l]^2 }} \times \nn\\&& \q\q a(j+2l-1,j+l-1,l)\,a'(j,j+l-1,l)\q .
\ea

\section{ $6j$ symbols for recursion relations} \label{appC}

The $[6j]$ symbols appearing in the sum (\ref{47}) are given by
\ba
 \Fj{1}{j}{j-1}{j+l}{m-1}{m} \sqrt{\frac{d_m}{d_{j-1}}} \!\!\!&\underset{l\leq m-2}{=}&\!\!\!
D_m \sqrt{\frac{[2j+m+l+1][2j+m+l][m-l][m-l-1]}{[2j+1][2j][2j-1]}} \nn\\
  \Fj{1}{j}{j}{j+l}{m-1}{m} \sqrt{\frac{d_m}{d_{j}}}  \!\!\!&\underset{l\leq m-1}{=}&\!\!\!\!
  D_m\sqrt{\frac{[2][2j+m+l+1][2j-m+l+1][m-l][m+l]}   {[2j+2][2j+1][2j]}} \nn\\
   \Fj{1}{j}{j+1}{j+l}{m-1}{m} \sqrt{\frac{d_m}{d_{j+1}}} \!\!\!&=&\!\!\!\!\!\!\!\!\!
D_m  \sqrt{\frac{[2j-m+l+2][2j-m+l+1][m+l][m+l-1]}   {[2j+3][2j+2][2j+1]}}\,. \nn\\
\ea
with
\ba
D_m=\sqrt{\frac{[2m+1]}{[2m][2m-1]}} \q .
\ea

For $l=m$ we get a  simplification as we only have one term in the sum (\ref{36}) and the $[6j]$ symbol involved is equal to one. 
Hence
\ba
a'(m,j,j+m)\,a(m,m-1,1)\,= \,\sqrt{\frac{d_m}{d_{j+1}}}\,\, a'(1,j,j+1) \,a(j+m,m-1,j+1)   \q .
\ea
For $l=m-1$ we still obtain a relatively simple relation
\ba
&&a'(m,j,j+m-1)\,a(m,m-1,1)\,= \, \nn\\  &&  \q\q
\sqrt{\frac{ [2m+1] [2m-2][2j] }{ [2m][2j+2][2j+3]}} \,\, a'(1,j,j+1) \,a(j+m-1,m-1,j+1) + \nn\\ &&\q\q  \sqrt{\frac{[2m+1][2] [2j+2m]} {[2m][2j+1][2j+2]} }\,\, a'(1,j,j) \,a(j+m-1,m-1,j) \q .
\ea

For the sum (\ref{36b}) we need
\ba
 \Fj{1}{j+l}{j+l-1}{j}{m-1}{m} \sqrt{\frac{d_m}{d_{j+l-1}}} \!\!\!&\underset{}{=}&\!\!\!
 D_m\sqrt{\frac{[2j+m+l+1][2j+m+l][m+l][m+l-1]}{[2j+2l+1][2j+2l][2j+2l-1]}} \nn\\
  \Fj{1}{j+l}{j+l}{j}{m-1}{m} \sqrt{\frac{d_m}{d_{j+l}}} \!\!\!&\underset{l\leq m-1}{=}&\!\!\!\!\!\!
  D_m\sqrt{\frac{[2][2j+m+l+1][2j-m+l+1][m-l][m+l]}   {[2j+2l+2][2j+2l+1][2j+2l]}} \nn\\
    \Fj{1}{j+l}{j+l+1}{j}{m-1}{m} \sqrt{\frac{d_m}{d_{j+l+1}}} \!\!\!&\underset{l\leq m-2}{=}&\!\!\!\!\!\!\!
D_m  \sqrt{\frac{[2j-m+l+2][2j-m+l+1][m-l][m-l-1]}   {[2j+2l+3][2j+2l+2][2j+2l+1]}}\,. \nn\\
\ea

\section*{Acknowledgements}


We are thankful for discussions with Benjamin Bahr, Aristide Baratin, Valentin Bonzom, Laurent Freidel, Mercedes Martin-Benito, Frank Hellmann, Sebastian Steinhaus and  Guifre Vidal and acknowledge especially   Oliver Buerschaper and Xiao-Gang Wen for very helpful remarks.
We thank Pablo Picasso for inspiring figure \ref{itmodel}.
This research was supported by Perimeter Institute for Theoretical Physics. Research at Perimeter Institute is supported by the Government of Canada through Industry Canada and by the Province of Ontario through the Ministry of Research and Innovation.

\end{document}